\documentclass[twocolumn]{aastex631}    
\shorttitle{Spatially Resolved Environmental Quenching at $z>1$}
\shortauthors{Henry et al.} 

\usepackage[varg]{txfonts}
\usepackage{graphicx}
\usepackage{hyperref}

\bibpunct{(}{)}{;}{a}{}{,} 

\begin{document}

\title{The GOGREEN Survey: AI Powered Deconvolution Lifts The Veil on Outside-in Environmental Quenching at $z > 1$}

\author[0000-0002-9466-2763]{Aur{\'e}lien Henry}
\affiliation{Department of Physics, University of California, Merced, 5200 Lake Road, Merced, CA 92543, USA}

\author[0000-0002-6572-7089]{Gillian Wilson}
\affiliation{Department of Physics, University of California, Merced, 5200 Lake Road, Merced, CA 92543, USA}

\author[0000-0001-5851-1856]{Gregory Rudnick}
\affiliation{Department of Physics and Astronomy, The University of Kansas, 1251 Wescoe Hall Drive, Lawrence, KS 66045, USA}

\author[0000-0002-9655-1063]{Pascale Jablonka}
\affiliation{Laboratory of Astrophysics, Ecole Polytechnique Fédérale de Lausanne (EPFL), Observatoire de Sauverny, CH-1290 Versoix,
Switzerland.}

\author[0000-0002-8474-2575]{Utsav Akhaury}
\affiliation{Laboratory of Astrophysics, Ecole Polytechnique Fédérale de Lausanne (EPFL), Observatoire de Sauverny, CH-1290 Versoix,
Switzerland.}

\author[0000-0003-1514-9652]{Craig Brooks}
\affiliation{Department of Physics and Astronomy, The University of Kansas, 1251 Wescoe Hall Drive, Lawrence, KS 66045, USA}

\author[0000-0003-4849-9536]{Michael Balogh}
\affiliation{Department of Physics and Astronomy, University of Waterloo, Waterloo, ON N2L 3G1, Canada}

\author[0000-0001-6003-0541]{Ben Forrest}
\affiliation{Department of Physics and Astronomy, University of California, Davis, One Shields Avenue, Davis, CA 95616, USA}

\author[0000-0002-9330-9108]{Adam Muzzin}
\affiliation{Department of Physics and Astronomy, York University, 4700, Keele Street, Toronto, ON M3J 1P3, Canada}

\author[0000-0003-0780-9526]{Visal Sok}
\affiliation{Department of Astrophysical and Planetary Sciences, University of Colorado, 2000 Colorado Ave, Boulder, CO 80309, USA}

\author[0000-0003-3595-7147]{Mohamed H. Abdullah}
\affiliation{Department of Physics, University of California, Merced, 5200 Lake Road, Merced, CA 92543, USA}

\author[0000-0002-6505-9981]{M.E. Wisz}
\affiliation{Department of Physics, University of California, Merced, 5200 Lake Road, Merced, CA 92543, USA}

\author[0009-0009-4795-0912]{Elias Works}
\affiliation{Department of Physics and Astronomy, The University of Kansas, 1251 Wescoe Hall Drive, Lawrence, KS 66045, USA}

\begin{abstract}

A powerful probe of the physical processes that quench star formation in dense environments is determining where within galaxies star formation is suppressed.
At high redshift, the spatial resolution of multi-band imaging limits such measurements.
We use deep-learning-based deconvolution to recover spatially resolved optical and near-infrared photometry for galaxies in nine GOGREEN clusters at $1<z<1.4$, using customized models trained on \textit{HST} and \textit{JWST} imaging.
Using resolved rest-frame $UVJ$ colors, we classify galaxies by the star-forming states of their inner and outer regions into predominantly star-forming, predominantly quiescent, inside-quenched, or outside-quenched.
We find that 24\% of galaxies classified as quiescent from their integrated colors retain significant star formation.
The predominantly quiescent fraction increases with stellar mass and is higher in clusters than in the field while the cluster quenched fraction excess is, when limiting to predominantly quenched galaxies, approximately 20\%. 
Contrary to previous GOGREEN studies using integrated colors, we find this excess to be independent of stellar mass, demonstrating that partially quenched galaxies can bias measurements based on integrated colors.
Among galaxies retaining significant star formation, outside-quenched galaxies are substantially more common than inside-quenched galaxies and have a fraction excess of $(22.8\pm7.0)\%$ in clusters relative to the field at low masses.
This provides evidence that clusters preferentially suppress star formation in the outskirts of low-mass galaxies. 
Our results demonstrate the importance of spatially resolved classifications for interpreting environmental quenching at $z\sim1$ and the potential of deep-learning-based deconvolution to recover such information from large ground-based imaging datasets.

\end{abstract}



\section{Introduction} \label{sec:intro}

Decades of multi-passband imaging surveys have transformed our understanding of galaxy evolution, revealing a clear bimodality in the galaxy population. Galaxies are broadly divided into red, quiescent, early-type systems with little ongoing star formation and blue, actively star-forming, late-type systems (\citealt{strateva_color_2001,kauffmann_host_2003,baldry_quantifying_2004}). At intermediate redshift ($z \sim 0.7$), these populations contribute roughly equally to the cosmic stellar mass density, however, the quiescent population has grown by a factor of ten in the last 9.4 Gyrs and now dominates in the local Universe (\citealt{bell_nearly_2004,baldry_galaxy_2012, muzzin_evolution_2013,tomczak_galaxy_2014}). 
One of the main questions that has been driving research for years is how do galaxies transition from these blue spirals and actively star-forming galaxies to massive, red and passive elliptical galaxies.

A wide range of mechanisms have been proposed to explain the transition from star-forming to quiescent galaxies (e.g. \citealt{man_star_2018}). 
The main factors responsible for quenching are known to be mass and/or environment dependent (\citealt{sobral_dependence_2011,darvish_cosmic_2017,mao_revealing_2022}). Indeed, dense environments are believed to accelerate galaxy evolution, producing both more massive and more quenched systems (e.g., \citealt{de_lucia_buildup_2004, gao_age_2005, maulbetsch_dependence_2007, shankar_size_2013, kuchner_effects_2017, alberts_clusters_2022}). 
For galaxies in a dense environment, the quenching mechanisms are typically grouped into processes affecting mostly massive central galaxies, such as AGN (e.g. \citealt{bourne_black_2014,sampaio_co-evolution_2023}) or stellar feedback (e.g. \citealt{spilker_fast_2018,davies_multiphase_2026}), and those acting mostly on lower-mass satellite galaxies, including strangulation (\citealt{larson_evolution_1980}), ram-pressure stripping (\citealt{gunn_infall_1972}), and harassment (\citealt{farouki_simulations_1982}).

At high redshifts ($z > 1$), surveys like the IRAC Shallow Cluster Survey (ISCS), the FourStar Galaxy Evolution
(ZFOURGE), the Observations of Redshift Evolution in Large Scale Environments (ORELSE) and The Gemini Observations of Galaxies in Rich Early Environments (GOGREEN) demonstrated that dense environments already had a high quenching efficiency (\citealt{alberts_evolution_2014,balogh_gemini_2017,kawinwanichakij_effect_2017,lemaux_persistence_2019,balogh_gogreen_2020}). 
Moreover, while observations at lower redshift suggests that environmental quenching efficiency is independent of stellar mass (e.g. \citealt{baldry_galaxy_2006,peng_mass_2010,de_lucia_environmental_2012}), the GOGREEN and ZFOURGE surveys demonstrated that, at $z>1$, environmental quenching efficiency is strongly mass dependent (\citealt{balogh_evidence_2016,kawinwanichakij_effect_2017,van_der_burg_gogreen_2020}), and halo mass dependent (\citealt{reeves_gogreen_2021}). This implies that the mechanisms responsible for environmental quenching at high redshift are different from those affecting lower-$z$ clusters.
Additionally, \cite{chan_gogreen_2021} found morphological evidence that galaxies quenching in dense environments were being affected by different quenching mechanisms than those affecting galaxies quenching in the field.
However, while these aforementioned studies show a strong direct impact of the environment on galaxies, \cite{webb_gogreen_2020} and \cite{werner_satellite_2021} found evidence that a significant fraction of quiescent galaxies in clusters might actually have quenched before cluster infall, through preprocessing.

All of these findings demonstrate that, understanding environmental quenching at $z>1$ requires looking into what is happening inside of the galaxies that are both already in the clusters and in their infalling regions.

The goal of our study is to exploit the GOGREEN survey to gather a large number of cluster members, including in the infalling region, and study their spatially resolved $U - V$ and $V-J$ rest-frame colors to identify where within galaxies quenching occurs. 
However, while the center-most parts of these clusters have been imaged using \textit{Hubble Space Telescope (HST)} in at least one band, their infalling regions have only been imaged using ground-based telescopes with spatial resolutions incompatible with the study of the internal structure of galaxies.

The main reason for the lack of high-resolution imaging is that ground-based observations are significantly affected by atmospheric blurring, which limits spatial resolution. Even though the progress made in recent years with adaptive optics has helped mitigate the atmospheric blurring, there is still a wide gap between the spatial resolutions of ground-based optical/near-IR images ($\sim 0.5-0.8$ arcsecond) and optical/near-IR space images ($\sim 0.03 - 0.16$ arcsecond).
For decades, alternative routes that did not require expensive specialized instrumentation have been studied to improve the quality of ground-based images, especially deconvolution (e.g., \citealt{lucy_iterative_1974,hook_image_1994,bobin_sparse_2013,scherzer_starlet_2015,cantale_firedec_2016,michalewicz_starred_2023}). 

In recent years, the development of deep learning techniques has offered substantial gains in deconvolution quality and in computational efficiency by learning the properties of the sources using high-resolution space images. Particularly, the \textsc{UNet} architecture (\citealt{ronneberger_u-net_2015}) combined with a Learnlet-based approach (\citealt{ramzi_wavelets_2023}) led to the development of \textsc{SUNet} (\citealt{akhaury_deep_2022,akhaury_ground-based_2024}). The deep-learning based \textsc{SUNet} software has been shown to be successful at deconvolving real ground-based optical data, leading to a spatial resolution close to that of \textit{HST}, allowing detection of small clumps (\citealt{akhaury_ground-based_2024}). 
In this study, we trained new models for the \textsc{SUNet} deconvolution software to deconvolve and spatially resolve optical and near-IR ground-based images of the GOGREEN cluster members. For additional details on the deconvolution technique and the training process, we refer the reader to Section \ref{sec:deconv} and Appendix \ref{app:training} of this paper.


The paper is organized as follows. We describe the GOGREEN data in Section \ref{sec:gogreen} and present the selection of our samples of cluster members and field galaxies in Section \ref{sec:sample_selec}. 
As mentioned above, the deconvolution process is detailed in Section \ref{sec:deconv}. 
For readers interested in the scientific analysis, we describe the photometric analysis of our galaxies in Section \ref{sec:color_meas} while our results are described in Section \ref{sec:results}.
More precisely, in Section \ref{subsec:spatial_colors} we describe how we found that a large portion of unresolved quiescent galaxies actually show signs of active star-formation in their core. In Section \ref{subsec:quench_frac} we explain how, while we find an excess of quiescent galaxies in clusters, this excess appears to be independent from stellar mass. Finally, in Section \ref{subsec:env_quench} we describe how we found an excess of galaxies showing signs of star-formation in their core while having their outskirts quenched in clusters, especially at lower stellar mass. 
In Section \ref{sec:discussion}, we discuss how our results impact our understanding of environmental quenching at $z>1$ and how impactful deep-learning based deconvolution techniques could be in the era of big-data astronomy.
Finally, we refer the reader to Section \ref{sec:summary} for a summary of this paper.


Throughout this paper, we assumed a \cite{chabrier_galactic_2003} initial mass function (IMF), as well, a flat $\Lambda CDM$ cosmology with $H_{0} = 70$ km s$^{-1}$ Mpc$^{-1}$, $\Omega_m = 0.3$, and $\Omega_{\Lambda} = 0.7$.

\section{DATA}\label{sec:gogreen}

\subsection{The GOGREEN Survey}\label{subsec:gogreen_data}

This study is based on the Gemini Observations of Galaxies in Rich Early Environments (GOGREEN) survey. This survey was designed to provide imaging and deep spectroscopy on a wide halo mass range of groups and clusters of galaxies in order to better understand the impact of the environment on the evolution of galaxies at $1.0<z<1.5$. 
GOGREEN imaged a total of 21 groups and clusters from the SpARCS, SPT, COSMOS and SXDS surveys. We refer the reader to \cite{balogh_gemini_2017} for the detailed survey description. In this paper, we focused on the clusters only, and out of the 12 clusters, we had to exclude three.
Because the set of available observed filters did not correspond to the same rest-frame colors as for the lower redshift clusters, the $z=1.46$ SpARCS-1033 cluster was excluded (see Section \ref{subsec:select_bands}). We also had to exclude two additional clusters (SPTCL-0205 and SpARCS-0035) because they each had at least one image with a quality incompatible with a good deconvolution. Specifically, the low signal-to-noise ratio in these clusters led to catastrophic failures in at least one band ($J$ for SPTCL-0205 and $I$ for SpARCS-0035) as explained in Section \ref{subsec:deconv}. 

Hence, we studied a total of 9 clusters in the redshift range $1.0<z<1.4$, including two from the SPT survey and seven from the SpARCS survey (see Table \ref{tab:instruments} for more details).

We used images probing rest-frame $U$, $V$, and $J$ bands as detailed in Section \ref{subsec:select_bands}. We used images from the VLT (VIMOS.I and HAWK-I.Ks filters), Subaru (HSC.z and SuprimeCam.i filters), Magellan (FourStar.J, FourStar.J1 and FourStar.K filters) and CFHT (WIRCam.J and WIRCam.Ks filters) telescopes. For details on the reduction of these images, we refer the reader to the GOGREEN first data release\footnote{Technically, the most recent data release is DR2, however for the clusters of interest in this paper, there are no differences between DR1 and DR2.} (\citealt{balogh_gogreen_2020}). In the GOGREEN DR1, all of the images have a matched pixel scale of 0.2 arcsec per pixel. 
However, to avoid introducing any bias in the deconvolution process due to prior extrapolation (when the pixel size was reduced) or due to lost information (when the pixel size was increased), we used the images produced before the pixel-size matching step. See Section \ref{sec:deconv} for additional details on the deconvolution requirements and training. 

We also used the publicly available \textit{HST} images from the GOGREEN survey (\citealt{balogh_gemini_2017, chan_gogreen_2021}) to visually compare them to the deconvolved images to help flag catastrophic failures and understand the image quality limit compatible with a good deconvolution as detailed in in Section \ref{sec:deconv}. When available, we used images from the filters ACS\_WFC.F814W, WFC3\_IR.F105W, WFC3\_IR.F140W and WFC3\_IR.F160W.

\subsection{Redshifts, Stellar masses and SFHs}\label{subsec:z_m_sfhs}

For this study, we used the redshifts, stellar masses and integrated rest-frame colors publicly available in the GOGREEN DR1 (\citealt{balogh_gogreen_2020}).

The spectroscopic redshifts were measured by \cite{balogh_gogreen_2020} using a custom and interactive software based on the Manual and Automatic Redshifting Software (\textsc{marz}, \citealt{hinton_marz_2016}). The photometric redshifts were estimated using \textsc{eazy} (\citealt{brammer_eazy_2008}), as described in \cite{van_der_burg_gogreen_2020}.

The stellar masses were estimated by fitting the SEDs ($U$ through IRAC data) using the FAST SED-fitting code (\citealt{kriek_ultra-deep_2009}) with \cite{bruzual_stellar_2003} stellar population models. The analysis assumed a \cite{chabrier_galactic_2003} IMF, solar metallicity, and the \cite{calzetti_dust_2000} dust law. Star formation histories were modeled as exponentially declining. We refer the reader to \cite{van_der_burg_gogreen_2020} for additional details on the SED-fitting.

We probed the integrated star-formation histories of the galaxies by using the $U-V$ and $V-J$ integrated rest-frame colors to separate between star-forming and quiescent galaxies (see Section \ref{subsec:select_bands}). These colors were also measured by \cite{van_der_burg_gogreen_2020} using \textsc{eazy}. 
The magnitudes were derived from the best-fit SEDs where they used the cluster spectroscopic redshift for the cluster members and using the best redshift available for the field galaxies.

\begin{table*}[ht]
    \caption{Filters used to probe the rest-frame $U$, $V$ and $J$ colors in our 9 clusters.}
    \label{tab:instruments}
    \hspace{-60pt}
        \begin{tabular}{|c|c|ccc|ccc|ccc|}
            \hline \hline
            Cluster ID & Redshift & Filter $U$ & Depth $U$ & Quality $U$ & Filter $V$ & Depth $V$ & Quality $V$ & Filter $J$ & Depth $J$ & Quality $J$\\
             & $z_{spec}$ & & $mag_{AB}$ & $arcsec$ &  & $mag_{AB}$ & $arcsec$ & & $mag$ & $arcsec$\\
            \hline
            SPTCL-2106 & $1.132$ & I$^a$ & 25.3 & 0.65 & J1$^d$ (J$^d$) & 24.4 (24.1) & 0.62 (0.86) & Ks$^f$ & 23.6 & 0.38\\
            SPTCL-0546 & $1.067$ & I$^a$ & 25.0 & 0.61 & J1$^d$ (J$^d$) & 24.1 (23.9) & 0.71 (0.79) & Ks$^d$ & 23.9 & 0.67\\
            SpARCS-0219 & $1.325$ & I$^a$ & 25.2 & 0.56 & J$^d$ & 24.3 & 0.70 & Ks$^d$ & 24.0 & 0.78\\
            SpARCS-0335 & $1.368$ & I$^a$ & 25.5 & 0.57 & J$^d$ & 24.3 & 0.63 & Ks$^d$ & 23.7 & 0.70\\
            SpARCS-1034 & $1.386$ & z$^b$ (i$^c$) & 25.4 (25.5) & 0.55 (0.85) & J$^e$ & 24.5 & 0.70 & Ks$^e$ & 24.0 & 0.69\\
            SpARCS-1051 & $1.035$ & i$^c$ (z$^b$) & 25.6 (25.4) & 0.63 (0.57) & J$^e$ & 24.5 & 0.79 & Ks$^e$ & 24.1 & 0.77\\
            SpARCS-1616 & $1.156$ & i$^c$ (z$^b$) & 25.7 (25.6) & 0.65 (0.63) & J$^e$ & 24.2 & 0.80 & Ks$^e$ & 23.8 & 0.78\\
            SpARCS-1634 & $1.177$ & i$^c$ & 25.8 & 0.65 & J$^e$ & 24.2 & 0.72& Ks$^e$ & 23.8 & 0.70\\
            SpARCS-1638 & $1.194$ & i$^c$ & 25.6 & 0.58 & J$^e$ & 24.1 & 0.79 & Ks$^e$ & 23.6 & 0.74\\
            \hline
        \end{tabular}
        \\
        \vspace{2pt}
        \hspace{12pt}\footnotesize \textbf{Notes.} The values in parenthesis show the different filters used to study the field galaxies when different from the filters used for the cluster members.\\
        \vspace{1pt}
        \hspace{12pt}\footnotesize The depths are median 5-$\sigma$ limits measured on the PSF-homogenized stacked images in circular apertures with a diameter of $2''$ (\citealt{van_der_burg_gogreen_2020}).\\
        \vspace{1pt}
        \hspace{12pt}\footnotesize The image qualities are the full width at half maximum of the PSFs.\\
        \vspace{1pt}
        \hspace{12pt}\footnotesize $^a$ The $I$ filter from VLT/VIMOS.\hspace{77pt} $^b$ The $z$ filter from Subaru/HSC. \hspace{44pt} $^c$ The i filter from Subaru/SuprimeCam.\\
        \vspace{1pt}
        \hspace{12pt}\footnotesize $^d$ The $J$, $J1$ or $Ks$ filter from Magellan/FOURStar.\hspace{20pt} $^e$ The $J$ or $Ks$ filter from CFHT/WIRCam.\hspace{10pt} $^f$ The $Ks$ filter from VLT/HAWK-I.\\
\end{table*}

\section{SAMPLE SELECTION}\label{sec:sample_selec}

In this section, we describe how we built our spectroscopic and photometric samples of cluster members in Section \ref{subsec:select_cluster} and of field galaxies in Section \ref{subsec:select_field}. In Section \ref{subsec:select_bands}, we detail our strategy to use observed colors to probe rest-frame $U-V$ and $V-J$ colors in order to identify star-forming and quiescent regions.

\begin{figure}[htb]
    \centering
    \includegraphics[width=0.99\linewidth]{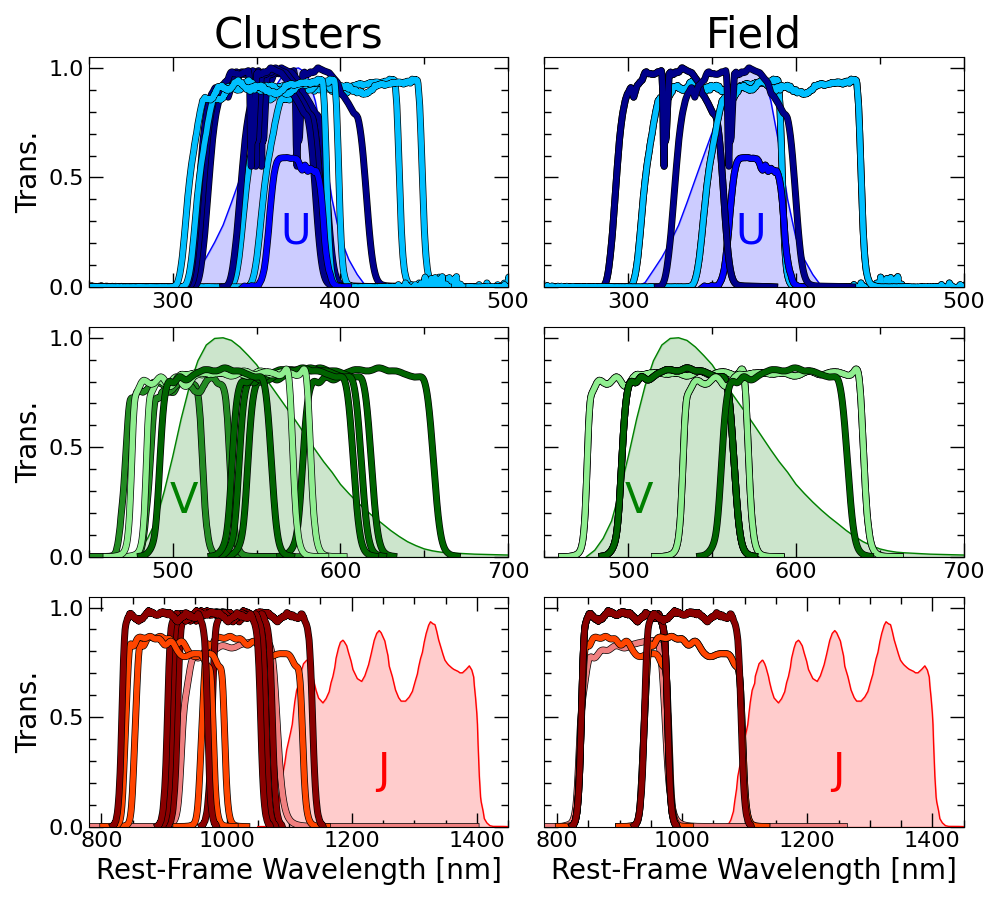}
    \caption{Observed filters used to probe rest frame $U$, $V$ and $J$ colors for the cluster members (left) and the field galaxies (right). The filled transmission curves corresponds to the rest-frame $U$, $V$ and $J$ bands while the thick lines are the transmission curves of the filters we used to probe these bands. Each shade of the lines represent the transmission curves from a different observatories and/or instruments and/or filters. For the field, some of the 9 curves in each panel are overlapping since we used only two different redshifts (see Section \ref{subsec:select_field}). See Table \ref{tab:instruments} and Section \ref{subsec:select_bands} for more details on the filters and how we chose them.}
    \label{fig:filters}
\end{figure}

\subsection{Cluster Membership} \label{subsec:select_cluster}


The spectroscopic cluster members were selected from \cite{abdulshafy_dynamical_2025}, who applied the GalWeight membership technique (\citealt{abdullah_galweight_2018}) to the GOGREEN cluster sample. Among the nine clusters analyzed in this work, seven overlapped directly with the \cite{abdulshafy_dynamical_2025} sample, therefore, we adopted their identified cluster memberships. Two clusters, SPARCS-0219 and SpARCS-1034, were not included in that study. 
Hence, we independently applied the GalWeight technique and identified the cluster members in SPARCS-1034 in a manner consistent with \cite{abdulshafy_dynamical_2025}.
However, for SPARCS-0219, the spectroscopic sample was not large enough to apply this technique so we used the membership flags of the GOGREEN DR1 catalog (\citealt{balogh_gogreen_2020, biviano_gogreen_2021}). 
While the low amount of spectroscopic data available for this cluster could imply that it is not a robust cluster candidate and that including it in or study could introduce a bias, we found no significant impact on our results when running our analysis with and without this cluster. That is why, as we had limited statistics, we decided to include it in our study. 
We restricted ourselves to the galaxies with $M_{*} > 10^{10.2}M_{\odot}$ as it is the mass limit for completeness in the spectroscopic sample (\citealt{balogh_gogreen_2020}). We note that for galaxies with $M_{*} > 10^{10.2}M_{\odot}$, the uncertainties on the integrated $U-V$ and $V-J$ rest-frame colors from \cite{van_der_burg_gogreen_2020} become smaller (less than $0.1$mag). The same stellar mass limit was applied when selecting the photometric cluster members.

Since our goal was to study a large sample of cluster members, especially members sitting at a larger distance from their center that were not covered by \textit{HST} or spectroscopic surveys, we added photometric members to the sample.
To select photometric cluster members, we used the available photometric information in the GOGREEN DR1: the peak, the $16^{th}$ and the $84^{th}$ percentile of the redshift probability distribution function ($z_{peak}$, $z_{16^{th}}$ and $z_{84^{th}}$ respectively). 
These values were obtained by \cite{van_der_burg_gogreen_2020} as output of the \textsc{eazy} software. 
For each galaxy, we estimated the probability distribution function to be a normalized Gaussian centered on $z_{peak}$ with a width $\sigma$ defined as the mean of $z_{peak} - z_{16^{th}}$ and $z_{84^{th}} - z_{peak}$. This was justified as $z_{peak}$ was always close to the central position between the $16^{th}$ and the $84^{th}$ percentiles.

Then, we adapted a technique from \cite{rudnick_rest-frame_2009,pello_photometric_2009} to select photometric cluster members. A galaxy was considered a photometric cluster member if:
\begin{equation}
    \int _{z_{cl} - \Delta z} ^{z_{cl} + \Delta z} N(z_{peak}, \sigma)dz > P_{thresh}
\label{eq:selection}
\end{equation}
where $N(z_{peak}, \sigma)$ is the Gaussian function describing the galaxy's redshift probability distribution function, $z_{cl}$ is the spectroscopic redshift of the cluster, $\Delta z$ is the median absolute deviation of the differences between the photometric redshifts and the redshift of the cluster for spectroscopically confirmed cluster members. Due to low statistics, we calculated $\Delta z$ on the whole sample rather than calculating an individual value for each cluster. We obtained $\Delta z = 0.114$.

$P_{thresh}$ was chosen to maximize the completeness while minimizing the interloper fraction in the whole sample ($P_{thresh}$ is the same for all clusters). The completeness of the sample was defined as:
\begin{equation}
    C = \frac{N_{C,spec,select}}{N_{C,spec,tot}}
\end{equation}
where $N_{C,spec,select}$ is the number of spectroscopically confirmed cluster members selected with this method when using their photometric redshifts and $N_{C,spec,tot}$ the total number of spectroscopically confirmed cluster members.
The interloper fraction was defined as:
\begin{equation}
    I = \frac{N_{F,spec,select}}{N_{C,spec,select}}
\end{equation}
where $N_{F,spec,select}$ is the number of spectroscopically confirmed field galaxies selected with this method when using their photometric redshifts.
We found that the best compromise ($P_{thresh} = 0.19$) led to a completeness $\approx 92\%$ for an interloper fraction of $\approx 22\%$. 

Compared to a method where we would have used a hard cut on the redshift distribution, this method allowed us to keep galaxies that would have fallen outside of the $\Delta z$ interval while having a significant probability of being a cluster member and to remove galaxies with $z_{peak}$ in the interval but having a high probability of not being a cluster member (i.e. having large photometric uncertainties).

Our sample of cluster members consisted of 960 galaxies with $M_{*} > 10^{10.2}M_{\odot}$, including 204 spectroscopically confirmed members.

\subsection{Field Sample} \label{subsec:select_field}

In order to probe the effect of the environment on the spatial quenching of galaxies, we needed to build a sample of field galaxies spanning the same redshift range as our sample of cluster members. We built this sample from the GOGREEN data rather than using larger surveys as they would have brought in different observation conditions, which would have potentially affected the homogeneity of the deconvolved sample.
On the contrary, using the same images ensured that the differences observed in our analysis were not due to our deconvolution process.

To avoid including galaxies with an unclear membership in the field sample, for each cluster we selected field galaxies with a redshift that was significantly different from the cluster's redshift. 
Practically, we divided the clusters into two classes. The lower redshift class included the six clusters with $1.0 \leq z \leq 1.2$ while the high redshift class included the three clusters with $1.3 \leq z < 1.4$. 
The lower and higher redshift classes had a weighted average of $z_{low} = 1.114$ and $z_{high} = 1.368$ respectively. 
We then used the catalogs of the low redshift clusters to select higher-$z$ field galaxies and vice-versa. 

For the spectroscopic sample, we selected the galaxies with $|z_{low} - z| < \Delta z$ if selected in the catalog of a higher redshift cluster or $|z_{high} - z| < \Delta z$ if selected in the catalog of a lower redshift cluster. We required that the selected field galaxies were confirmed as not being a cluster member. As our analysis was based on observed colors in this study, we chose a small redshift range to ensure the consistency of probed rest-frame colors for all the field galaxies observed in the same cluster catalog. For consistency with the cluster members selection, we used $\Delta z = 0.114$.
For the photometric sample, we used the same method as described in Section \ref{subsec:select_cluster} replacing the redshift of the cluster ($z_{cl}$ in Eq. \ref{eq:selection}) by $z_{low}$ when selecting field galaxies in the catalog of a higher redshift cluster or by $z_{high}$ when selecting field galaxies in the catalog of a lower redshift cluster. We made sure to photometrically select only galaxies in the field that were not already considered photometric cluster members.
Since all the field galaxies were selected in the same catalogs as the cluster members, their redshift data had similar uncertainties, so we decided to keep $\Delta z = 0.114$ and $P_{thresh} = 0.19$ in Eq. \ref{eq:selection}) for consistency.

Our sample of field galaxies consisted of 376 galaxies with $M_{*} > 10^{10.2}M_{\odot}$, including 30 galaxies with a spectroscopic redshift.

\subsection{Probing Rest-Frame UVJ using Observed Colors} \label{subsec:select_bands}

The main goal of our study was to probe the quenching spatial dependency in our galaxies. Achieving this required to establish, for each galaxy, if star-formation was happening either in their core or their outskirts.
Since we only had access to photometry for most of the galaxies in our samples, we used the rest-frame ($U-V$, $V-J$) color-color diagram to probe our galaxies as it is a well known tool to discriminate star-forming and quiescent regions. 

For every cluster, we had access to images from the $u$ band all the way to the IRAC bands. However, we were not able to use the IRAC data as IRAC large PSFs rendered the images incompatible with a good deconvolution.
In this pilot study, since we tested a new deconvolution technique, rather than looking to deconvolve all the images and fit spatially resolved spectral energy distributions (SED), we chose to probe the rest-frame $U$, $V$ and $J$ bands using the three closest observed bands.
Hence, for each cluster, depending on its redshift, we selected the 3 bands that were the closest to probe rest-frame $U$, $V$ and $J$. For most of the clusters, these observed bands were $I$, $J$ and $Ks$. The same method was applied to select the observed bands to probe the rest-frame $U$, $V$ and $J$ bands of the field galaxies. 
As shown on Figure \ref{fig:filters} we were able to identify filters that aligned very well with rest-frame $U$ and $V$ filters (Johnson/Bessell curves, \cite{bessell_ubvri_1990}) in every case. 
However, by using the Ks band, we probed a slightly bluer rest-frame color than rest-frame $J$ filter (2MASS filter, \cite{cutri_2mass_2003}). In Table \ref{tab:instruments}, we detail for each cluster image what filters were used to probe the $UVJ$ rest-frame colors for cluster members and field galaxies and we give the depth and resolution of the images.

Two consequences of using observed colors were that, first, we were not probing the exact rest-frame $UVJ$ colors, and then, since all clusters had a different redshift and we used different filters, we were probing slightly different rest-frame colors for each cluster.
This means that if we wanted to use the quiescent wedge usually used in the $U-V$ and $V-J$ color-color diagram to discriminate between star-forming and quiescent galaxies we had to adapt it to our ($I-J$, $J-K$)\footnote{Here and in the rest of the paper, when we write $I$, $J$ or $K$, it stands for ($I$, i or $z$), ($J$ or $J1$) and $Ks$ respectively depending on the cluster (see Table \ref{tab:instruments}).} color-color diagram.
To do so, after comparing the positions of each galaxy in the integrated rest-frame ($U-V$,$V-J$) and observed ($I-J$, $J-K$) color-color diagrams, we were able to identify and constrain the region in the observed ($I-J$, $J-K$) color-color plane containing the quiescent galaxies.

Practically, we used the rest-frame $U-V$ and $V-J$ integrated colors that were derived by \cite{balogh_gogreen_2020} using SED fitting in the GOGREEN DR1 to classify every galaxy in our sample as star-forming or quiescent using the quiescent wedge defined in \cite{whitaker_newfirm_2011}:

\begin{equation}
    \label{eq:uvj_1}
    U - V > a * (V - J) + b
\end{equation}
\begin{equation}
    \label{eq:uvj_2}
    V - J < c
\end{equation}
\begin{equation}
    \label{eq:uvj_3}
    U - V > d
\end{equation}
with $(a, b, c, d) = (0.8, 0.7, 1.6, 1.3)$.

Then, we looked at the observed $I-J$ and  $J-K$ colors for star-forming and quiescent galaxies. However, as aforementioned, since each cluster was at a different redshift, the observed bands were probing slightly different rest-frame colors. This resulted in the quiescent and star-forming galaxies of each cluster being located in a slightly different region of the observed ($I-J$, $J-K$) color-color plane. 
To remove this redshift bias, we normalized the colors by shifting each cluster's colors towards the same point of the color-color plane.
We chose to normalize the colors using the population of quiescent galaxies of each clusters as they are known to occupy a more concentrated region of the color-color plane contrary to star-forming galaxies that are spread due to dust or specific star-formation rate differences.
Practically, for each cluster, we measured the median position of the population of quiescent galaxies and found that these medians were on average separated by 0.37 mag along the $I - J$ axis and by 0.48 mag along the $J - K$ axis.
We estimated the necessary shift to get the medians of the populations of quiescent galaxies to (0.84, 2.20) in ($I-J$, $J-K$) coordinates. These coordinates were chosen so that the colors would be similar to the rest-frame ($U-V$, $V-J$) colors. This shift was applied to the whole sample of members in the cluster.
In order to estimate the uncertainty introduced by these shifts, we compared these measured shifts with what we would expect from redshifted star-forming or quiescent SEDs. We simulated the SED of a quiescent and a star-forming galaxy and redshifted it to the redshift of each cluster before convolving it with the relevant I, J and Ks filters. This allowed us to estimate the shifts necessary to bring the colors of the quiescent SED to the aforementioned coordinates and star-forming SED colors to its expected location based on their rest-frame UVJ colors. After measuring the difference between these simulated shifts and the observed shifts, we found them to be consistent with a median difference less than 0.1mag and measured the typical uncertainty on the shift to be 0.11mag on both axis.

\begin{figure}[htb]
    \centering
    \includegraphics[width=0.95\linewidth]{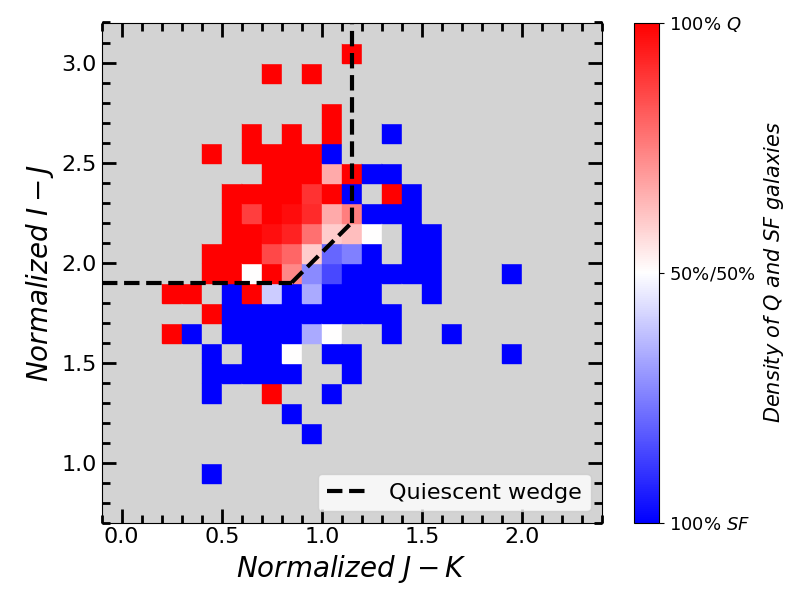}
    \caption{Calibration of the observed $IJK$ color-color diagnostic used for spatially resolved classification. The distribution of galaxies in the normalized observed $(I-J, J-K)$ plane is shown, with each galaxy classified as quiescent or star-forming from its conventional integrated rest-frame $UVJ$ colors. Each 0.1-mag cell is colored according to the fraction of galaxies classified as quiescent or star-forming, ranging from dark blue (entirely star-forming) to dark red (entirely quiescent), with white indicating equal fractions. The clear separation of the two populations allows us to define an observed-color quiescent wedge and apply the same diagnostic to the spatially resolved photometry.}
    \label{fig:quies_wedge}
\end{figure}

Figure \ref{fig:quies_wedge} is used to calibrate the observed ($I-J$, $J-K$) color-color diagnostic used for spatially resolved classification. 
More precisely, this figure shows the 2D-histogram of the distribution of quiescent and star-forming cluster member galaxies using the normalized $I-J$ vs $J-K$ colors resulting from the procedure described above. The star-forming status of the galaxies was determined using the conventional integrated $U-V$ and $V-J$ rest-frame colors. 
The ($I-J$, $J-K$) plane was divided in small individual cells with a size of $0.1$mag (representative of the typical uncertainty). 
The color of each cell is representative of the fraction of star-forming (bluer cells) and quiescent (redder cells) galaxies in the cell. The white cells contains an equal amount of star-forming and quiescent galaxies. 
In this figure, one can see that the quiescent and star-forming galaxies are occupying different regions of the plane, very similarly to the rest-frame ($U-V$, $V-J$) color-color diagram that we were trying to reproduce. Based on this figure, we delimited an observed-color wedge for quiescent galaxies. 
This wedge was chosen to maximize the completeness when isolating quiescent galaxies (we were able to reach a completeness of $> 92.5\%$) while also keeping the star-forming interlopers to an acceptable level of $\sim 10\%$. The quiescent wedge was defined as follows:
\begin{equation}
    \label{eq:ijk_1}
    I - J > a * (J - K) + b
\end{equation}
\begin{equation}
    \label{eq:ijk_2}
    J - K < c
\end{equation}
\begin{equation}
    \label{eq:ijk_3}
    I - J > d
\end{equation}
with $(a, b, c, d) = (1.00, 1.05, 1.15, 1.90)$.

Using the normalized ($I-J$, $J-K$) colors and this quiescent wedge, we were able to classify each region of a cluster member galaxy as either star-forming or quiescent.

For the field galaxies, we decided to apply the same technique to estimate the necessary shifts and used the same reference point for the quiescent galaxies and the same quiescent wedge as for cluster members. However, the sample of field galaxies has a smaller number of quiescent galaxies and a broader redshift distribution. Consequently, while our shifts and quiescent wedge allowed us to keep a completeness for quiescent galaxies $> 90\%$ and an interloper fraction $\sim 10\%$, the uncertainties introduced by the shifts are larger with a 0.17mag typical uncertainty on the shifts for field galaxies (0.145mag for quiescent galaxies and 0.19mag for star-forming galaxies).

In the following section, we detail how we used an innovative deep-learning based technique to deconvolve the images in order to spatially resolve our galaxies.

\section{Deconvolution using \textsc{SUNet}} \label{sec:deconv}

In this section, we detail the deep-learning based technique we used to deconvolve the images of our galaxies. The readers that are only interested in the scientific analysis of the data can skip this section. In Section \ref{subsec:sunet} we explain what is \textsc{SUNet} and how it works, in Section \ref{subsec:new_models} we give some details on the models that we used and trained, additional details on the training is given in Appendix \ref{app:training}. Then, in Section \ref{subsec:deconv}, we explain how we deconvolved our galaxies and evaluate the deconvolution success. Finally, in Section \ref{subsec:final_sample} we give the details of the final sample of deconvolved cluster members and field galaxies that we used in our analysis.

\subsection{The \textsc{SUNet} deconvolution} \label{subsec:sunet}

The Swin Transformer UNet (\textsc{SUNet}) deconvolution framework (\citealt{fan_sunet_2022,akhaury_ground-based_2024}) was developed to solve one of the main limitations of observational astronomy: producing images reaching high spatial resolution and high signal-to-noise for a large number of distant galaxies. 
Indeed, while space telescopes like \textit{HST} or \textit{JWST} have achieved high spatial resolution by bypassing the atmospheric turbulence, obtaining multi-band photometry over large portions (contiguous or not) of the sky remains extremely challenging due to their small field of view.

The \textsc{SUNet} deconvolution is a two-step process: it starts with a Tikhonov deconvolution step, which is a linear inversion in Fourier space.
This first step produces deconvolved images containing correlated additive noise. The second step consists of removing the noise from these deconvolved images using the \textsc{SUNet} neural network. 
The framework uses an innovating technique to efficiently remove noise without introducing any bias. The debiasing is achieved by using multi-resolution supports on neural networks. 
We refer the reader to \cite{akhaury_ground-based_2024} for further information on \textsc{SUNet} and the deep-learning-based deconvolution technique.

\textsc{SUNet} was developed to be easily adapted to different datasets. Running \textsc{SUNet} requires three things: a trained model (see Section \ref{subsec:new_models} and Appendix \ref{app:training}), an image cutout of the galaxy to deconvolve, and the corresponding PSF. 
For this study, we built custom PSFs for each image and tested their spatial dependence. We refer the reader to Appendix \ref{app:psfs} for a more detailed description of how we built our PSFs.
\textsc{SUNet} produces a deconvolved image with a pixel scale matching the sampling of the input PSF, which in turn controls the resolution of the deconvolved image.
By design, the \textsc{SUNet} deconvolution is the most efficient when there is a single galaxy in the image cutout.

The two most important factors in reaching a good quality deconvolution are the use of a customized model trained on a dataset that has properties (wavelength, depth, spatial resolution and pixel scale) as close as possible to the dataset one wants to deconvolve and the choice of the deconvolution factor; defined as the ratio between the sizes of the output and input images in pixels. 
In the next section, we detail the characteristics of the models we trained and used for this study.

\subsection{\textsc{SUNet} models} \label{subsec:new_models}

As mentioned in the previous section, the choice of the model one uses has a significant impact on the quality of the deconvolution. Since the images we had were obtained from different telescopes, most had different properties (seeing, pixel scale, depth, etc). This led us to train customized models.
To estimate the number of models we needed, we first separated our images between optical and near-IR. Then, we gathered them by their native pixel scales. While all the optical images have the same pixel scales, the near-IR images pixel scales depends on the instrument they were observed with. Finally, we compared their seeings, which were not different enough to justify several models.
Hence, in total, we used 4 models, one for the optical bands ($I$, $i$, $z$) and three models for the near-IR bands, one for each instrument: HAWK-I ($Ks$), FOURStar ($J$, $J1$, $Ks$) and WIRCam ($J$, $Ks$).

For the optical bands, we used the model trained by \cite{akhaury_ground-based_2024}. This model was trained on \textit{HST} images from CANDELS (\citealt{grogin_candels_2011,koekemoer_candels_2011}). They extracted $128\times128$ pixels cutouts in the F606W filter. To simulate ground-based data, these cutouts were first convolved with a 15-pixel FWHM Gaussian PSF (0.75"), which is consistent with the spatial resolution of the GOGREEN images. They were then injected white Gaussian noise.
The level of noise was the same for all the cutouts and was chosen to ensure that the faintest objects in the training set had a peak signal-to-noise ratio close to one. The model was then trained on at least $20,000$ galaxies. The details of the training can be found in \cite{akhaury_ground-based_2024}.

\begin{table}[ht]
    \caption{Properties of the ground-based image cutouts and of the PSFs the \textsc{SUNet} models were trained with.}
    \label{tab:models}
    \hspace{-40pt}
        \begin{tabular}{|c|cc|c|cc|}
            \hline \hline
            Model & $Px_{GB}$$^a$ & $N_{GB}$$^b$ & DF$^{c}$ & $Px_{PSF}$$^a$ & $N_{PSF}$$^b$\\
            \hline
            Optical & 0.20"/px & 32 & 4 & 0.05"/px & 128\\
            HAWK-I & 0.11"/px & 32 & 2 & 0.055"/px & 64\\
            FOURStar & 0.16"/px & 32 & 4 & 0.04"/px & 128\\
            WIRCam & 0.30"/px & 16 & 4 & 0.075"/px & 64\\
            \hline
        \end{tabular}
        \\
        \vspace{2pt}
        \hspace{10pt}\footnotesize $^{a}$ Size in arcseconds of the pixels.\\
        \vspace{1pt}
        \hspace{10pt}\footnotesize $^{b}$ The cutout is a square with a side length of $N$ pixels.\\
        \vspace{1pt}
        \hspace{7pt}\footnotesize $^{c}$ The deconvolution factor DF describes the ratio between the pixels scales ($Px$) of the input and output images. It also influences the size in pixel ($N$) of the output images. See Equations \ref{eq:px_dec} and \ref{eq:n_dec} in Section \ref{subsec:new_models}. for additional details.\\
\end{table}

For the near-IR models, we trained our own models. We used the same recipe as in \cite{akhaury_ground-based_2024} using publicly available \textit{JWST}/NIRCam.F277W images and the photometric catalog from the COSMOSWeb survey (\citealt{casey_cosmos-web_2023,shuntov_cosmos2025_2025,franco_cosmos-web_2025}). The F277W image was chosen as it provided the deepest near-IR image consistent with the $J$ and/or $Ks$ bands as well as a target resolution close to the optical model. We selected a total of $\sim 20,000 - 25,000$ galaxies for each model with $mag_{AB} < 24$ and $0.1<z<3.0$ that were not point-like sources. 
We convolved the NIRCam images with Gaussian PSFs that were representative of the image quality of the GOGREEN images (0.5" for HAWK-I, 0.7" for FOURStar and 0.75" for WIRCam). 
Even though the \textit{JWST} images have their own PSF, they are so much narrower than any of the ground-based instruments that we considered the \textit{JWST} images to be fully ideal.
After the convolution, we added Gaussian noise to simulate realistic ground-based images signal-to-noise ratios. 
The training ran on a cluster of parallelized GPUs and was completed in $\sim 3-4$ days per model. 

Table \ref{tab:models} summarizes the characteristics of the four models we used and/or trained. For each of these models, we give the pixel scales and the sizes in pixel of the input cutouts ($Px_{GB}$ and $N_{GB}$ respectively) and of the PSFs ($Px_{PSF}$ and $N_{PSF}$ respectively) it was trained with. The $Px_{GB}$ values were specifically chosen to match the pixel scale of the GOGREEN images. The size of the input cutouts ($N_{GB}$) were chosen to minimize the risk of having more than one galaxy per cutout. However, we made sure that the cutouts were not too small so that all the galaxies would fully fit inside.
Then, the deconvolution factor $DF$, defined by the ratio between the pixels scales of the input and output images, gave us the pixel scales and sizes of the deconvolved images: 
\begin{equation} \label{eq:px_dec}
    Px_{deconv} = \frac{Px_{GB}}{DF}
\end{equation}
\begin{equation}    \label{eq:n_dec}
    N_{deconv} = N_{GB} \times DF
\end{equation}
where $Px_{deconv}$ is the output image pixel scale and $N_{deconv}$ is the size of the output image.
For each model, the deconvolution factor was chosen so that the deconvolved image had a pixel scale close to the pixel scale of the \textit{JWST} image it was trained with and was imposed using the sampling of the PSF. We note that while the deconvolution factor determines the target resolution, it does not entirely determine the finite resolution. Indeed, the actual resolution of the deconvolved image depends on a variety of factors including the ground-based image quality and signal-to-noise ratio.


We give additional details on how to build a customized training set, choose the size of the cutouts and set the $DF$ for training models for \textsc{SUNet} in Appendix \ref{app:training}.

While \citealt{akhaury_ground-based_2024} demonstrated that the optical model they trained was efficient at deconvolving optical images and recovering morphologies, they did not measure gradients. 
In order to verify that our deconvolution technique was able to recover quenching gradients accurately, we used publicly available \textit{HST} and \textit{JWST} data from the COSMOS field. 
We selected galaxies from the \cite{shuntov_cosmos2025_2025} catalog with a stellar mass and redshift consistent with our sample. 
We made cutouts in the ACS\_WFC.F814W, NIRCam.F115W and NIRCam.F277W and PSF-matched the three bands for each galaxy. 
We then measured their (F814W $-$ F115W) and (F115W $-$ F277W) gradients between two apertures defined by ellipses encompassing 50\% and 90\% of their total flux in the F277W band. 
To simulate ground-based images, we used he same method we did when building our training sets: convolve the cutouts with a representative PSF, add gaussian noise and change the pixel scale. 
We then used our optical and near-IR models to deconvolve these galaxies and measured the same color gradient using the same apertures. 
The comparison between the two measurements showed that our deconvolution process was able to recover colors with a 0.035 mag uncertainty and gradients with a 0.05 mag uncertainty. 
We refer the reader to Appendix \ref{app:test_grad} for additional details and figures regarding this tests.

\subsection{Deconvolution of GOGREEN Galaxies}\label{subsec:deconv}

For every galaxy that we selected in Section \ref{sec:sample_selec}, we made cutouts in the three images probing rest-frame $U$, $V$ and $J$ as described in Section \ref{subsec:select_bands} and Table \ref{tab:instruments}. The size of the cutouts were determined by the \textsc{SUNet} model to be used for the deconvolution as shown in Table \ref{tab:models}.

\begin{figure}[htb]
    \centering
    \includegraphics[width=0.99\linewidth]{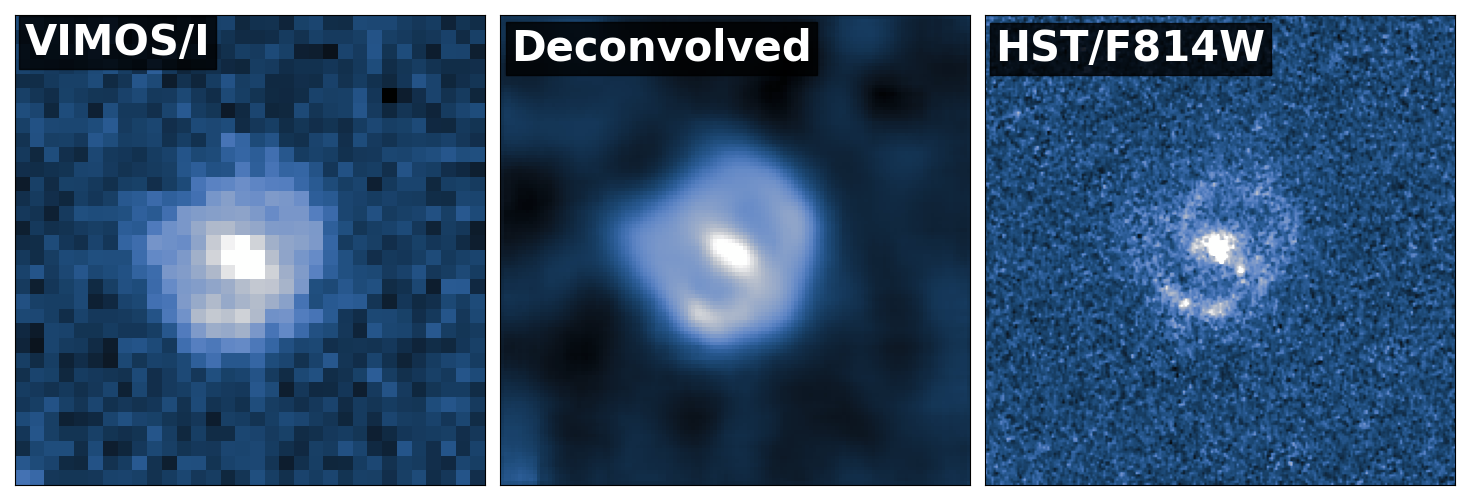}
    \includegraphics[width=0.99\linewidth]{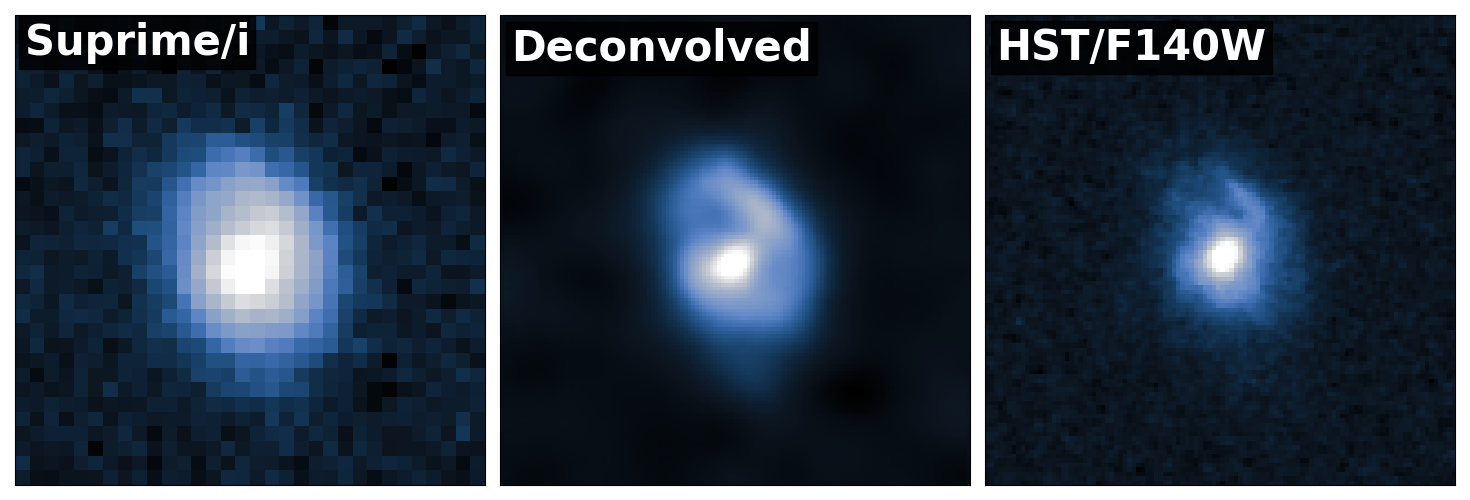}
    \includegraphics[width=0.99\linewidth]{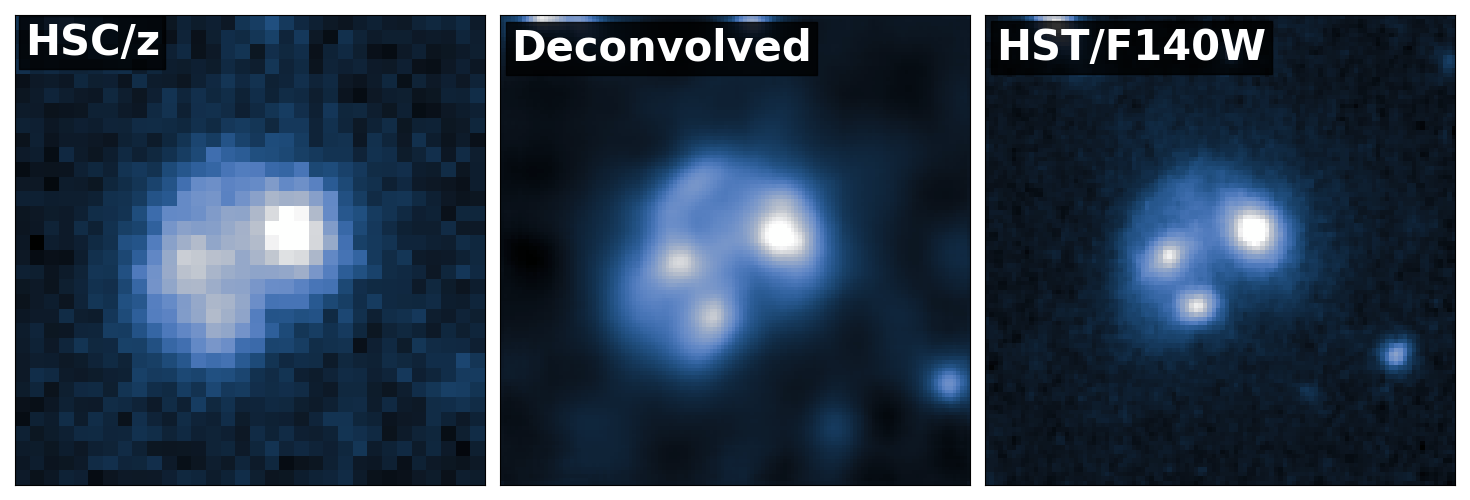}
    \caption{Examples of galaxies deconvolved using \textsc{SUNet} with the optical model. Each row shows a galaxy observed with a different instrument. The three cutouts per galaxy show, from left to right, the ground-based GOGREEN image, the deconvolved image produced by \textsc{SUNet} and the \textit{HST} image. This last image is for reference only and was not used during the deconvolution process. Each cutout is 6.4" wide. One can notice how our deconvolution process is able to recover the internal structures of the galaxies.}
    \label{fig:dec_Opt}
\end{figure}

After running \textsc{SUNet} using the cutouts, the models and the PSFs we built, we visually assessed the quality of the deconvolutions. In the very large majority of cases ($> 95\%$ of our total sample), the deconvolution performed extremely well, leading to a much improved spatial-resolution, getting close to \textit{HST}-like resolution for the most favorable cases (optical, HAWK-I and FOURStar models). In all cases, the deconvolution process was efficient at removing the spurious deconvolution noise while preserving and deconvolving fainter structures like spiral arms or clumps. 

\begin{figure}[htb]
    \centering
    \includegraphics[width=0.99\linewidth]{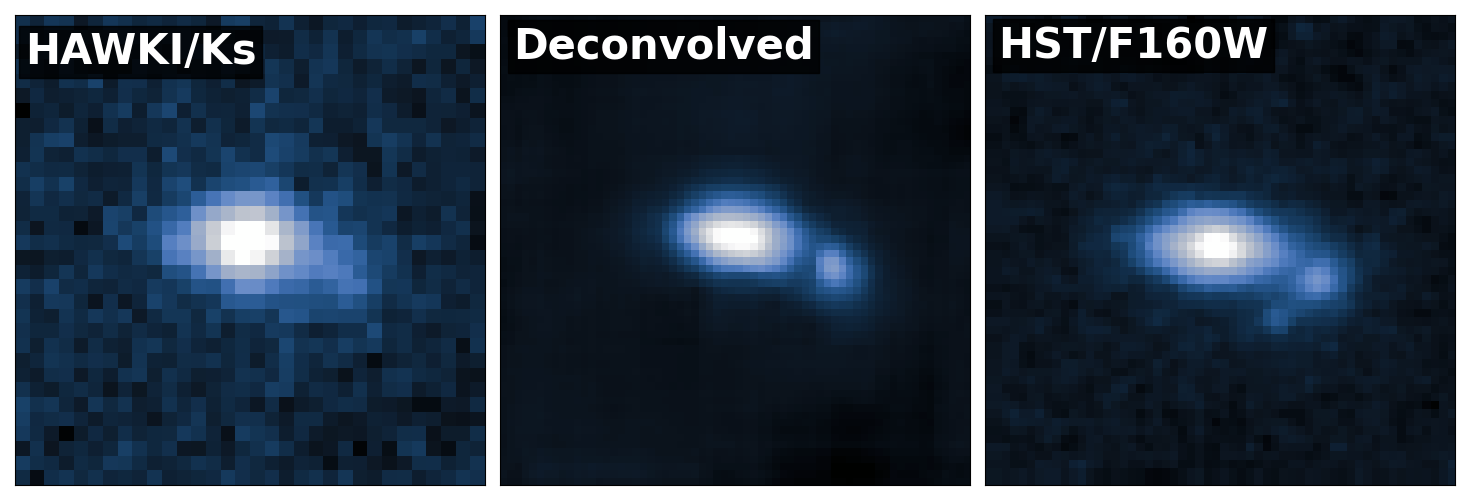}
    \includegraphics[width=0.99\linewidth]{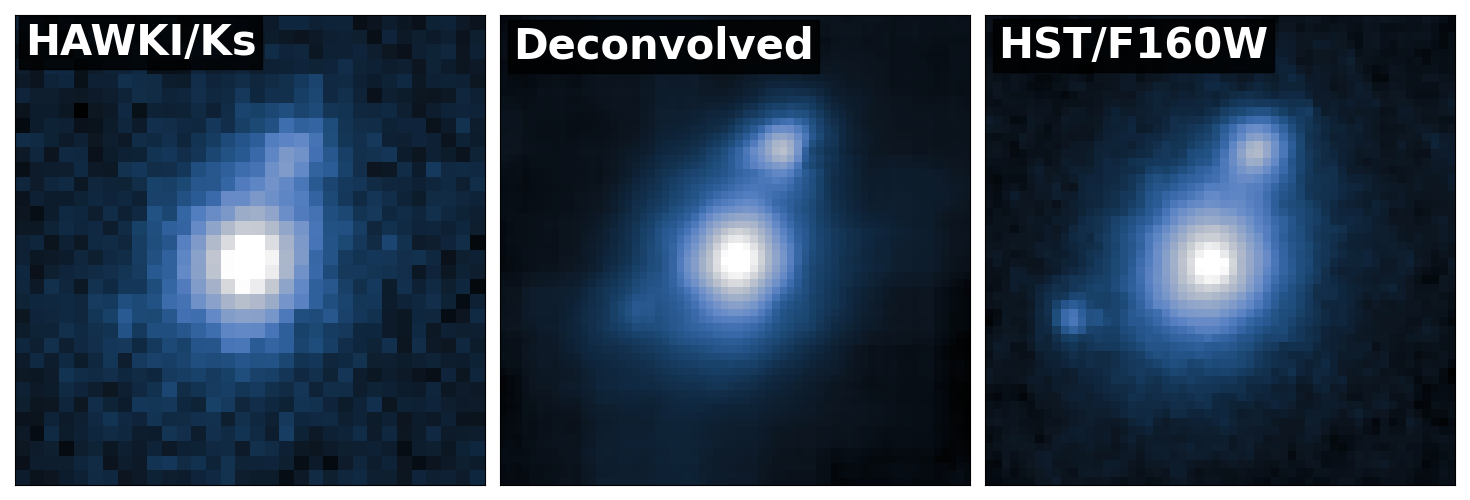}
    \caption{Examples of galaxies deconvolved using \textsc{SUNet} with the HAWK-I model. Similar to Figure \ref{fig:dec_Opt}. Each cutout is 3.52" wide. One can notice how our deconvolution process is able to recover the internal structures of the galaxies.}
    \label{fig:dec_HA}
\end{figure}

\begin{figure}[htb]
    \centering   
    \includegraphics[width=0.99\linewidth]{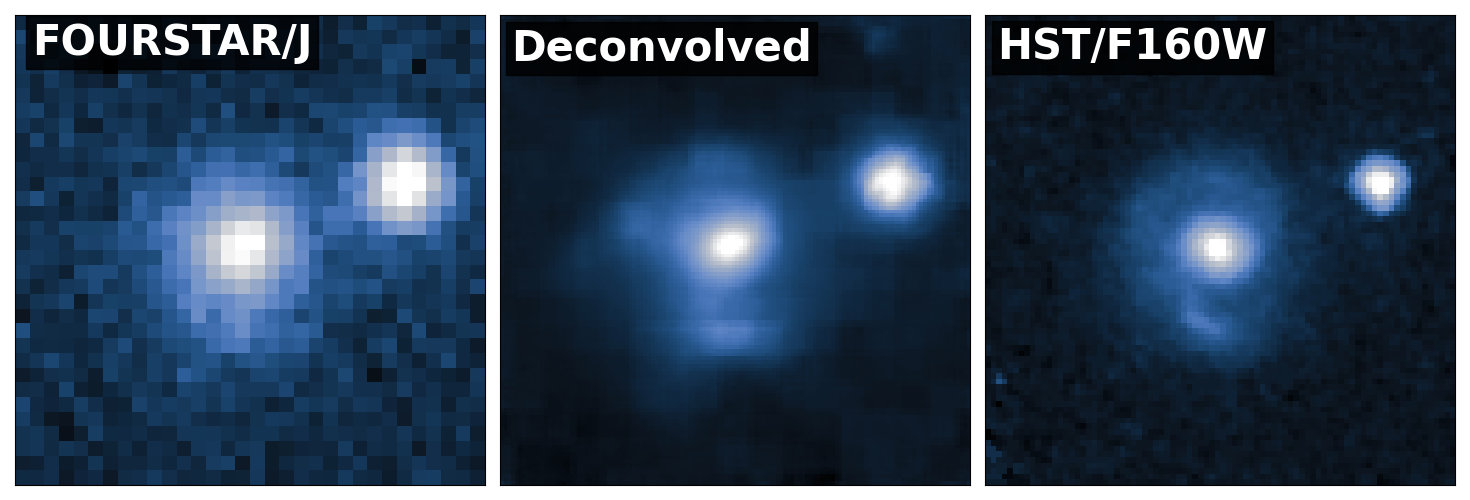}
    \includegraphics[width=0.99\linewidth]{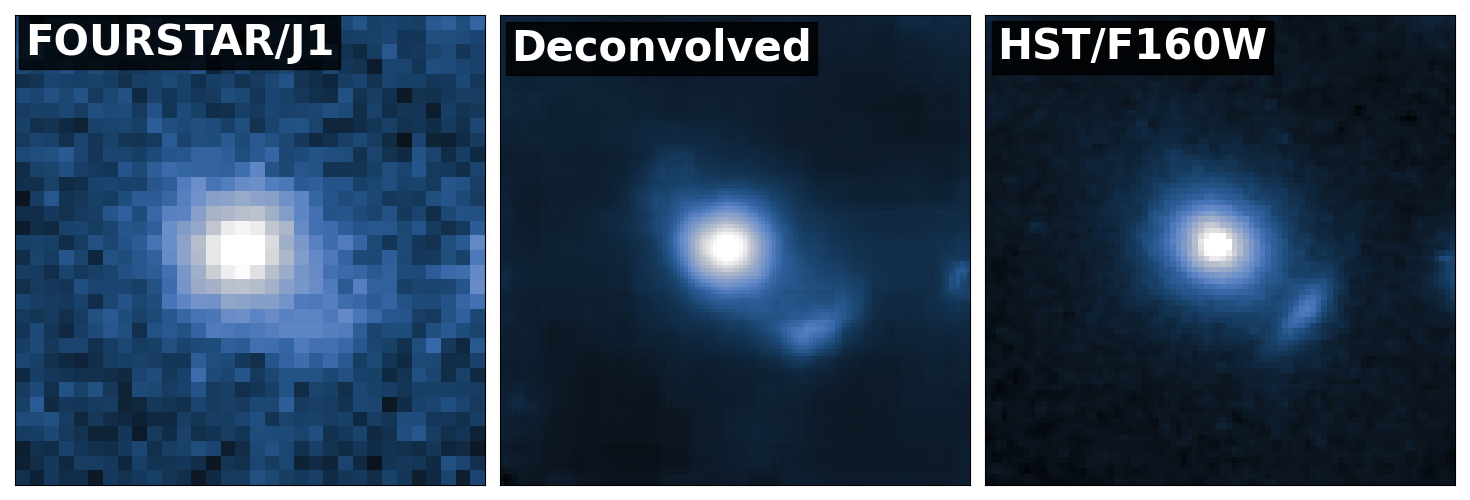}
    \includegraphics[width=0.99\linewidth]{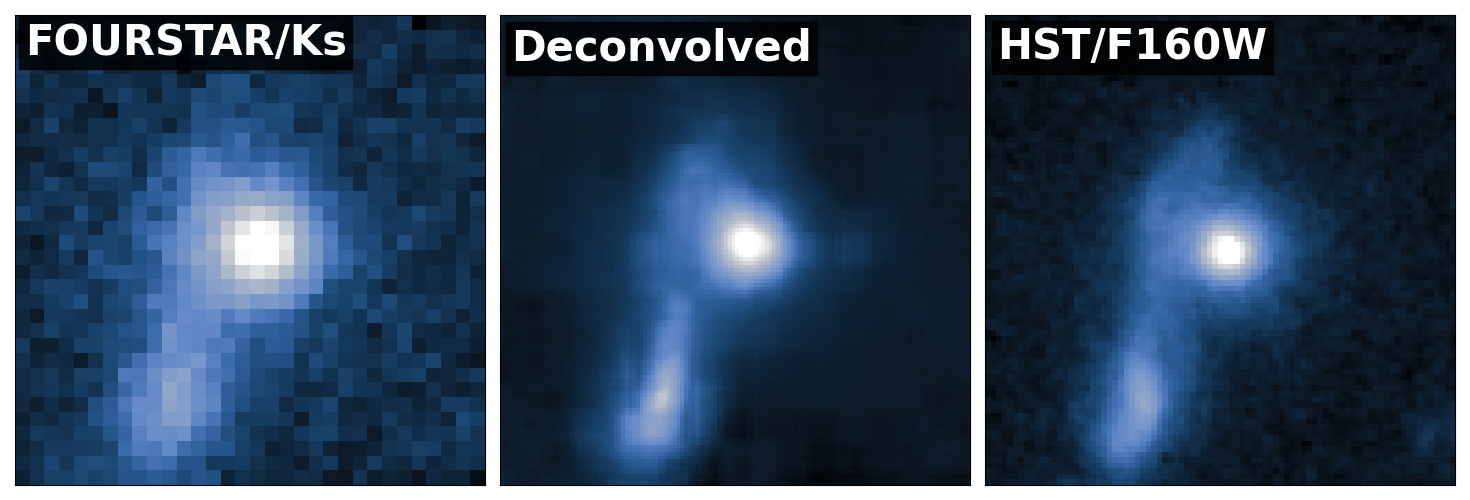}
    \caption{Examples of galaxies observed through different filters and deconvolved using \textsc{SUNet} with the FOURStar model. Similar to Figure \ref{fig:dec_Opt}. Each cutout is 5.12" wide. One can notice how our deconvolution process is able to recover the internal structures of the galaxies.}
    \label{fig:dec_4S}
\end{figure}

\begin{figure}[htb]
    \centering
    \includegraphics[width=0.99\linewidth]{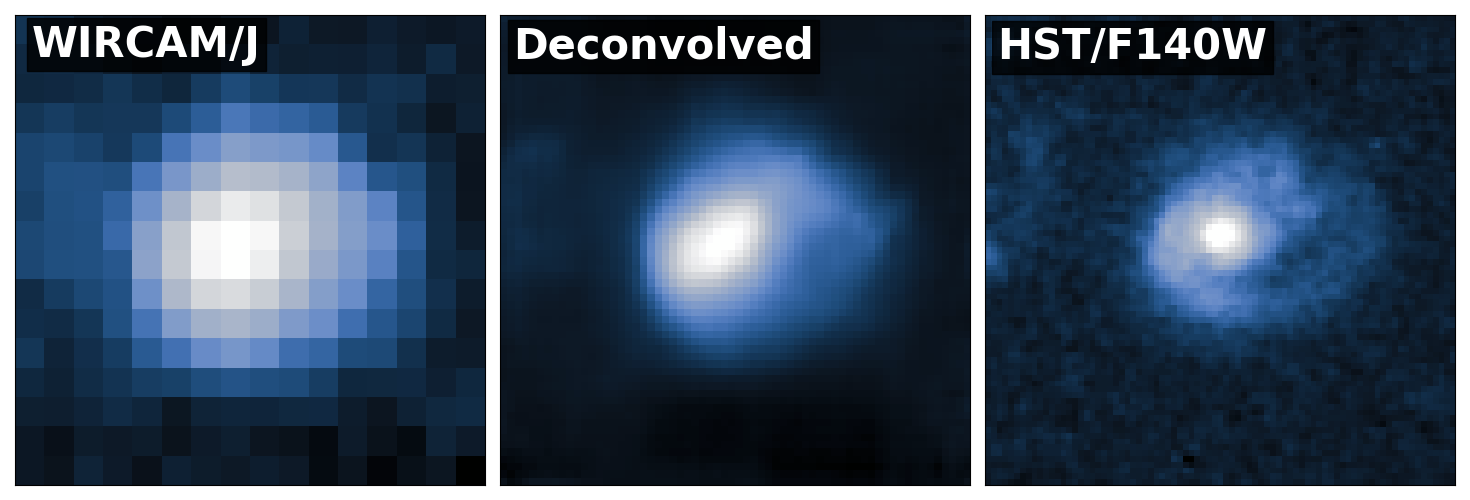}
    \includegraphics[width=0.99\linewidth]{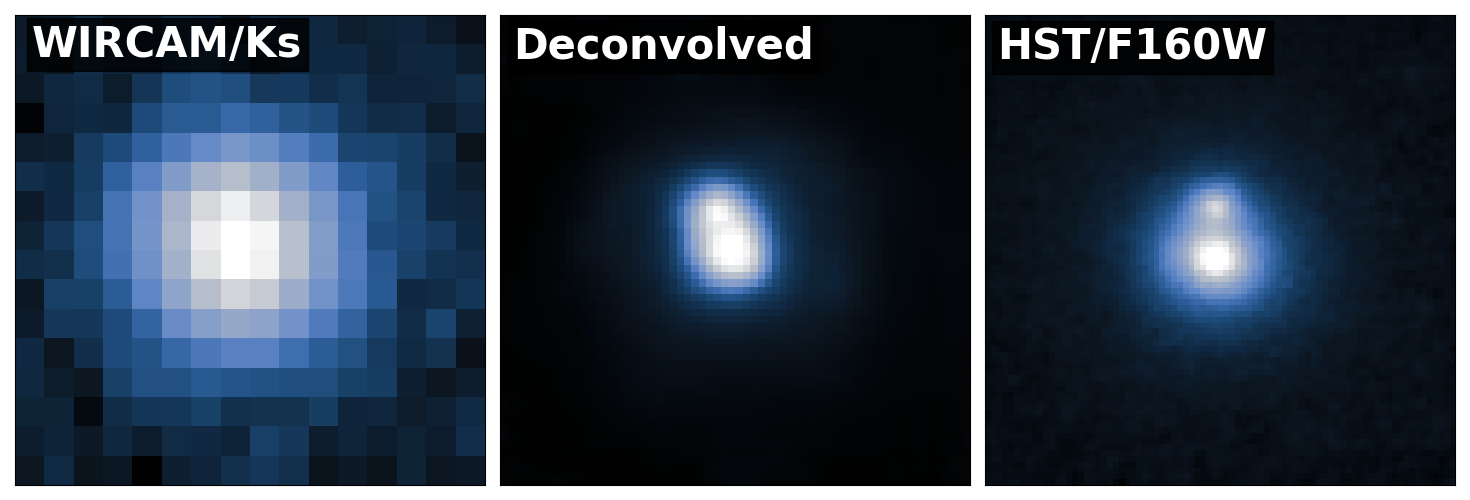}
    \includegraphics[width=0.99\linewidth]{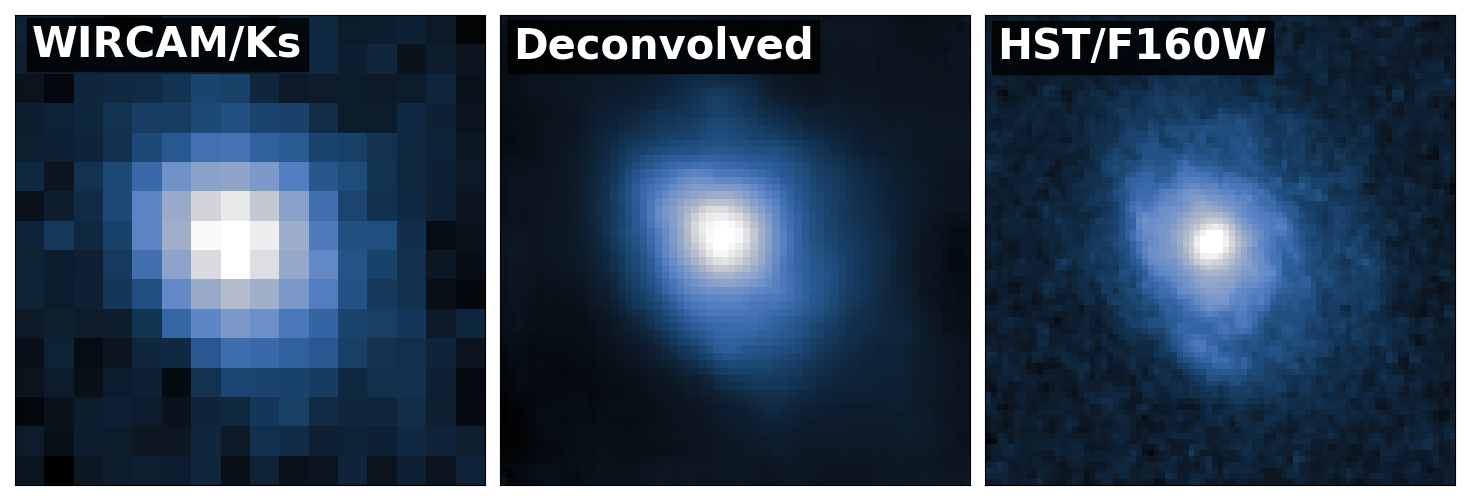}
    \caption{Examples of galaxies observed through different filters and deconvolved using \textsc{SUNet} with the WIRCam model. Similar to Figure \ref{fig:dec_Opt}. Each cutout is 4.8" wide. One can notice how our deconvolution process is able to recover the internal structures of the galaxies.}
    \label{fig:dec_WC}
\end{figure}

We show a sample of two or three successful deconvolutions using each of the optical, HAWK-I, FOURStar and WIRCam models in Figure \ref{fig:dec_Opt}, \ref{fig:dec_HA}, \ref{fig:dec_4S} and \ref{fig:dec_WC} respectively where the improvement from the original to the deconvolved images is striking. For reference, we show the \textit{HST} image of these galaxies, however, they were not used at any point in the deconvolution process. We decided to show these specific galaxies as they have clearly identifiable structures, they nonetheless are representative of the efficiency of \textsc{SUNet} to deconvolve our galaxies.

\begin{figure*}[htb]
    \centering
    \includegraphics[width=0.99\linewidth]{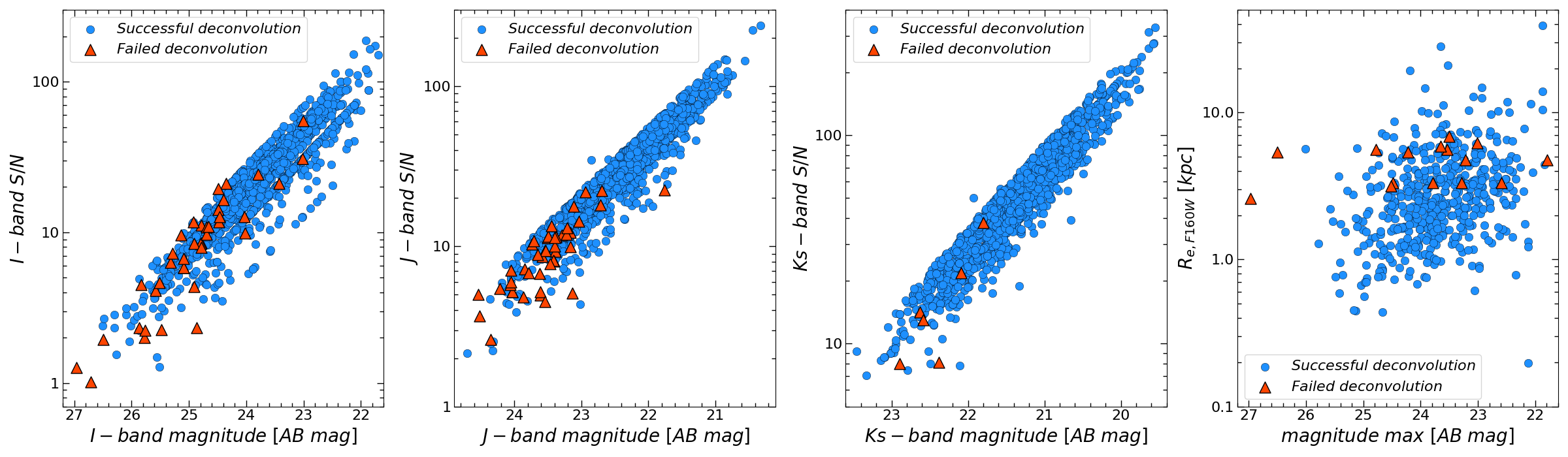}
    \caption{Surface brightness of \textsc{SUNet} deconvolution successes and failures. Left three panels, from left to right: flux S/N of each galaxy vs AB magnitude in the \textit{I}, \textit{J} and \textit{Ks} bands, circular blue markers correspond to successful deconvolutions, the triangular red markers denote failed deconvolutions. Right panel: Half-light radii measured in the F160W band (when \textit{HST} coverage was available, radii obtained from \cite{chan_gogreen_2021}) vs AB magnitude. For each galaxy, we chose the largest magnitude between the three bands. Most of the failed deconvolution did not have HST imaging. The deconvolution has a very high success rate, with most catastrophic failures linked to a low surface brightness. We note that for all the failed deconvolutions, the catastrophic failures only occurred in one or two bands.}
    \label{fig:Success_rate}
\end{figure*}

The remaining 5\% of galaxies were flagged and removed as they did not meet the requirements for a good quality deconvolution. 
These were identified as cases where in at least one band, the deconvolution resulted in catastrophic failure, meaning that either no structure or shape was distinguishable in the output image or a huge noise pattern made any measurement objectively unrealistic. 
In total, we removed 70 galaxies because of catastrophic failures.
To investigate the reason of the catastrophic failures of some deconvolutions, for every galaxy, we looked at its flux signal-to-noise ratio (S/N) in the three deconvolved bands. These ratios were obtained from the GOGREEN DR1 catalog (\citealt{balogh_gogreen_2020}). The left three panels of Figure \ref{fig:Success_rate} show these S/N plotted against the corresponding magnitudes in each band for every galaxy in our sample. Most of the catastrophic failures are concentrated at low S/N and high magnitudes. One can notice that the \textit{Ks} band has a lower number of catastrophic failures as the \textit{Ks} fluxes have on average a higher S/N compared to the other two bands. Most of the galaxies that were removed from the sample have catastrophic failures in a single band. This means that the failures are not linked to the galaxies or their physical properties but to the image quality.
This is confirmed by the right panel of Figure \ref{fig:Success_rate} where we show the effective radii measured in the \textit{HST} F160W band when available (\citealt{chan_gogreen_2021}) of our galaxies versus there highest magnitudes between the three bands. The failed deconvolutions have an average radius that is twice the average radius of our total sample, leading to lower average surface brightness. However, as the panel shows, only a low fraction of deconvolution fails.
As there is no systematics, in order to minimize any bias of our sample, we decided to only remove the galaxies with bad image quality data from the sample and to keep the galaxies that were successfully deconvolved, including at lower S/N or high magnitude.
However, we chose to limit our sample to galaxies with a stellar mass greater than $10^{10.2} M_{\odot}$, which corresponds to the mass completeness limit for the GOGREEN spectroscopic sample. 

Additionally, for some galaxies, while the deconvolution was successful in every band, the source of interest appeared as a single object in at least one band and as separated into two or more objects/clumps in at least one other band (some might be mergers) making multi-band and gradient measurements complicated. In order to be conservative, we decided to reject these 77 galaxies from our sample, they represented $\sim 5\%$ of the total sample.
To be conservative, we looked at the colors of all the rejected galaxies and verified that removing them did not result in any bias in our results.

\subsection{Final Samples of Cluster Members and Field Galaxies}\label{subsec:final_sample}

After completing the deconvolution process, we obtained our final samples of 895 cluster members and 294 field galaxies. Table \ref{tab:samples} details how these galaxies are distributed among the cluster images.

\begin{figure}[htb]
    \centering
    \includegraphics[width=0.85\linewidth]{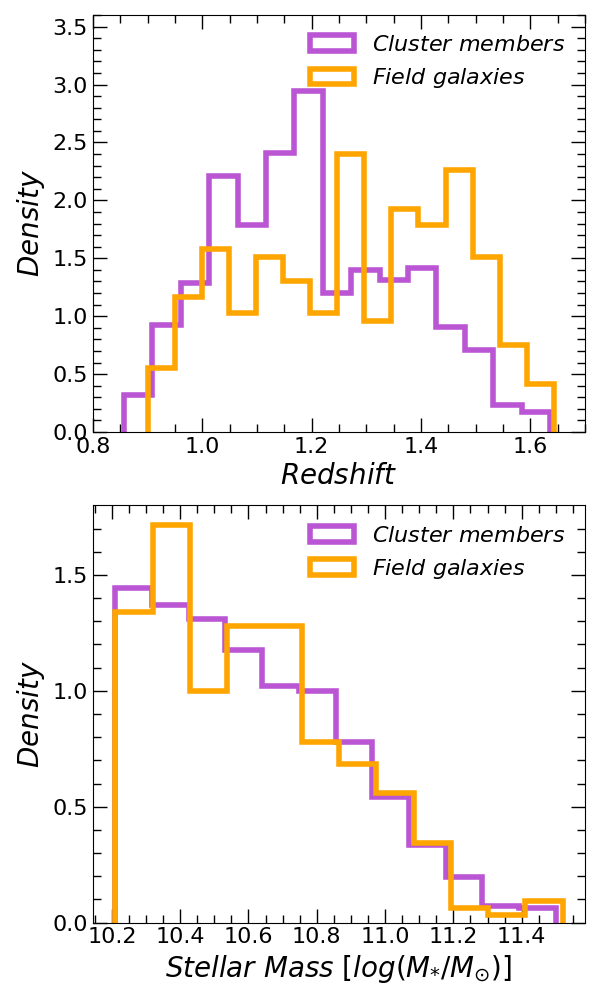}
    \caption{Comparison of our final samples of cluster members (magenta) and field galaxies (orange). Upper panel: Redshift distributions, \textit{p}-value = $1.10e-7$. Lower panel: Stellar masses distributions, \textit{p}-value=0.664. While the redshift distributions are statistically different, we explain in Section \ref{subsec:final_sample} why this is not a source of bias.}
    \label{fig:field_clust_samples}
\end{figure}

To avoid any bias in our results, we tested the similarity of the field and cluster samples by comparing their stellar mass and redshifts distributions. Figure \ref{fig:field_clust_samples} shows the distributions of redshifts and stellar masses for our two samples. We ran a Kolmogorov-Smirnov (K-S) test on those distributions. 
For the stellar masses, we obtained a \textit{p}-value of 0.664 ($\gg 0.05$) meaning that these two distributions are very similar. 
For the redshift distributions, in order to take into account the uncertainties on the photometric redshifts, we ran the K-S test $1,000$ times, while perturbing the photometric redshifts between each run. We obtained a median \textit{p}-value of $4.0e-10$ with a $16^{th}$ and $84^{th}$ percentile of $5.5e-12$ and $3.0e-8$ respectively meaning that these distributions were significantly different (\textit{p}-value $\ll 0.05$). 
This can be explained by the way we built our sample of field galaxies, as explained in Section \ref{subsec:select_field}.  
However, both samples cover the same redshift range, the only difference is that the cluster sample has a redshift distribution peaking at $z\sim 1.1$ while the field sample distribution peaks at $z\sim 1.4$. Since we do not expect galaxies to have a different behavior between these two redshifts, we estimated the (potentially) introduced bias to be negligible.

\begin{table}[ht]
    \caption{Number of galaxies in our final samples.}
    \label{tab:samples}
    \hspace{-10pt}
        \begin{tabular}{|c|c|c|}
            \hline \hline
            Cluster ID & Cluster members & Field galaxies\\
            \hline
            SPTCL-2106 & 144 (26) & 21 (1)\\
            SPTCL-0546 & 177 (24) & 25 (4)\\
            SpARCS-0219 & 64 (6) & 22 (2)\\
            SpARCS-0335 & 68 (9) & 42 (9)\\
            SpARCS-1034 & 164 (9) & 49 (2)\\
            SpARCS-1051 & 73 (23) & 62 (0)\\
            SpARCS-1616 & 97 (35) & 44 (6)\\
            SpARCS-1634 & 57 (39) & 8 (0)\\
            SpARCS-1638 & 51 (31) & 21 (4)\\
            \hline
            TOTAL & 895 (202) & 294 (28)\\
            \hline
        \end{tabular}
        \\
        \vspace{2pt}
        \hspace{21pt}\footnotesize \textbf{Notes.} The numbers in parenthesis are the numbers of spectroscopically confirmed cluster members or field galaxies in each sample.\\
\end{table}

\section{Probing Spatially resolved rest-frame UVJ colors} \label{sec:color_meas}

In this section, we explain how we probed the quenching spatial dependency by measuring magnitudes in two inner and outer apertures in the bands probing rest-frame $U$, $V$ and $J$. We analyzed the cluster and field samples in the same way.
Section \ref{subsec:photo} presents how we chose to divide our spatially resolved galaxies into an inner and an outer aperture to probe the quenching spatial dependency.
Then, Section \ref{subsec:quiescent_aper} explains how, using the normalized observed $IJK$ colors of the apertures, we separated our galaxies into four classes based on the radial location of where quenching is (or is not) occurring.

\subsection{Multi-band Spatially-Resolved Photometry} \label{subsec:photo}

After running \textsc{SUNet}, we obtained deconvolved images of all our galaxies in the three bands probing rest-frame $U$, $V$ and $J$ (see Section \ref{sec:deconv} for more details on the deconvolution process and Table \ref{tab:instruments} for the three observed bands used to probe rest-frame $U$, $V$ and $J$ for each cluster). 

In order to probe the quenching spatial dependency of our galaxies, we measured fluxes in two elliptical apertures. In order to probe consistent colors, for each source, we used the same apertures to measure the fluxes in all three bands. We defined the apertures based on the Kron aperture of the $Ks$ band as it probes the rest-frame near-IR ($\sim 1\mu m$) that is known to be a good tracer of the center of mass of galaxies (\citealt{van_der_wel_stellar_2023}). For each galaxy, we built a smaller aperture defined as an ellipse with the same shape as the $Ks$ Kron aperture containing 50\% of the Kron flux and a similar larger aperture containing 90\% of the Kron flux.

Practically, for each source and in each image, we first estimated and removed any remaining spurious post-deconvolution background noise. Then, we used the $Ks$ band to identify the position of the source of interest in the cutout. To take into account the fact that some galaxies were asymmetric, we used two source detection thresholds to measure the position and extent of the galaxy: a lower threshold to detect the global shape of the galaxy, this detection was used to build the segmentation map for the Kron and larger apertures, and a higher threshold to detect the position of the center of mass of the galaxy, used for the smaller apertures. As most of our galaxies were symmetric in the $Ks$ band, these two positions were identical (or with a difference less than a pixel) in $>90\%$ of cases. 
In the other cases, the few pixels of a difference allowed to take into account the asymmetry of the galaxies and avoid having the central luminosity peak divided between the two apertures.
We used the same apertures for the two other bands probing rest-frame $U$ and $V$, adjusting their sizes in pixel to match the different pixel scales. 
However, we decided to adjust the pixel coordinates of the apertures in these images. 
This was required as each cutout had a different pixel scale, was deconvolved using a different PSF, had small astrometry differences and were covering different portions of the sky based on the model used to deconvolve them. 
To estimate the significance of these adjustments, we used the WCS to convert the pixel coordinates into sky coordinates. We found these adjustments to be very small, typically no larger than 0.05 arcseconds ($\sim1$ pixel).
Finally, we defined the inner region as the smaller elliptical aperture and the outer region as the annuli encompassed between the smaller and larger apertures.
Additionally, we defined the integrated flux as the total flux in the larger aperture. 

\begin{figure}[htb]
    \centering
    \includegraphics[width=0.99\linewidth]{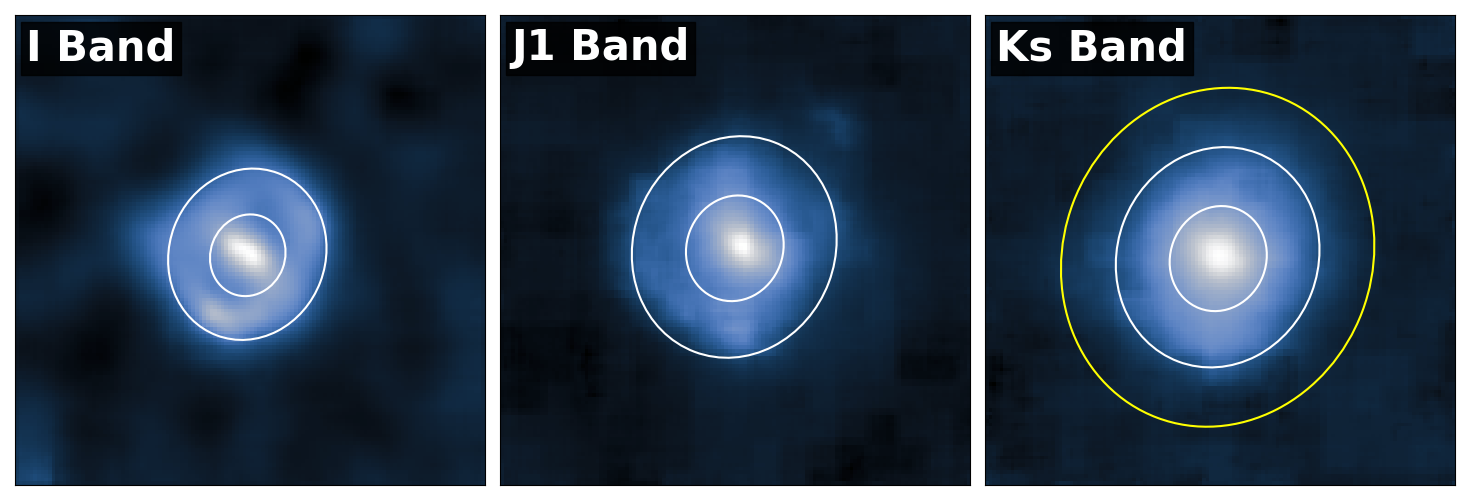}
    \includegraphics[width=0.99\linewidth]{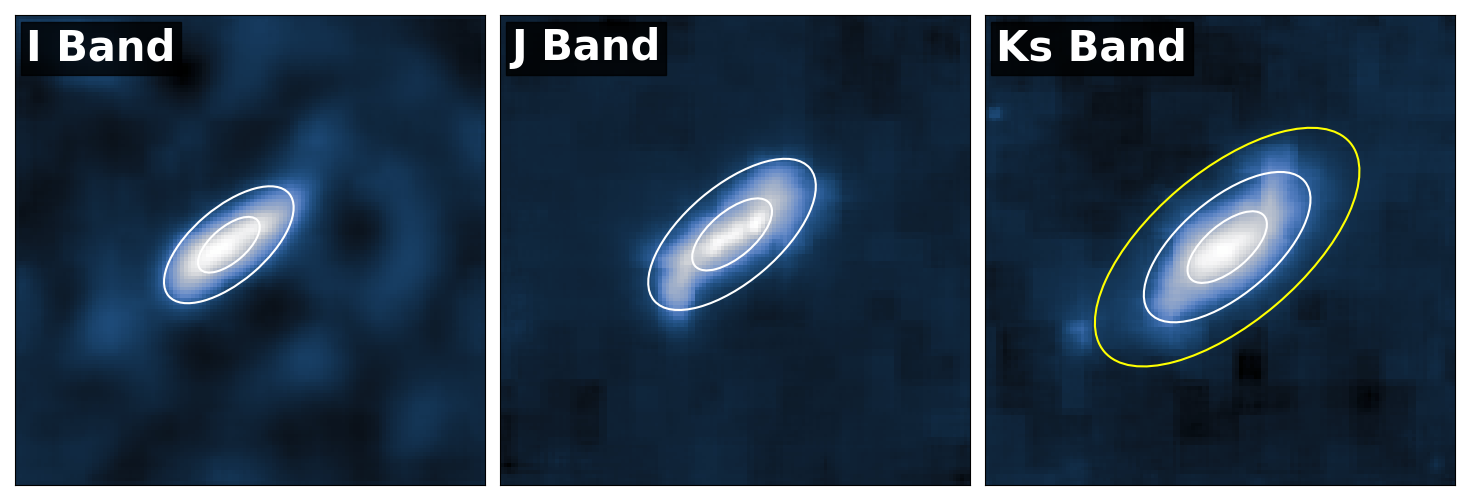}
    \caption{Deconvolved images probing rest-frame $UVJ$ of two galaxies. Upper row: a face-on galaxy, member of the cluster SPTCL-0546. Lower row: an edge-on galaxy, member of the cluster SpARCS-0219. From left to right, we show the image in the $I$, $J1$/$J$ and $Ks$ bands. The white ellipses show the inner and outer apertures encompassing $50\%$ and $90\%$ of the flux in the Kron aperture of the $Ks$ band, dividing the galaxy into an inner region and an outer annuli. The Kron aperture of the $Ks$ band is shown in yellow in the right panels. These two examples show that our color apertures consistently encompass most of the light in the target galaxies.}
    \label{fig:cutouts_aper}
\end{figure}
 
In Figure \ref{fig:cutouts_aper} we show the deconvolved images in the three bands for two galaxies with the inner and outer apertures appearing in white. This figure shows that our way of drawing the apertures works well for both face-on and edge-on galaxies as well as for different pixel scales.

The choice of dividing the galaxies into two apertures encompassing 50\% and 90\% of the \textit{Ks} Kron aperture flux was made to avoid any bias and to reduce uncertainties linked to small resolution differences. Indeed, when performing multi-band analysis, it is crucial that all the bands have a similar spatial resolution. This is usually achieved by PSF-matching the images. 
With SUNet, the target resolution is the resolution of the training sample, however, the actual resolution depends on the quality of the ground based images ($\sim 0.5^"-0.8^"$ for the GOGREEN images, see Table \ref{tab:instruments}) and on the S/N. 
This means that two galaxies extracted from the same original image could be deconvolved to a slightly different resolution based on their S/N making measuring the exact spatial resolution of these images extremely challenging. 
However, since we made sure that the target resolution for each model was similar (0.11" for the optical model and 0.09" for the near-IR models) and that the spatial resolution of the original ground-based images are close (see Table \ref{tab:instruments}), we assumed that the deconvolved images had very similar spatial resolutions.
Moreover, if we assume that the deconvolution achieved to reach $\sim 2\times$ the target resolution (as estimated by \cite{akhaury_ground-based_2024} for the optical model), we can assume that our deconvolved images have a resolution $\sim 0.2"$. Since the typical diameter of the inner apertures of our galaxies was $0.54"$, or $\sim 2.7\times$ the 0.2" resolution limit, we can consider the bias or uncertainties due to small resolution differences to be negligible. To be conservative, we made some tests by assuming that the resolution differences after deconvolution were the same as before the deconvolution. We found that "matching'' the resolutions following this assumption led to estimating the uncertainties linked to spatial resolution differences on the colors to be 0.035mag on average.

\subsection{Identifying Quenched regions} \label{subsec:quiescent_aper}

After measuring the fluxes in the $I$, $J$ and $K$ bands, we normalized the $I-J$ and $J-K$ observed colors, as described in Section \ref{subsec:select_bands} and used the wedge that we defined with Eq. \ref{eq:ijk_1}, \ref{eq:ijk_2}, and \ref{eq:ijk_3} to identify the quiescent regions (or lack thereof) in each galaxy.

\begin{figure}[htb]
    \centering
    \includegraphics[width=0.85\linewidth]{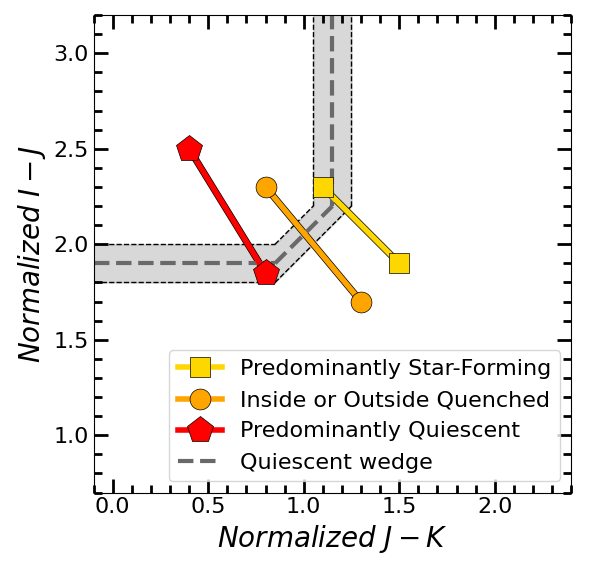}
    \caption{Definition of a significant spatial quenching gradient. Each pair of markers represents the inner and outer colors of a schematic galaxy in the normalized $(I-J, J-K)$ plane. The thick gray region defines the transition boundary between the star-forming and quiescent classifications. Galaxies are classified as having a significant quenching gradient only when their inner and outer regions lie on opposite sides of this boundary, as illustrated by the orange circles. Galaxies whose resolved colors remain on the same side or land on the boundary are classified as predominantly star-forming (yellow squares) or predominantly quiescent (red pentagons). Galaxies with significant gradients are subsequently classified as inside- or outside-quenched according to which region lies in the quiescent portion of the diagram. See Section \ref{subsec:quiescent_aper} for additional details.}
    \label{fig:quies_grad_ex}
\end{figure}

We then divided the galaxies into four classes based on the star-forming status of their inner and outer regions; the galaxies that had both regions star-forming were classified as \textit{predominantly star-forming galaxies} (pSF) and the galaxies with both regions quiescent were classified as \textit{predominantly quiescent galaxies} (pQ). This classification was applied independently from the original classification based on their integrated rest-frame $UVJ$ colors mentioned in Section \ref{subsec:select_bands}.
The galaxies that had one region that was classified as star-forming and the other as quiescent were divided into \textit{outside quenched galaxies} (OQ) and \textit{inside quenched galaxies} (IQ). 
In order to account for the uncertainties on the colors and to ensure that the galaxies classified as outside or inside quenched had a significant quenching gradient, we replaced the quiescent-star forming dividing line with a thick boundary that extended $0.1$mag on either side of the dividing line (see Figure \ref{fig:quies_grad_ex}).
The $0.1$mag thickness was chosen as it was representative of the typical uncertainty on the colors. 

Indeed, the main sources of uncertainties on the colors are 1) the deconvolution process color recovery efficiency (see Section \ref{subsec:new_models} and Appendix \ref{app:test_grad}) estimated at 0.035mag on average, 2) the uncertainty resulting from the normalization shifts (see Section \ref{subsec:select_bands}) estimated at 0.11mag on average for cluster members and 0.17mag for field galaxies and 3) the uncertainty linked to the small resolution differences between the three bands (see Section \ref{subsec:photo}) estimated at 0.035mag on average. 
This means that the typical uncertainty on colors is 0.121mag for cluster members and 0.177mag for field galaxies, hence a minimum of 0.2mag for a gradient (as required by our boundary) corresponds to a $1.65\sigma$ significance in clusters and $1.13\sigma$ significance in the field.
Moreover, when looking at the gradients of our selected OQ and IQ galaxies using this boundary, we found that 90\% of them are greater than 0.42mag, meaning that the vast majority of galaxies with a quenching gradient have a gradient that is at least twice the width of our boundary and has a $\geq 3.5\sigma$ ($\geq 2.4\sigma$) significance in clusters (in the field).
The borders inside and outside of the resulting thick quiescent wedge are defined using Eq. \ref{eq:ijk_1}, \ref{eq:ijk_2} and \ref{eq:ijk_3} with $(a, b, c, d) = (1.00, 1.15, 1.05, 2.00)$ and $(1.00, 0.95, 1.25, 1.80)$ respectively.

Only the galaxies that had one region on each side of the thick wedge were considered outside or inside quenched. The galaxies that had a region with their colors falling on the thick wedge were classified based on their integrated colors.
As an example, Figure \ref{fig:quies_grad_ex} shows the inner and outer colors of three schematic galaxies (we do not mention which marker is the inner or outer region as it is not relevant here), where the colors of the two regions fall on a different side of the wedge defined in Section \ref{subsec:select_bands}. 
This means that these galaxies should be considered as having a quenching gradient. However, two of them have the colors of at least one region falling on the thick wedge, meaning that there is an uncertainty on the star-forming status of these regions. Hence, to be conservative, we classified them as predominantly star-forming (yellow galaxy with square markers) or predominantly quiescent (red galaxy with pentagon markers) depending on their integrated colors. 
On the contrary, the orange galaxy with circular markers has a region on each side of the thick wedge, hence should indeed be classified has either outside or inside quenched depending on which region has its colors inside the quiescent wedge.

\begin{table}[ht]
    \caption{Number of cluster member galaxies.}
    \label{tab:Ngal_cluster}
    \hspace{-20pt}
        \begin{tabular}{|c|c|c|c|c|}
            \hline \hline
            Cluster ID & $N_{pSF}$$^*$ & $N_{IQ}$$^*$ & $N_{OQ}$$^*$ & $N_{pQ}$$^*$\\
            \hline
            SPTCL-2106 & 34 & 5 & 28 & 77\\
            SPTCL-0546 & 33 & 8 & 22 & 114\\
            SpARCS-0219 & 27 & 4 & 7 & 26\\
            SpARCS-0335 & 36 & 3 & 6 & 23\\
            SpARCS-1034 & 61 & 3 & 38 & 62\\
            SpARCS-1051 & 19 & 0 & 20 & 34\\
            SpARCS-1616 & 24 & 0 & 30 & 43\\
            SpARCS-1634 & 13 & 0 & 10 & 34\\
            SpARCS-1638 & 11 & 2 & 17 & 21\\
            \hline
            TOTAL & 258 & 25 & 178 & 434\\
            \hline
        \end{tabular}
        \\
        \vspace{2pt}
        \hspace{25pt}\footnotesize $^*$ See Section \ref{subsec:quiescent_aper} for acronym definitions.\\
\end{table}

In Table \ref{tab:Ngal_cluster} and \ref{tab:Ngal_field}, we show the number of galaxies in each class for every cluster for cluster members and field galaxies respectively.

\begin{figure*}[htb]
    \centering
    \includegraphics[width=0.99\linewidth]{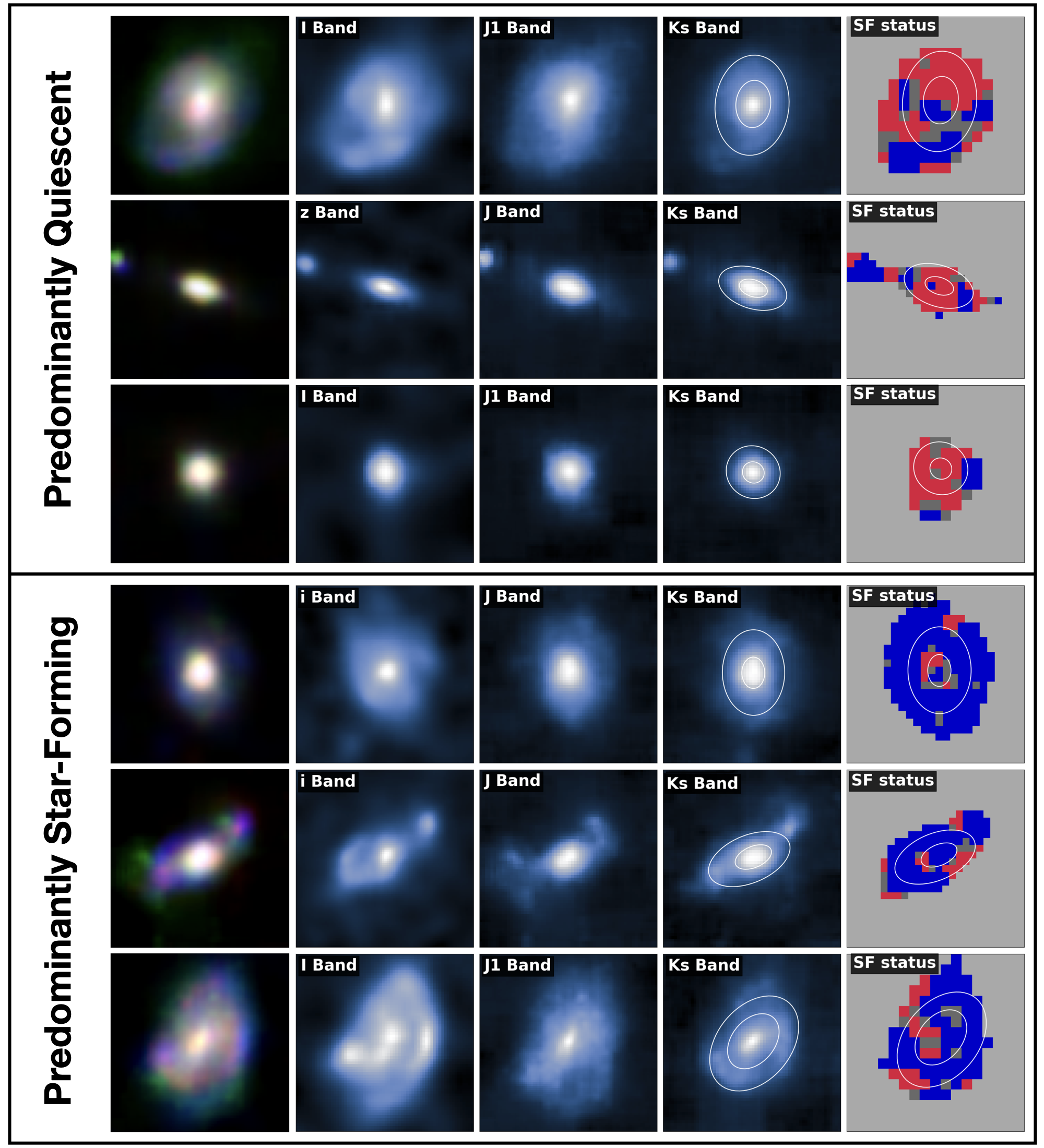}
    \caption{Three examples of deconvolved galaxies representative of the predominantly quiescent (upper three rows) and predominantly star-forming classes (lower three rows). For each galaxy, the panels show, from left to right, the RGB image, the deconvolved cutouts in the $I$, $i$ or $z$ bands, the deconvolved cutouts in the $J$ or $J1$ bands and the deconvolved cutouts in the $Ks$ band. The panels on the right show the spatially resolved star-forming status of the galaxies with red pixels being quiescent, blue pixels being star-forming and gray pixels undetermined (meaning that their colors fall on the thick region defined in section \ref{subsec:quiescent_aper} and Figure \ref{fig:quies_grad_ex}), the pixels are 0.2" wide, corresponding to the resolution limit mentioned in Section \ref{subsec:photo}. The RGB and SF status panels show clearly how predominantly quiescent galaxies are dominated by redder light and quiescent pixels while predominantly star-forming galaxies are dominated by star-forming pixels. See Section \ref{subsec:quiescent_aper} for additional details.}
    \label{fig:Deconv_exQpSF}
\end{figure*}

\begin{figure*}[htb]
    \centering
    \includegraphics[width=0.99\linewidth]{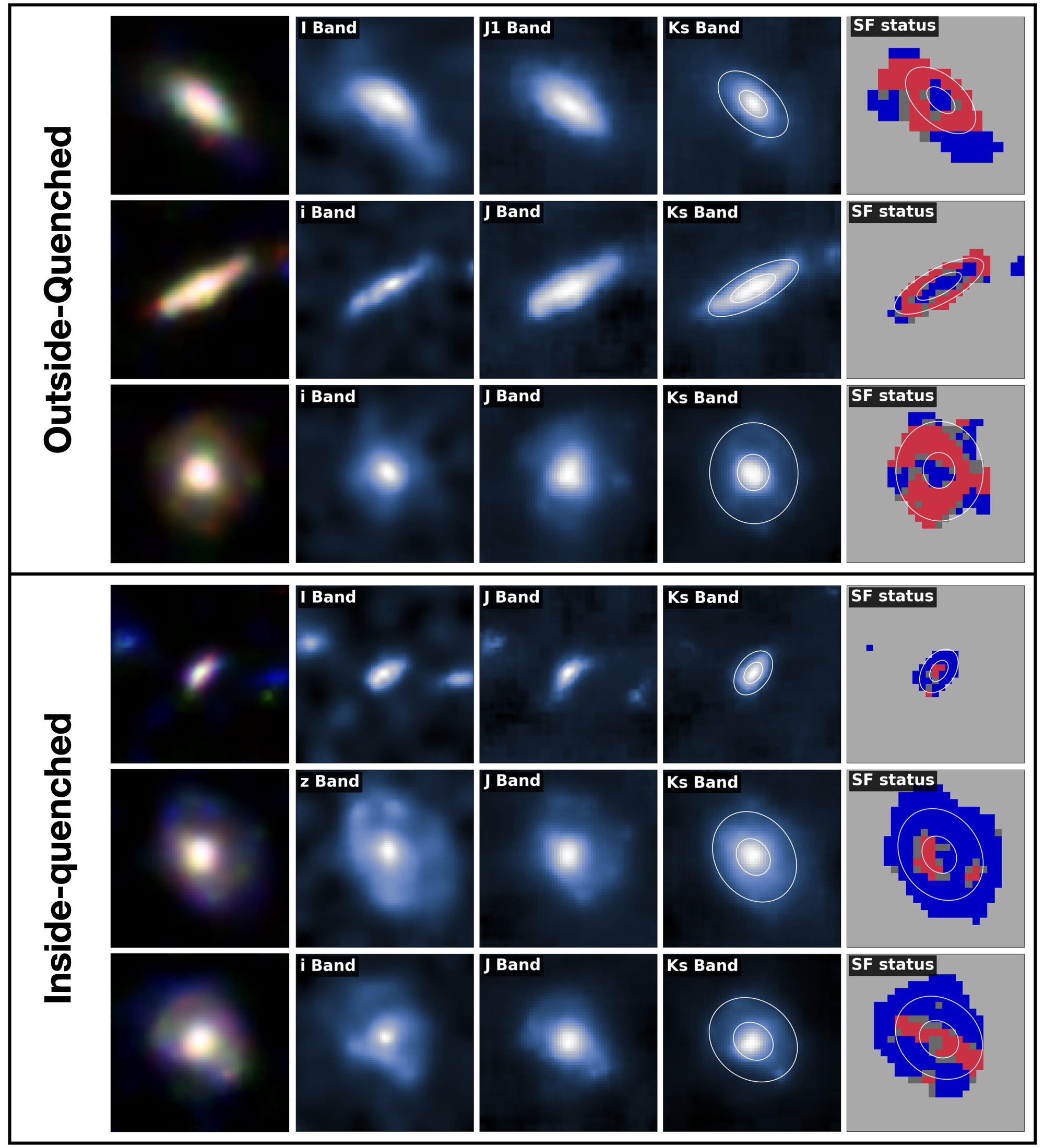}
    \caption{Three examples of deconvolved galaxies representative of the predominantly outside quenched (upper three rows) and inside quenched classes (lower three rows). Similar to Figure \ref{fig:Deconv_exQpSF}. The RGB and SF status panels show clearly how outside quenched galaxies have disks dominated by redder light and quiescent pixels with cores dominated by star-forming pixels, while inside quenched galaxies have disks dominated by star-forming pixels with cores dominated by quiescent pixels. See Section \ref{subsec:quiescent_aper} for additional details. }
    \label{fig:Deconv_exOQIQ}
\end{figure*}

In Figure \ref{fig:Deconv_exQpSF} and \ref{fig:Deconv_exOQIQ}, we show cutouts of three representative galaxies for each class. 
For every galaxy, we show their cutouts in the three bands we used to probe the rest-frame $U - V$ and $V - J$ colors after deconvolution. 
To guide the eye, we show in the $Ks$ cutout the inner and outer apertures used to define the two regions we studied. We also resized and aligned the cutouts. 
Additionally, we matched the pixel scales of the three deconvolved bands and normalized them to build the RGB images that we show on the left panels of Figure \ref{fig:Deconv_exQpSF} and \ref{fig:Deconv_exOQIQ}. Finally, we resampled the images in 0.2" pixels to probe the spatially resolved colors. We chose 0.2" as it corresponded to the estimated resolution limit of the deconvolved images as explain in Section \ref{subsec:photo}. 
In the right panels of Figure \ref{fig:Deconv_exQpSF} and \ref{fig:Deconv_exOQIQ}, we color coded each of these 0.2" pixels based on their position in the observed normalized $IJK$ color-color diagram. We calculated the colors of every pixel that had a magnitude lower than 30 in all three bands, fainter pixels are showed in light gray. 
Then, the pixels falling in the quiescent wedge are showed in red and those falling in the star-forming region are showed in blue. 
Finally, the pixels with colors falling on the thick limit defined in Figure \ref{fig:quies_grad_ex} are showed in dark gray as uncertainties did not allow us to conclude on their star-forming activity.

\begin{table}[ht]
    \caption{Number of field galaxies.}
    \label{tab:Ngal_field}
    \hspace{-20pt}
        \begin{tabular}{|c|c|c|c|c|}
            \hline \hline
            Cluster ID & $N_{pSF}$$^*$ & $N_{IQ}$$^*$ & $N_{OQ}$$^*$ & $N_{pQ}$$^*$\\
            \hline
            SPTCL-2106 & 14 & 1 & 4 & 2\\
            SPTCL-0546 & 12 & 3 & 3 & 7\\
            SpARCS-0219 & 5 & 0 & 3 & 14\\
            SpARCS-0335 & 16 & 2 & 8 & 16\\
            SpARCS-1034 & 28 & 2 & 7 & 12\\
            SpARCS-1051 & 23 & 0 & 19 & 20\\
            SpARCS-1616 & 17 & 2 & 5 & 20\\
            SpARCS-1634 & 3 & 0 & 0 & 5\\
            SpARCS-1638 & 7 & 0 & 5 & 9\\
            \hline
            TOTAL & 125 & 10 & 54 & 105\\
            \hline
        \end{tabular}
        \\
        \vspace{2pt}
        \hspace{25pt}\footnotesize $^*$ See Section \ref{subsec:quiescent_aper} for acronym definitions.\\
\end{table}

From the RGB images of Figure \ref{fig:Deconv_exQpSF}, one can see that the predominantly quiescent galaxies are dominated by the redder band while the predominantly star-forming galaxies show more extended and bluer disks. 
When looking at the star-forming status panels, we confirm that predominantly quiescent galaxies are dominated by red quiescent pixels and predominantly star-forming galaxies are dominated by blue star-forming pixels. 
The fact that we see blue pixels in the predominantly quiescent galaxies and red pixels in the predominantly star-forming galaxies is not unexpected. Indeed, from the way we defined the classes, to be considered inside quenched or outside quenched, galaxies needed to have a large enough gradient to go over the thick limit defined in Figure \ref{fig:quies_grad_ex}. If they did not, they were considered predominantly star-forming or quiescent. Hence, these latter classes will include some galaxies with a small number of pixels of the other color.

From the RGB images of Figure \ref{fig:Deconv_exOQIQ}, one can see that the outskirts of the outside quenched galaxies are dominated by the redder band while the outskirts of inside quenched galaxies are dominated by the blue and green bands.
When looking at the star-forming status panels, we confirm the classes of the galaxies. Indeed, outside quenched galaxies have their inner apertures filled with star-forming blue pixels while their outer annuli are filled with quiescent red pixels while the inside quenched galaxies show the opposite: quiescent red pixels in their inner apertures and star-forming blue pixels in their outer annuli.

\section{Results} \label{sec:results}

In this section, we present how getting access to resolved colors allowed us to better probe the quenching processes affecting galaxies by identifying which regions of the galaxies were quiescent. 
More precisely, in Section \ref{subsec:spatial_colors}, we show how we identified galaxies with a clear quenching gradient independently of environment and how a large fraction of galaxies classified as quiescent when looking at their integrated colors actually show signs of star-formation when looking at their resolved colors.

Then, in Section \ref{subsec:quench_frac}, we investigate the stellar mass dependence of the quenching fraction in clusters and measure the quenched fraction excess compared to the field sample. 

Finally, in Section \ref{subsec:env_quench}, we detailed how we observed an excess of outside quenched galaxies in clusters, especially at lower stellar masses.

\begin{figure}[htb]
    \centering
    \includegraphics[width=0.8\linewidth]{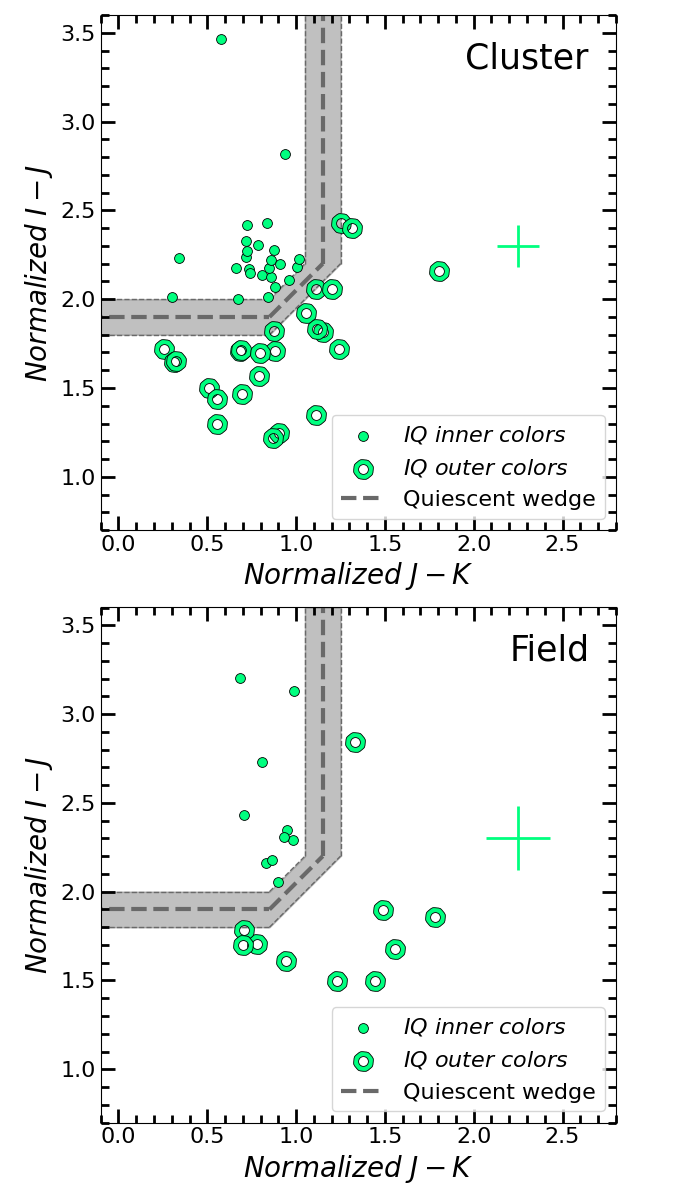}
    \caption{Spatially resolved colors of inside-quenched galaxies (See Section \ref{subsec:quiescent_aper}). Normalized observed $(I-J, J-K)$ resolved colors of galaxies classified as inside quenched, shown separately for cluster members (upper panel) and field galaxies (lower panel). The cross in each panel shows the typical uncertainty on the colors. For each galaxy, the inner region lies within the quiescent portion of the diagram while the outer region lies in the star-forming region, demonstrating a clear inside-out quenching gradient. The gray region shows the transition boundary used to identify a significant color difference between the inner and outer classifications. Inside-quenched galaxies are found in both environments, with no clear difference in their resolved color distributions.}
    \label{fig:IOQ_res_colors}
\end{figure}

\subsection{Identifying Partially Quiescent Galaxies}\label{subsec:spatial_colors} 

\begin{figure*}[htb]
    \centering
    \includegraphics[width=0.99\linewidth]{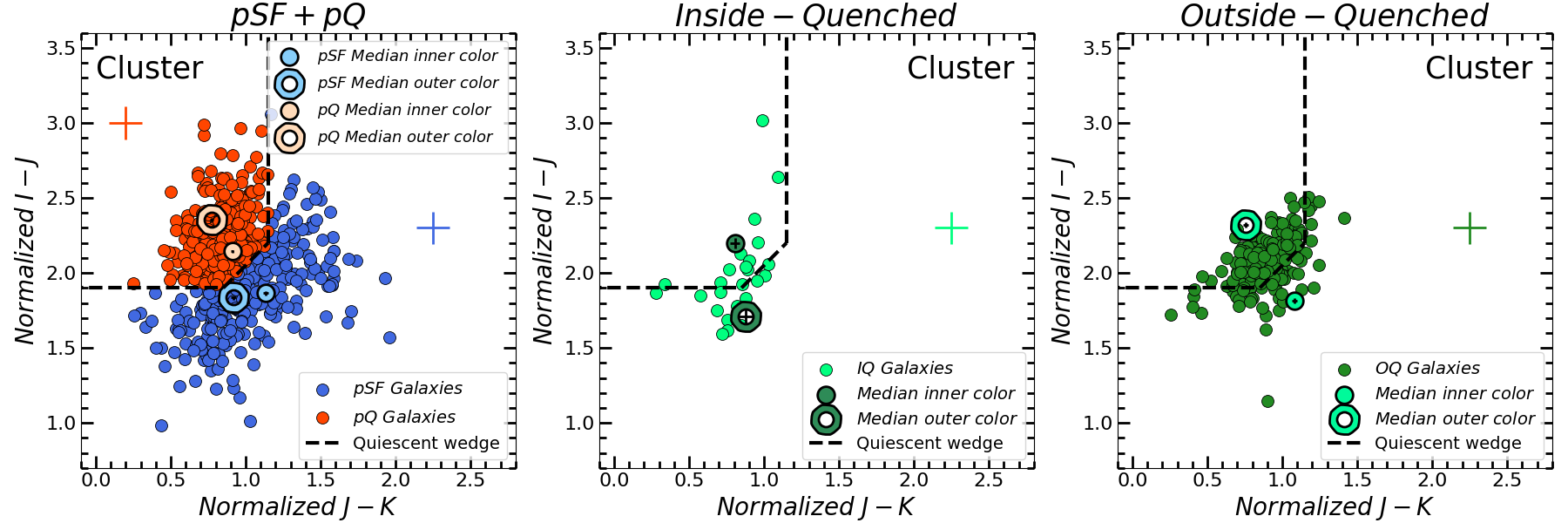}
    \includegraphics[width=0.99\linewidth]{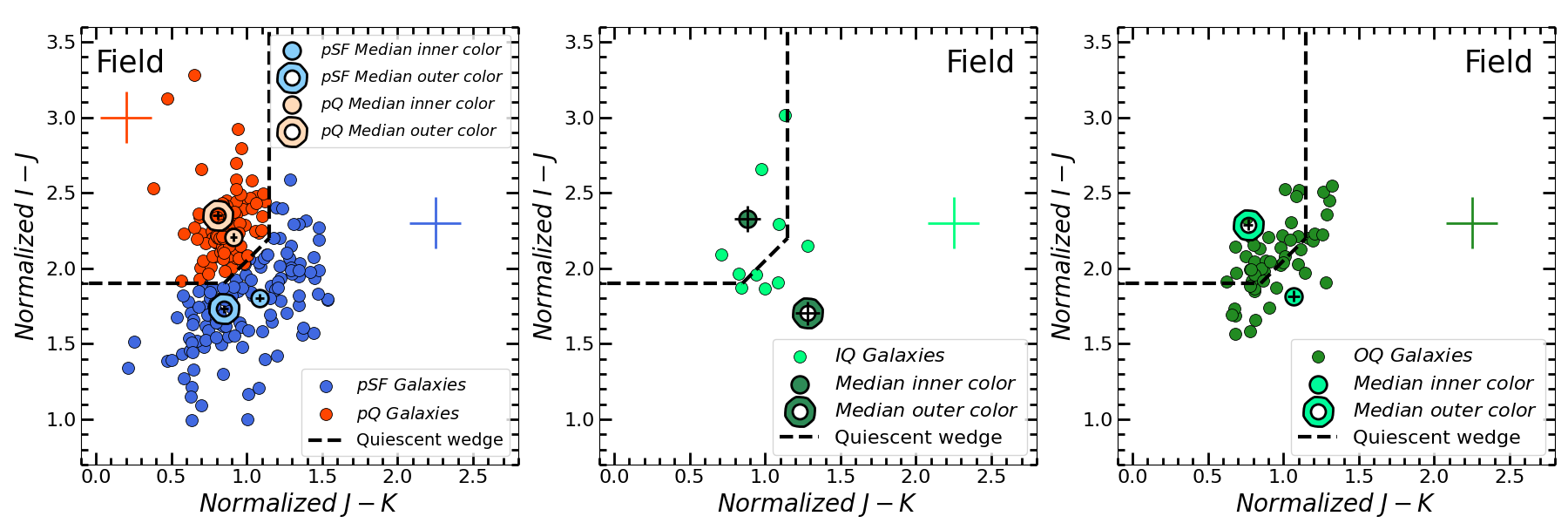}
    \caption{Integrated colors do not uniquely identify the resolved star-forming state of galaxies. Normalized observed integrated ($I - J$, $J - K$) color-color diagram for the four classes of galaxies defined from the spatially resolved photometry (see Section \ref{subsec:quiescent_aper}). Predominantly quiescent and predominantly star-forming galaxies are shown in the left panels, inside quenched galaxies in the middle panels and outside quenched galaxies in the right panels. The crosses in each panels shows the typical uncertainties on the colors. Cluster members are shown in the upper panels and field galaxies in the lower panels. The bold circle and ring in each panel show the median resolved colors of the inner and outer regions, respectively. A substantial fraction of inside- and outside-quenched galaxies have integrated colors within the quiescent wedge despite retaining star formation in part of the galaxy. This is evidenced by the clear separation of the spatially resolved median colors and demonstrates that integrated colors can misclassify partially quenched galaxies as fully quiescent. See Section \ref{subsec:spatial_colors} for additional details.}
    \label{fig:IJK_int}
\end{figure*}

In Section \ref{subsec:quiescent_aper}, we showed how the deconvolution of our galaxies allowed us to identify galaxies with clear quenching gradients. 
Being able to identify these galaxies was crucial as they hold information on the quenching processes occurring in their environments. These galaxies were divided in two classes.

In the first class, we had the inside quenched galaxies where the colors of the inner region were typical of a quenched galaxy and the colors of the outskirts were typical of a star-forming galaxy. We found these galaxies in all environments.
Figure \ref{fig:IOQ_res_colors} shows the inner and outer colors of these galaxies for cluster members and field galaxies in the upper and lower panels respectively. All of these galaxies show a significant quenching gradient with their inner colors well inside the quiescent wedge and their outer colors away from the wedge. 
Despite the low statistics (especially in the field sample), we ran a 2D K-S test to probe the similarities of the color distributions between environments. We used an N-dimensional generalization of the two-sample K-S test described by \cite{fasano_multidimensional_1987} and developed by \cite{chow_decameter-sized_2025}\footnote{The python code for the 2D K-S test can be found at \url{https://github.com/wmpg/fasano-franceschini-test}}. 
After running the test 100 times, the mean \textit{p}-values were 0.33 and 0.09 for the inner and outer color distributions respectively. While the \textit{p}-value for the outer color distributions is close to the 0.05 threshold, we obtained a \textit{p}-value $< 0.05$ in less than 10\% of the cases. This means that we can not rule out the null hypothesis that these samples are derived from a common parent sample.
Hence, we did not observe any significant differences in terms of colors between inside quenched galaxies in clusters and in the field.
Additionally, we compared the inner colors of IQ galaxies to the colors of the pQ galaxies as well as their outer colors to the colors of the pSF galaxies. We found them to be indistinguishable, confirming again the quenching gradient in these galaxies.

In Figure \ref{fig:IJK_int} the middle panels show the normalized integrated $IJK$ colors of the IQ galaxies. 
They all lie close to the border of the quiescent wedge, equally divided between each side, as expected since roughly 50\% of the light is in each region of the $IJK$ diagram. 
We also show how the median inner and outer colors are distinct and, as expected, lie on a different side of the quiescent wedge. We note that, consistently with the resolved colors, we can't reject the null hypothesis that the integrated colors distributions are the same between clusters and field (\textit{p}-value $= 0.14$).

\begin{figure}[htb]
    \centering
    \includegraphics[width=0.8\linewidth]{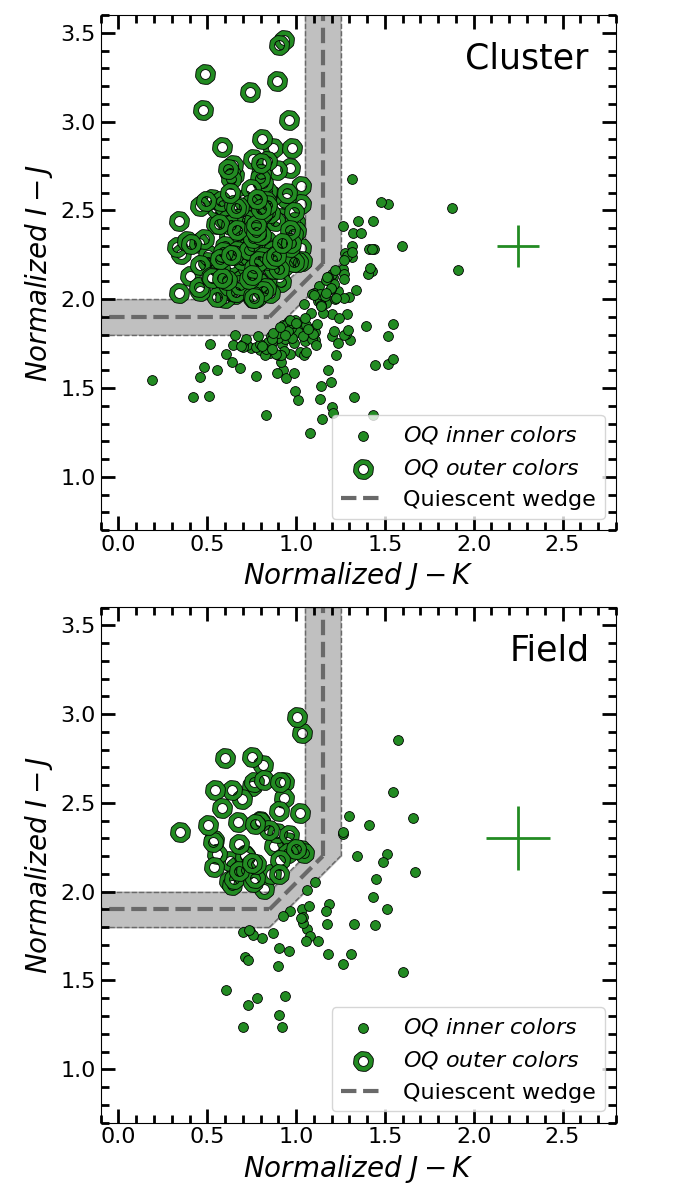}
    \caption{Spatially resolved colors of outside quenched galaxies (see Section \ref{subsec:quiescent_aper}). Normalized observed $(I-J,J-K)$ colors of galaxies classified as outside quenched, shown for cluster members (upper panel) and field galaxies (lower panel). The cross in each panel shows the typical uncertainty on the colors. For each galaxy, the outer region lies within the quiescent portion of the diagram while the inner region lies in the star-forming region, producing the resolved color gradient used to define this class. The gray region shows the transition boundary used to identify a significant difference between the inner and outer classifications. Outside quenched galaxies are found in both environments, with no significant difference in their resolved color distributions.}
    \label{fig:OIQ_res_colors}
\end{figure}

In the second class, we had the outside quenched galaxies, with a reverse configuration compared to the inside-quenched galaxies: the colors of the inner aperture were typical of a star-forming galaxy and the colors of their outskirts were typical of a quiescent galaxy. We also found these galaxies in all environment.
Similarly to the IQ galaxies, Figure \ref{fig:OIQ_res_colors} shows that all of these galaxies have a significant quenching gradient. 
We also ran 100 2D K-S tests to compare the color distributions between clusters and field. We found the average \textit{p}-values of 0.17 and 0.77 for the inner and outer color distributions respectively meaning that there is no significant differences between these two samples. 
Here again, we compared the inner and outer colors of the OQ galaxies to the colors of the pSF and pQ galaxies respectively. We found that the outer colors of OQ galaxies are similar to the colors of pQ galaxies, however, their inner colors are different from the colors of the pSF galaxies. The difference between the median position of the inner colors of OQ galaxies and pSF galaxies is 0.1mag along the $J-K$ axis. This means that the average inner color of an OQ galaxy is closer to the quiescent wedge than the average inner color of a pSF galaxy.

In Figure \ref{fig:IJK_int} the right panels show the normalized integrated $IJK$ colors of these galaxies. Similarly to the IQ galaxies, they all lie near the border of the quiescent wedge. 
We noted that, while for cluster members a larger fraction of these galaxies lies inside the quiescent region (2/3) of the color-color diagram when the field sample is nearly equally divided, the distributions of integrated colors are similar in all environment (\textit{p}-value $= 0.25$). Hence, we do not observe any significant differences in terms of colors between outside quenched galaxies in clusters and in the field.
We also show how the median inner and outer colors are distinct and, as expected, lie on a different side of the quiescent wedge. 

In the left panels of Figure \ref{fig:IJK_int} we show the normalized integrated $IJK$ colors of the predominantly star-forming galaxies in blue and predominantly quiescent galaxies in red. 
Similarly to the intermediate galaxies, we ran 100 2D K-S tests to probe a potential environmental effect. 
We obtained the mean \textit{p}-values of 0.17 and 0.32 for the integrated predominantly star-forming and quiescent color distributions respectively, demonstrating that these distributions are not statistically different. The environment does not affect the color distributions of predominantly star-forming or quiescent galaxies.
We also show the median of the colors of the inner and outer regions for these two populations of galaxies. One can notice that, since these galaxies do not show any evidence of quenching gradients, the mean color difference between inner and outer regions (distance between the bold shaded ring and circular markers in Figure \ref{fig:IJK_int}) is smaller than for the partially quenched galaxies ($\lesssim 0.3$~mag for pSF and pQ galaxies versus $\gtrsim 0.6$~mag for IQ and OQ galaxies). 

As mentioned above and illustrated in Figure \ref{fig:IJK_int}, the galaxies that we identified as partially quiescent and partially star-forming do not necessarily have integrated colors consistent with star-forming galaxies. 
In Table \ref{tab:class_comp_int_res} we show how the quiescent and star-forming galaxies, as defined by their integrated $IJK$ colors, are distributed among the different classes. 
We note that $(21.5\pm1.6)\%$ of the globally quiescent galaxies in all environments actually show signs of star-formation in their center ($(21.6\pm1.8)\%$ and $(20.9\pm3.8)\%$ for cluster and field galaxies respectively) and $(2.4\pm0.6)\%$ of them show signs of star-formation in their outskirts ($(2.1\pm0.7)\%$ and $(3.6\pm1.9)\%$ for cluster and field galaxies respectively).
Similarly, we note that $(16.6\pm1.8)\%$ of globally star-forming galaxies are partially quenched in their outskirts ($(16.9\pm2.1)\%$ and $(16.1\pm3.3)\%$ for cluster and field galaxies respectively) and $(3.7\pm1.0)\%$ are quenched in their core ($(4.0\pm1.2)\%$ and $(3.2\pm1.7)\%$ for cluster and field galaxies respectively).

\begin{table}[ht]
    \caption{Classes of Globally Star-forming (SF) and Quiescent (Q) galaxies.}
    \label{tab:class_comp_int_res}
    \hspace{-15pt}
        \begin{tabular}{|c|c|c|c|c|}
            \hline \hline
            Class & SF & Q & SF & Q\\
             & Clusters & Clusters & Field & Field\\
            \hline
            ${N_{pSF}}^*$ & 258 & 0 & 125 & 0\\
            ${N_{IQ}}^*$ & 13 & 12 & 5 & 5\\
            ${N_{OQ}}^*$ & 55 & 123 & 25 & 29\\
            ${N_{pQ}}^*$ & 0 & 434 & 0 & 105\\
            \hline
            TOTAL & 326 & 569 & 155 & 139\\
            \hline
        \end{tabular}
        \\
        \vspace{3pt}
        \hspace{31pt}\footnotesize \textbf{Notes.} Each column shows how the galaxies with integrated $I - J$ and $J - K$ colors consistent with being globally quiescent or star-forming are distributed among our four classes after probing their resolved colors.\\
        \vspace{1pt}
        \hspace{31pt}\footnotesize $^*$ See Section \ref{subsec:quiescent_aper} for acronym definitions.\\
\end{table}

We calculated the fraction of galaxies belonging to each class: predominantly star-forming, inside quenched, outside quenched or predominantly quiescent in the clusters and in the field. These fractions were calculated from the numbers in Table \ref{tab:Ngal_cluster} and \ref{tab:Ngal_field} on the whole sample of galaxies, with $M_{*} > 10^{10.2}M_{\odot}$. We show the fractions in Table \ref{tab:group_frac} where the uncertainties are the binomial 68\% confidence interval (we note that for the rest of this paper, all uncertainties on fractions are binomial 68\% confidence intervals). 

\begin{table}[ht]
    \caption{Prevalence of each class of galaxies.}
    \label{tab:group_frac}
    \hspace{10pt}
        \begin{tabular}{|c|c|c|}
            \hline \hline
            Group & Cluster & Field\\
            \hline
            pSF$^*$ & $(28.8 \pm 1.6)\%$ & $(42.5 \pm 3.0)\%$\\
            IQ$^*$ & $(2.8\pm 0.6)\%$ & $(3.4\pm 1.2)\%$\\
            OQ$^*$ & $(19.9 \pm 1.4)\%$ & $(18.4\pm2.4)\%$\\
            pQ$^*$ & $(48.5 \pm 1.7)\%$ & $(35.7 \pm 2.9)\%$\\
            \hline
            \end{tabular}
        \\
        \vspace{3pt}
        \hspace{42pt}\footnotesize $^*$ See Section \ref{subsec:quiescent_aper} for acronym definitions.\\
        
\end{table}

From these fractions, we noted some trends.
First, the predominantly quiescent fraction is significantly larger ($3.8\sigma$ difference) in the clusters than in the field. Indeed, nearly half of the cluster members are predominantly quiescent while only $\sim 35\%$ of the field galaxies are. 
Inversely, the predominantly star-forming fraction is significantly higher ($4\sigma$ difference) in the field than in clusters as we had more than $40\%$ of predominantly star-forming galaxies in the field while they represented less than $30\%$ of cluster members. 
Finally, we noted that the fractions of partially quenched galaxies were similar in all environments ($0.5\sigma$ differences) with the outside quenched fraction being around $20\%$ and the inside quenched fraction around $3\%$.
This means that, despite a higher fraction of predominantly quiescent galaxies in clusters, we did not observe an excess of partially quenched galaxies when looking at the whole population of galaxies.

In the next sections, we investigate the prevalence of galaxies with different quenching profiles in clusters and the field in different stellar mass bins.

\subsection{Quiescent Fraction vs Stellar Mass} \label{subsec:quench_frac}


To investigate the dependence of the prevalence of the classes at different stellar masses, we divided our cluster and field samples into 4 stellar mass bins defined as $\log(M_{*}/M_{\odot}) \in$ [10.2; 10.4], [10.4; 10.6], [10.6; 10.8] and [10.8; 11.4] respectively.

\begin{figure}[htb]
    \centering
    \includegraphics[width=0.99\linewidth]{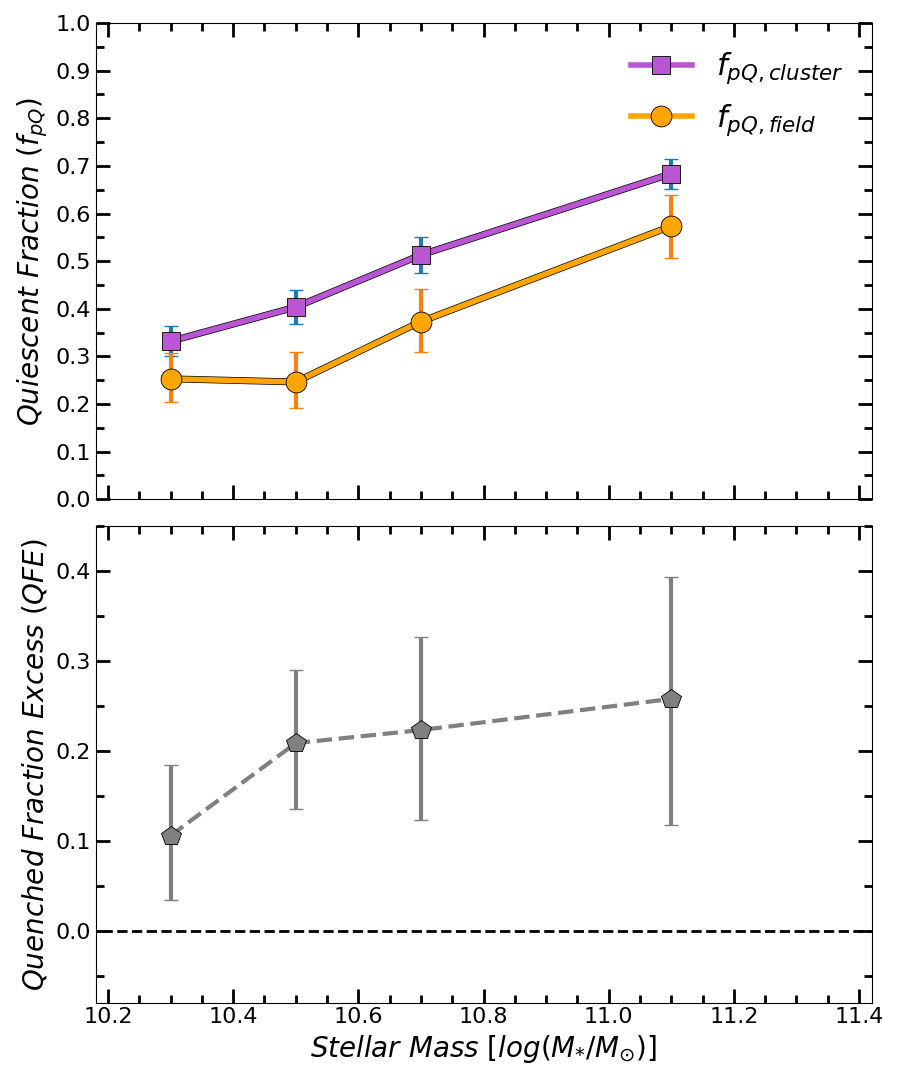}
    \caption{Predominantly quiescent fraction and Quenched Fraction Excess as a function of stellar mass.  Upper panel: Fraction of predominantly quiescent galaxies $f_{pQ}$, defined as galaxies with both inner and outer regions classified as quiescent (see Section \ref{subsec:quiescent_aper}), for cluster members (magenta squares) and for the field (orange circles). The predominantly quiescent fraction increases strongly with stellar mass in both environments and is systematically higher in clusters. Lower panel: Quenched Fraction Excess (QFE, Eq. \ref{eq:qfe}) in clusters relative to the field. Despite the strong stellar-mass dependence of the quiescent fractions in the upper panel, the additional quenching associated with the cluster environment shows little to no dependence on stellar mass. This differs from previous measurements based on integrated-color classifications, as discussed in Section \ref{subsec:dis_quenc_frac}. For the definition of the stellar mass bins and additional details, see Section \ref{subsec:quench_frac}.}
    \label{fig:quench_frac}
\end{figure}

In Figure \ref{fig:quench_frac} we looked at the correlation of the predominantly quiescent fraction ($f_{pQ}$) in clusters and in the field with stellar mass. In this figure, the predominantly quiescent fraction is calculated based on the predominantly quiescent sample defined in Section \ref{subsec:quiescent_aper}, that contains only galaxies that have both regions quiescent. We found that in all environments, $f_{pQ}$ increases with stellar mass and that for every mass bin, $f_{pQ}$ is higher in clusters than in the field. To quantify the excess of predominantly quiescent galaxies in clusters, we calculated the quenched fraction excess (QFE, \citealt{van_der_burg_gogreen_2020}) defined as:

\begin{equation}\label{eq:qfe}
    QFE = \frac{f_{pQ,cluster} - f_{pQ,field}}{1 - f_{pQ,field}}
\end{equation}
with $f_{pQ,cluster}$ and $f_{pQ,field}$ the predominantly quiescent fractions in clusters and in the field respectively. This excess represents the proportion of cluster member galaxies that would have been star-forming if they were in the field, but were quenched by their environment. The lower panel of Figure \ref{fig:quench_frac} shows the QFE for each stellar mass bin. While there is a hint for a (non statistically significant) decrease in the lowest mass bin, the QFE seem to be stable across all stellar mass bins at a level $\sim 20\%$ meaning that the excess of predominantly quiescent galaxies in clusters is independent of stellar mass. When taking the whole sample, we obtain a statistically significant excess: QFE $= (19.9\pm4.5)\%$.

\subsection{Probing the Environmental Quenching vs Stellar Mass} \label{subsec:env_quench}

\begin{figure}[htb]
    \centering
    \includegraphics[width=0.99\linewidth]{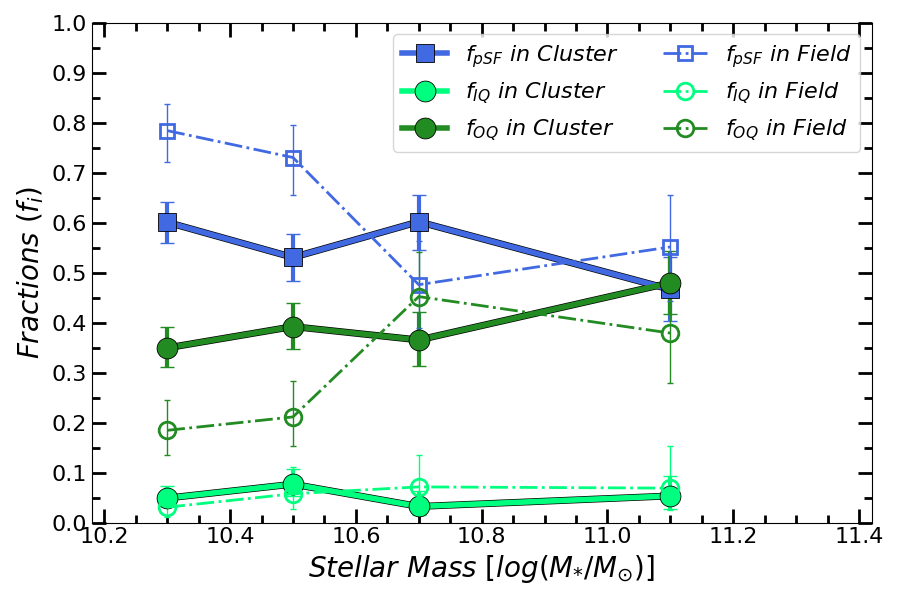}
    \caption{Prevalence of the resolved quenching classes among galaxies with ongoing star formation. Fractions of pSF (blue), IQ (light green) and OQ (dark green) galaxies as a function of stellar mass (See Section \ref{subsec:quiescent_aper} for acronym definitions.), normalized by the total number of galaxies in these three classes ($N_{SF}^{tot} = N_{pSF} + N_{OQ} + N_{IQ}$), as defined in Eq. \ref{eq:sf_frac}). Solid lines show cluster members and dashed lines show field galaxies. In clusters, the relative prevalence of the three classes is approximately independent of stellar mass. In the field, low-mass galaxies are mostly pSF, while the OQ fraction increases toward higher stellar mass. Consequently, at low stellar mass, clusters contain fewer pSF and more OQ galaxies than the field. The IQ fraction remains low in both environments at all stellar masses. For the stellar mass bins and additional details, see Section \ref{subsec:quench_frac}.}
    \label{fig:quench_frac_vs_mass}
\end{figure}

We then investigated the star-forming galaxies, among which we had the predominantly star-forming galaxies, the outside quenched galaxies and the inside quenched galaxies. For each stellar mass bin, we measured $N_{pSF}$, $N_{OQ}$ and $N_{IQ}$; the number of predominantly star-forming, outside quenched and inside quenched galaxies respectively. We then defined the prevalence of each class as:

\begin{equation} \label{eq:sf_frac}
    f_{i} = \frac{N_{i}}{N_{SF}^{tot}}
\end{equation}

where $i$ stands for ``$pSF$" (predominantly star-forming), ``$OQ$" (outside quenched) or ``$IQ$" (inside quenched) and $N_{SF}^{tot} = N_{pSF} + N_{OQ} + N_{IQ}$. In Figure \ref{fig:quench_frac_vs_mass} we show these fractions for the different stellar mass bins defined above. 

First, when looking at the prevalence in clusters (bold solid lines in Figure \ref{fig:quench_frac_vs_mass}) we observed that they were all independent of stellar mass.
The pSF galaxies were the most common, representing $(55.9\pm2.1)\%$ of the sample of star-forming galaxies in clusters, followed by the OQ galaxies, representing $(38.6\pm2.1)\%$, and finally, we had a few IQ galaxies, representing $(5.4\pm1.1)\%$.

Then, when looking at the prevalence in the field (thin dashed lines in Figure \ref{fig:quench_frac_vs_mass}), we noticed a stellar mass dependence for the fractions of predominantly star-forming galaxies and outside quenched galaxies. 
Indeed, these fractions seem to be changing rapidly around $log(M_{*}/M_{\odot}) \sim 10.6$. When combining the two lower mass bins together, we found that at low mass, the predominantly star-forming galaxies represent a large majority ($(75.8\pm4.6)\%$ of the star-forming sample) while the outside quenched galaxies are much less common (representing $(19.8\pm4.3)\%$ of that sample). 
When combining the two higher mass bins together, we found that at high stellar mass, the prevalence of outside quenched galaxies rises to a point where we observed a similar proportion of predominantly star-forming galaxies and outside quenched galaxies, representing $(51.4\pm6.9)\%$ and $(41.6\pm6.8)\%$ of the star-forming sample respectively ($1\sigma$ difference).
We note that the difference of the fraction of predominantly star-forming galaxies between the lower and higher stellar mass bins has a $2.9\sigma$ significance and that the difference for the fraction of outside quenched galaxies has a $2.7\sigma$ significance. 
While it appears that we have a mass dependency of these fractions, we did not have enough statistics to discriminate between a smooth change of the fractions with stellar mass or a bimodality of the fractions with two separate stellar mass regimes.
In the meantime, the proportion of inside quenched galaxies stayed constant and low at all stellar masses ($\sim 6\%$).

\begin{figure}[htb]
    \centering
    \includegraphics[width=0.99\linewidth]{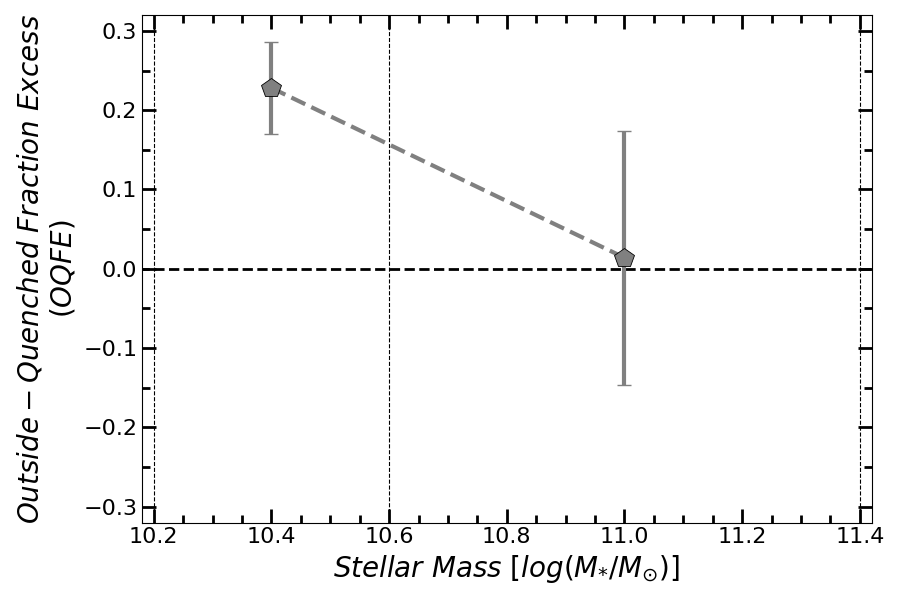}
    \caption{Outside Quenched Fraction Excess in the clusters relative to the field. Outside Quenched Fraction Excess (OQFE; Eq. \ref{eq:oqfe}) as a function of stellar mass. The black dotted vertical lines delimit the low- and high-mass bins defined in Section \ref{subsec:env_quench}. At low stellar mass, the OQFE of $22.8\pm7.0\%$ represents a significant ($3.3\sigma$) excess of outside quenched galaxies in clusters, providing evidence that the cluster environment preferentially suppresses star formation in galaxy outskirts. At high stellar mass, the larger uncertainty prevents a corresponding constraint on the OQFE. The current measurements therefore do not establish whether OQFE depends on stellar mass. See Section \ref{subsec:quench_frac} for additional details.}
    \label{fig:OQFE}
\end{figure}

Finally, when comparing the star-forming galaxies in clusters and in the field, we noticed differences between different stellar masses. At higher mass we did not observe any environmental effect on star-forming galaxies as the prevalence of the three classes were similar in all environments with no differences at the $>1\sigma$ level at $log(M_{*}/M_{\odot}) > 10.6$. 
However, at lower mass, we had a significant excess of outside quenched galaxies in clusters with a $3.3\sigma$ significance.
Since the inside quenched fraction was low and similar in all environments and in all mass bins ($\sim 5\%$ with a $0.6\sigma$ difference), we decided to define and compute the outside quenched excess fraction (OQFE) in clusters in two mass bins ($log(M_{*}/M_{\odot}) \in$ [10.2; 10.6] and [10.6; 11.4]) defined as:
\begin{equation}\label{eq:oqfe}
    OQFE = \frac{f_{OQ,cluster} - f_{OQ,field}}{f_{pSF,field}}
\end{equation}
with $f_{OQ,cluster}$ and $f_{OQ,field}$ the outside quenched fractions in clusters and in the field respectively and $f_{pSF,field}$ the fraction of predominantly star-forming galaxies in the field. 
As shown in Figure \ref{fig:OQFE}, we obtained OQFE $= (22.8\pm7.0)\%$ in the lower mass bin and OQFE $= (1.4\pm16.0)\%$ in the higher mass bin. 
We observed a significant ($3.3\sigma$ significance) excess of outside quenched galaxies in the lower mass bin.
This means that, at lower mass, $\sim 23\%$ of the galaxies that should have been predominantly star-forming if they were in the field, are being outside quenched by their environment.
Due to poor statistics in the higher mass bin, we cannot rule out an excess or default of outside quenched galaxies in the high-mass bin despite an OQFE close to zero and cannot conclude on the correlation of the OQFE with stellar mass.

To ensure that our results were not biased by the clusters halo masses, we investigated a possible halo mass dependence of the quenching fractions. To do so, we used the halo masses from \cite{hewitt_distinct_2025}. We did not find any significant correlation between the halo mass of the clusters and the quenching fractions.

\section{Discussion} \label{sec:discussion}

In this section, we discuss how the success of our deep-learning based deconvolution technique allowed us to probe the internal properties of galaxies in Section \ref{subsec:dis_deconv}.
In Section \ref{subsec:dis_quenc_frac}, we discuss how we can reconcile our results on the quenching fractions and the QFE with the previous studies on GOGREEN and in Section \ref{subsec:dis_env_quench}, we discuss the implication of the excess of outside quenched galaxies in clusters on our understanding of environmental quenching at $z>1$.
Finally, we discuss the potential impact of deep-learning based deconvolution for the future of research in astrophysics. 

\subsection{The success of deep-learning based deconvolution}\label{subsec:dis_deconv}

We have optimized the quality of the deconvolution of near-IR ground-based images of galaxies by using a deep learning technique (\textsc{SUNet}) trained on \textit{JWST}/NIRCam data.
While \textsc{SUNet} proved to be powerful to deconvolve optical images in \cite{akhaury_ground-based_2024}, our study demonstrates that it is just as powerful in the near-IR, despite the different noise properties and observation techniques used to obtain these images. 
Moreover, having access to large high spatial-resolution images in the near-IR from the publicly available COSMOS-Web \textit{JWST} data allowed us to train models for \textsc{SUNet} with a target resolution similar to the optical models, which made our study possible.
The impressive performance of these models allowed us to study the spatially resolved optical and near-IR stellar light of nearly 1200 galaxies at $z > 1$ in different environments and to probe their quenching gradients.
Without deconvolution, the only other way to obtain our results would have been by using space-based imaging, requiring multiple new \textit{HST} and \textit{JWST} surveys on each of these clusters.

Consequently, as we showed in Section \ref{subsec:spatial_colors}, when comparing integrated colors and resolved colors in Figure \ref{fig:IJK_int} and Table \ref{tab:class_comp_int_res}, we noticed that among the galaxies that were considered quiescent based on their integrated $UVJ$ colors, a large number actually showed signs of star-formation. 
Additionally, the spatially resolved colors allowed us to securely identify galaxies that had a clear quenching gradient and isolate them from the rest of the predominantly quiescent or star-forming galaxies, as illustrated in Figure \ref{fig:IOQ_res_colors} and \ref{fig:OIQ_res_colors}. Until now, isolating these partially quenched galaxies when only having access to integrated colors required to identify them through the so-called "Green Valley" (i.e. \citealt{martin_uvoptical_2007, salim_uv_2007,wyder_uvoptical_2007, salim_critical_2014, coenda_green_2018, mcnab_gogreen_2021}). However, it has been demonstrated that this selection was not ideal as only a fraction of these Green Valley galaxies are actually undergoing quenching (i.e. \citealt{schawinski_green_2014,mcnab_gogreen_2021}).
Additionally, isolating the partially quenched galaxies solely based on integrated colors through a Green Valley selection would have been impossible as some of these galaxies have integrated colors solidly in the quenched region of the observed $IJK$ color-color diagram (see middle and right panels of Figure \ref{fig:IJK_int}).

This demonstrates that the most effective way to accurately identify pure samples of predominantly quiescent or star-forming galaxies solely based on photometric data and understand where quenching occurs within galaxies is by probing their resolved rest-frame $UVJ$ colors. 

\subsection{Quenched fraction in clusters at $z>1$}\label{subsec:dis_quenc_frac}

As mentioned above, a major improvement brought by our deconvolution method is illustrated in Figure \ref{fig:IJK_int} where one can see that a large number of galaxies, in all environments, that have integrated colors consistent with quiescent galaxies actually show signs of star-formation in either their center or their outskirts.
These galaxies appear in the middle and right panels of Figure \ref{fig:IJK_int} as markers inside the quiescent wedge while they have star-formation happening in their center (middle panels) or in their outskirts (right panels).
This demonstrates that a galaxy with integrated rest-frame $UVJ$ colors inside the quiescent wedge is not necessarily completely quenched.

\begin{figure}[htb]
    \centering
    \includegraphics[width=0.99\linewidth]{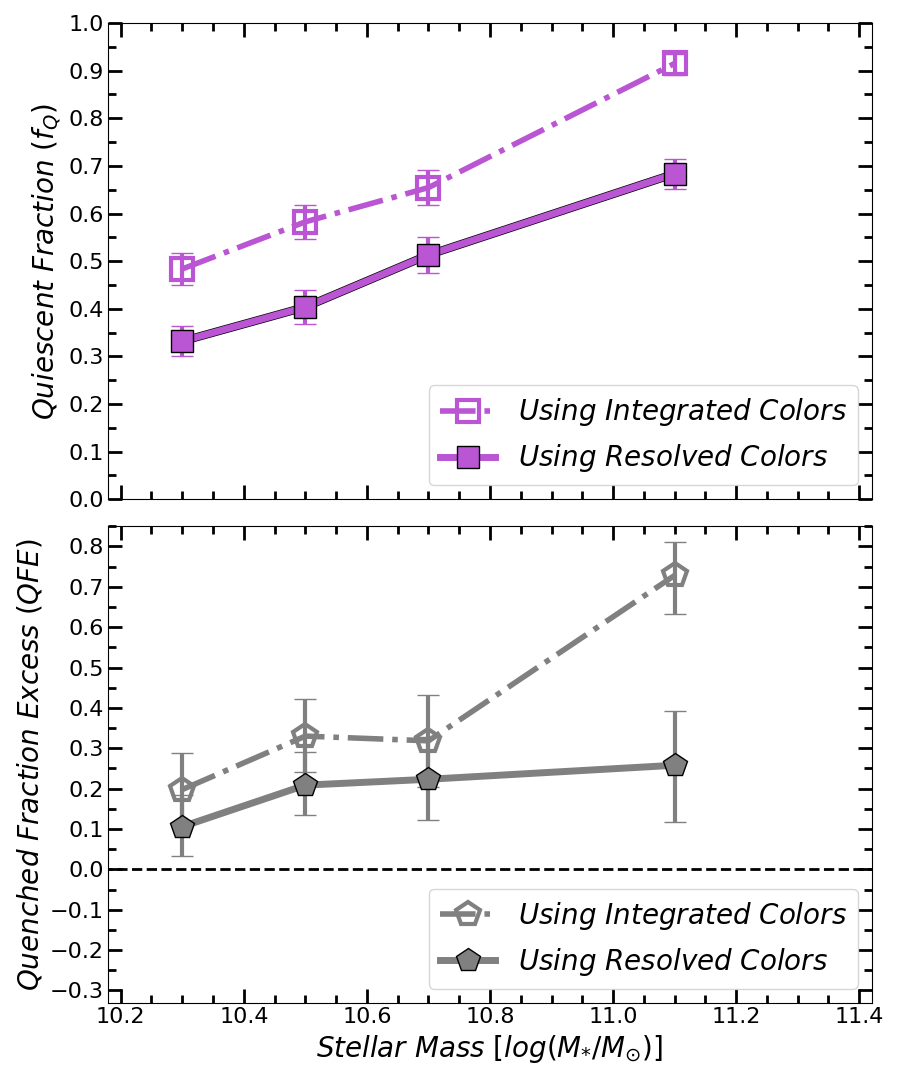}
    \caption{Effect of spatially resolved classification on the inferred quiescent fraction and environmental quenching efficiency. Upper panel: Cluster quiescent fraction as a function of stellar mass using the resolved classification adopted in this work (solid line) and conventional integrated colors (dotted line). Both recover the increase of the quiescent fraction with stellar mass, but the integrated-color classification systematically overestimates the quiescent fraction by including galaxies that retain spatially localized star formation. Lower panel: Quenched Fraction Excess (QFE, see Eq. \ref{eq:qfe})  derived using the same two classifications. The integrated-color classification produces a strong stellar-mass dependence of QFE consistent with previous GOGREEN measurements, whereas the resolved classification yields little to no mass dependence. This comparison demonstrates that partially quiescent galaxies, particularly at high stellar mass, can bias the inferred mass dependence of environmental quenching when classified using integrated colors. The stellar mass bins are the same as in Figure \ref{fig:quench_frac}. See Section \ref{subsec:dis_quenc_frac} for additional details.}
    \label{fig:comp_int_res}
\end{figure}

Additionally, in Section \ref{subsec:quench_frac}, we found that the quenched fraction increases with stellar mass and is higher than in the field. However, in contradiction with previous studies (\citealt{van_der_burg_gogreen_2020, reeves_gogreen_2021}), we found little to no correlation between the quenched fraction excess and stellar mass. This is important as the stellar mass dependency of the QFE implied that the GOGREEN progenitor (proto-) clusters were sites where massive quiescent galaxies were produced in large amounts.
To investigate the bias introduced by the contamination of partially star-forming galaxies, we decided to build a sample of quiescent galaxies solely based on integrated colors and compare it with our sample of predominantly quiescent galaxies. Practically, this new sample of quiescent galaxies contains all the galaxies from the upper panels of Figure \ref{fig:IJK_int} with integrated colors inside the quiescent wedge, including the galaxies in the middle and right panels even if they show signs of star-formation, as defined in the "Q Clusters" column of Table \ref{tab:class_comp_int_res}.
In Figure \ref{fig:comp_int_res} we show the quiescent fraction and QFE in clusters in several mass bins when using our filtered sample built based on resolved colors (solid lines, identical to Figure \ref{fig:quench_frac}) and when using the sample of quiescent galaxies built solely based on integrated colors (dotted lines).

From the quiescent fraction comparisons in the upper panels of Figure \ref{fig:quench_frac} and \ref{fig:comp_int_res}, we found that the two samples of quiescent galaxies follow the same trend with stellar mass, consistently with the well established facts that galaxies in denser environment have a higher chance of being quiescent than their field counterparts and that independently of environment, more massive galaxies have a higher chance of being quiescent (e.g, \citealt{baldry_galaxy_2006,peng_mass_2010,muzzin_gemini_2012,wetzel_galaxy_2013,haines_locuss_2015,fossati_galaxy_2017,van_der_burg_gogreen_2020}). 
Moreover, we found that the quiescent fractions calculated based on integrated colors (dotted line in Figure \ref{fig:comp_int_res}) are consistent with previous work on clusters at similar redshifts (\citealt{balogh_evidence_2016, lemaux_persistence_2019}).
However, we found that using the integrated colors would lead to overestimate the quiescent fraction in clusters by an average of $\Delta f_{Q} = f_{Q,int} - f_{Q,res} = 0.15 \pm 0.02$ over our stellar mass range. In other words, on average, $\sim 24\%$ of the galaxies that have integrated colors consistent with quenched galaxies are only partially quiescent, as mentioned in Section \ref{subsec:spatial_colors}. 


Additionally, when looking at the QFE (lower panel of Figure \ref{fig:comp_int_res}), we found that using a sample of quiescent galaxies based on integrated colors leads to overestimate the QFE, especially at higher mass. We also note that the QFEs measured with this sample and their strong stellar-mass dependence are perfectly consistent with the previous studies on the GOGREEN clusters (\citealt{van_der_burg_gogreen_2020, reeves_gogreen_2021,werner_satellite_2021,hewitt_distinct_2025}).
This means that, when using integrated colors to identify quiescent galaxies, the higher mass bins are heavily contaminated by partially quiescent galaxies.



\subsection{Constraints on Environmental Quenching at $z>1$}\label{subsec:dis_env_quench}

One of the main results of the GOGREEN and GCLASS surveys is that most of the massive quiescent galaxies in clusters were quenched at early times ($z>3$), when the clusters were still protoclusters (\citealt{webb_gogreen_2020, werner_satellite_2021,hewitt_distinct_2025}). However, in Section \ref{subsec:env_quench}, we observed a large fraction ($\sim 40\%$) of outside quenched galaxies at all mass. If we assume that these galaxies are actively quenching, this means that there must be two main quenching channels in clusters.
An early channel that produced the excess of massive quiescent galaxies and a late channel, happening in an outside-in fashion. 

Figure \ref{fig:OQFE}, shows that the excess of outside quenched galaxies compared to the field is only significant at lower stellar masses, meaning that this quenching channel is mostly efficient at lower mass. We note that the excess of partially quenched galaxies at low mass is consistent and confirms the work of \cite{mcnab_gogreen_2021} who found a (marginally) significant enhancement of transition galaxies (more specifically blue quiescent galaxies) at low stellar masses in the GOGREEN clusters.

The outside-in quenching signature we observe in clusters provides an important spatial constraint on the physical mechanisms driving environmental quenching. 
While starvation is widely invoked to explain long environmental quenching timescales ($\sim1 - 5$ Gyr) and mass-dependent quenching trends in clusters (e.g., \citealt{peng_mass_2010,balogh_gemini_2017, fossati_virgo_2018,hewitt_distinct_2025}), simply cutting off external gas replenishment does not naturally produce sharp truncation of the outer star-forming disk while leaving the underlying stellar distribution intact (\citealt{larson_evolution_1980}). 
By contrast, direct hydrodynamical removal of the interstellar medium via ram-pressure stripping (\citealt{gunn_infall_1972}) preferentially strips the more weakly bound gas at large radii first, confining residual star formation toward the galaxy center and producing an outside-in quenching profile consistent with our observations and observed in low- and intermediate-redshift clusters (e.g., \citealt{koopmann_h_2004,crowl_dense_2005, fossati_virgo_2018, finn_local_2018, schaefer_sami_2017, matharu_hstwfc3_2021, wang_sami_2022})
Nevertheless, starvation and ram-pressure stripping are not mutually exclusive: galaxies entering overdense environments may initially experience strangulation at larger halo-centric radii before encountering sufficiently dense intracluster gas to undergo outside-in ram-pressure stripping that completes the quenching process (e.g., \citealt{van_der_burg_gogreen_2020}).
Because our observations probe spatial gradients rather than direct quenching timescales, we cannot rule out a multi-stage or hybrid pathway in which both starvation and ram-pressure stripping play a role.


For the more massive galaxies, our statistics do not allow us to reject the possibility of the presence or of the absence of an excess of outside quenched galaxies. However, assuming that this excess is close to zero, as suggested by our study (see Figure \ref{fig:OQFE}), we can have two interpretations.
Either the outside-in environmental quenching channel is negligible for the most massive galaxies and they all quenched at earlier time as suggested by previous studies mentioned above. Or, we are missing some of the massive outside quenched galaxies. Indeed, \cite{lemaux_persistence_2019}  and \cite{baxter_importance_2025} found that massive galaxies tend to quench on shorter timescales than lower mass galaxies. 
And, if a massive galaxy quenches quickly enough, a significant quenching gradient would not have the time to settle in before the galaxy becomes predominantly quiescent. However, based on \cite{webb_gogreen_2020}, we know that on average, the stellar population of these massive quiescent galaxies are old, meaning that even if this later and quicker quenching channel is effective at high mass, it is not widespread in the GOGREEN clusters.

To summarize, our work suggests that, on top of the early stage quenching of the most massive galaxies, there is an outside-in quenching channel affecting galaxies in dense environment, especially at lower mass.

While this pilot study was based on only three bands for each galaxies, its success suggests that we should be able to go further and deconvolve many more bands and obtain access to the spatially resolved SED of these galaxies. Moreover, the technique having demonstrated its adaptability, it could be applied to a wider range of publicly available data to increase the statistics. This would allow us to probe the star-formation histories of these galaxies and put additional constraints on the quenching mechanism(s) responsible for this environmental quenching.


\subsection{The importance of deconvolution in the Era of big-data}\label{subsec:dis_imp_dec}


Having access to spatially-resolved photometry not only is crucial in clusters to understand how environmental quenching affects the galaxies, but it is also crucial to get a better understanding of galaxy formation and evolution in general. 
A recent example is the work of \cite{cutler_unbreaking_2026} who used \textit{JWST} photometry from the MINERVA survey (\citealt{muzzin_minerva_2025}) to demonstrate that by measuring color gradients in $z>3$ massive quiescent galaxies, they found less tensions with the models than other analysis using only slit spectroscopy probing the center-most region of the galaxies.

Until now, probing the spatially resolved stellar optical and near-IR light required space-based imaging. However, our study demonstrate that with \textsc{SUNet}, we can now probe this light from ground-based data. This is important as the large field of view of the ground-based telescopes allows to easily scan large portions of the sky while the much smaller field of view of the Hubble and James Webb space telescopes makes obtaining imaging for a large sample of galaxies extremely challenging.
Additionally, since there is already an enormous amount of publicly available multi-band ground-based data covering about a million of objects, deconvolution of ground-based data does not require any new observations.

One could argue that with the development of new space-based telescopes with larger mirrors and cameras like the Euclid or Roman space telescopes, the small field-of-view might not be such an issue anymore. 
However, the Euclid Space Telescope, with a spatial resolution of $\sim 0.4"-0.5"$ due to its $0.3"$ wide pixels in the near-IR bands\footnote{See \href{https://sci.esa.int/web/euclid/-/euclid-nisp-instrument}{ESA Euclid.NISP webpage}.} does not allow to spatially resolve high-$z$ galaxies and the $H$ band is not red enough to probe the rest-frame $J$ band at $z>1$. 
As for the Roman Space Telescope, while it will be able to take high-spatial resolution images of the rest-frame $U$, $V$, and $J$ bands of $z\sim1$ galaxies, it will miss the UV light necessary to constrain the shape of the SEDs to build spatially resolved SFR or $M_{*}$ maps. However, deconvolution can help bridge that gap.
Indeed, the ground-based NSF-DOE Vera C. Rubin Observatory will carry out the Legacy Survey of Space and Time (LSST, \citealt{ivezic_lsst_2019}), which will map the entire observable sky regions in 10 years, providing deep co-added images in the \textit{ugrizy} filters. 
Due to the atmosphere blurring, these images will have seeing of about 0.7 arcseconds. Using \textsc{SUNet} could allow to bring this resolution down to $\sim 0.2$ arcseconds. 
Together with the Roman data, this would allow to spatially resolve the SEDs of over a billion galaxies\footnote{See \href{https://www.stsci.edu/files/live/sites/www/files/home/roman/_documents/roman-capabilities-galaxies.pdf}{STScI Roman fact sheet}.} from the UV to the near-IR, including millions of $z>1$ galaxies.

\section{Summary} \label{sec:summary}


In this work, we used deep-learning-based deconvolution of ground-based optical and near-IR imaging to investigate spatially resolved environmental quenching in nine GOGREEN clusters at $1<z<1.4$. Our sample comprises approximately 900 cluster members and 300 field galaxies. In an effort to probe the rest-frame $UVJ$ colors of our sample, we used \textsc{SUNet} to recover spatially resolved observed ($I/i/z$), ($J/J1$), and ($Ks$) photometry, using custom trained deconvolution models including new near-IR models trained using high-resolution \textit{JWST}/NIRCam imaging. These data allowed us to distinguish the star-forming states of the inner and outer regions of individual galaxies.

Based on their resolved colors, we classified galaxies as predominantly star-forming (pSF), predominantly quiescent (pQ), inside quenched (IQ), or outside quenched (OQ).
This classification allowed us both to revisit conventional measurements of the quiescent fraction and Quenched Fraction Excess (QFE) and to directly investigate spatially dependent quenching through both the prevalence of IQ and OQ galaxies and the Outside Quenched Fraction Excess (OQFE). The OQFE quantifies the excess fraction of galaxies, compared to the total population of galaxies with significant star formation, that are quenched in their outskirts by the environment.


The main conclusions of our analysis are:
\begin{itemize}
    \item Deep-learning-based deconvolution successfully recovers spatial information that is inaccessible from the original ground-based imaging. In particular, customized models trained on \textit{JWST}/NIRCam data successfully deconvolve the ground-based near-IR imaging, demonstrating the potential of \textit{JWST} imaging as a training set for recovering spatial information from lower-resolution wide-field datasets.
    \item Integrated colors conceal a substantial population of partially quenched galaxies. Approximately 24\% of galaxies classified as quiescent from their integrated colors retain star formation in either their core or their outskirts. Consequently, integrated-color selection overestimates the cluster quenched fraction by an average of $\Delta f_{Q} = f_{Q,int} - f_{Q,res} = 0.15\pm0.02$.  Spatially resolved colors therefore provide a materially different and more accurate census of fully quiescent galaxies.
    \item The quiescent fraction increases with stellar mass and is higher in clusters than in the field, but the environmental QFE derived from resolved classifications shows little to no dependence on stellar mass. We measure a QFE of approximately 20\% across the stellar-mass range studied. When we instead classify the same galaxies using integrated colors, we recover the strong stellar-mass dependence of QFE reported in previous GOGREEN studies. This comparison demonstrates that contamination of integrated-color quiescent samples by partially quenched galaxies, particularly at high stellar mass, can bias the inferred stellar-mass dependence of environmental quenching.
    \item Outside-quenched galaxies dominate the partially quenched population in all environments, while inside-quenched galaxies are uncommon. Among galaxies retaining significant star formation, IQ galaxies account for less than 10\% in both clusters and the field. The principal environmental difference instead appears in the prevalence of OQ galaxies.
    \item We detect a significant excess of outside-quenched galaxies in clusters at low stellar mass. At $10.2<\log(M_{\ast}/M_{\odot})<10.6$, we measure an OQFE of $(22.8\pm7.0)\%$, corresponding to a $3.3\sigma$ excess relative to the field. This excess is accompanied by a deficit of predominantly star-forming galaxies and provides evidence that the cluster environment preferentially suppresses star formation in the outskirts of low-mass galaxies. At higher stellar mass, the current uncertainties do not allow us to determine whether an OQ excess or deficit is present, and we therefore cannot establish a stellar-mass dependence of the OQFE.
\end{itemize}

Taken together, these results demonstrate the importance of spatially resolved information for interpreting environmental quenching at $z\sim1$. Integrated photometry can miss ongoing star formation within partially quenched galaxies and consequently alter the inferred efficiency and stellar-mass dependence of environmental quenching. Our resolved analysis provides evidence for an environmentally driven outside-quenched pathway at low stellar mass, although the present data do not uniquely identify the physical mechanism responsible.

In a subsequent paper, we will examine the phase-space distributions of galaxies following different quenching pathways. Combining this information with constraints on the star-formation histories and age gradients of outside-quenched galaxies will help distinguish among the physical mechanisms responsible for environmental quenching.

More broadly, this work demonstrates that deep-learning-based deconvolution can recover spatially resolved information from existing and future wide-field ground-based imaging. Application of these techniques to large datasets, including those from the Rubin Observatory and in combination with space-based imaging, offers a path toward spatially resolved studies of galaxy evolution for samples far larger than are practical with high-resolution space-based imaging alone.


\section{Acknowledgment}
G. Wilson gratefully acknowledges support from the U.S. National Science Foundation (NSF) through grant AST-2347348.

G. Rudnick gratefully acknowledges support from the U.S. NSF through grant AST-2206473.  

The authors thank the International Space Sciences Institute in Bern, Switzerland, for hosting a team meeting during which work on this project was performed.

This work was supported by the Swiss National Science Foundation (SNSF) under funding reference 200021\_213076 ``Galaxy evolution in the cosmic web".

Part of this research was conducted using Pinnacles (NSF MRI, Grant Number 2019144) at the Cyberinfrastructure and Research Technologies (CIRT) at University of California, Merced. 

This work was enabled by observations made from the Gemini North telescope, CFHT, and Subaru. The scientific community is honored to have the opportunity to conduct astronomical research on Maunakea in Hawai‘i. We recognize and acknowledge the very significant cultural role and reverence of Maunakea to the Kanaka Maoli (Native Hawaiians) community.

The Scientific color map \textsc{oslo} (\citealt{crameri_scientific_2023}) is used in this study to prevent visual distortion of the data and exclusion of readers with color-vision deficiencies (\citealt{crameri_misuse_2020}).

The authors thank the anonymous referee for their thorough review and helpful suggestions that allowed to make this paper stronger.

While this text has been entirely handwritten, generative AI (Gemini, by Google) was used by the authors as a tool to help improve the formulation, language and clarity of some sentences.

\bibliography{GOGREEN_DEC}{}
\bibliographystyle{aasjournal}

\newpage 
\appendix

\section{Building accurate PSFs for GOGREEN} \label{app:psfs}

The stars used to build the PSFs were selected in a multistep process. First, the stars were obtained from the public GOGREEN DR1 (\citealt{balogh_gogreen_2020}). In this catalog, the sources were flagged as stars or galaxies based on a color-color UJKs diagram as described in \cite{van_der_burg_gogreen_2020}. 
Among these stars, we selected the brightest ones (with a $mag_{AB} < 21$) with a half-light radius less than 0.5 arcseconds in the $Ks$ band. This last condition allowed us to remove wrongly classified compact galaxies. 
The next step was to remove the stars that had saturated pixels or were in a crowded environment. The environment was considered crowded if the star had a neighbor within a 5 arcsecond radius with a flux greater than 3\% of the stellar flux. 
In the cases where there were more than 20 stars selected, we only kept the 20 brightest ones as we realized that adding more (fainter) stars damaged the quality of the PSF. 
The PSFs were built using the \texttt{EPSFBuilder} function from the \texttt{photutils} \textsc{python} module. The PSFs were then sampled to the desired sizes and pixel scales (see Table \ref{tab:models}).

As ground-based images potentially have spatially dependent PSFs (i.e. \citealt{sok_finite-resolution_2022}), we checked for any spatial variations. To do so, we built PSFs from sub-samples of stars and compared them to the global PSF. For each image, we divided the sample of stars in three ways: first, based on the distance from the center of the image, then by dividing the image into four quadrants and finally by randomly selecting the stars. As the differences between the local PSFs and the global PSF never significantly exceeded the differences measured when using the random sub-samples, we decided to use a single PSF for each image.

\section{Training models for \textsc{SUNet}}  \label{app:training}

In this appendix, we provide more details on the training of models for \textsc{SUNet} to deconvolve ground-based data. We focus on the three models we trained to deconvolve the near-IR images (see Section \ref{subsec:new_models}). 
Training a model requires three steps: 1) extract cutouts of extended galaxies from high-resolution space images, 2) Simulate ground-based images by convolving these cutouts with a Gaussian PSF and adding Gaussian noise, and 3) train the model using the simulated ground-based image and the space truth image of thousands of galaxies.
In Section \ref{app:train_set} we explain how we selected galaxies that we used for training. In Section \ref{app:train_prep} we provide details on how we prepared the cutouts of the training sample while in Section \ref{app:train_sim} we explain how we simulated ground-based images. Finally, in Section \ref{app:train_run} we explain how the training ran. 
We focus here on the technical details, for more information on how \textsc{SUNet} works, we refer the reader to \cite{fan_sunet_2022} and \cite{akhaury_ground-based_2024}.
\textsc{SUNet} is a publicly available software, it can be pulled/downloaded from \textsc{GitHub} at \url{https://github.com/utsav-akhaury/SUNet.git}.

\subsection{Selecting a training set} \label{app:train_set}

Training a model for a deep learning based algorithm requires a large training set. For a good quality training, we targeted a sample size of at least $\sim 20,000$ galaxies. This sample would then be augmented through random rotations and mirroring. 
To optimize the deconvolution, the cutouts used for the training should be selected 1) from high spatial resolution space images, 2) from images having a similar (ideally identical) spatial resolution and 3) observed at a similar wavelength as the images one wants to deconvolve.
For these reasons, we looked for a large imaging survey at $\sim 2\mu $m ($Ks$ band). We selected our training sets in the COSMOS field imaged by COSMOSWeb with \textit{JWST}/NIRCam F277W (\citealt{casey_cosmos-web_2023, franco_cosmos-web_2025}). We chose specifically the \textit{JWST}/NIRCam.F277W filter for three reasons: 1) as it was close to 2 $\mu$m, 2) as the F277W image was deeper than the F150W and F115W images, and, 3) as the F277W image had a similar spatial resolution to the HST image, this is important as it means all our models would have the same target resolution, hence avoiding a gap in resolution between the optical and near-IR bands.

Then, to select the sample of galaxies to extract from the large COSMOSWeb image, we used the \cite{shuntov_cosmos2025_2025} photometric catalog.
From this catalog, we first removed all the sources identified as stars or as blended sources and applied a $M_{*} > 10^8 M_{\odot}$ and $0.1 < z < 3.0$ stellar mass and redshift cuts. 
These cuts allowed us to remove low-mass galaxies that we did not have the depth to detect in the GOGREEN survey and to remove the low- and high-z galaxies that would be either too large and/or too different from our $z\sim 1$ galaxies or poorly resolved by NIRCam.
We allowed for a relatively large redshift range to ensure a large sample. Additionally, to ensure the good quality of the selected galaxies, we added a magnitude cut ($mag_{AB} < 24$) and a signal-to-noise ratio cut ($S/N > 200$). At the end of this step, we had a training set of $\gtrsim 60,000$ galaxies. In the next section, we explain how we refined the selection based on direct measurements we made on the individual galaxies.

\subsection{Preparing the training sample} \label{app:train_prep}

Preparing the sample for training required two crucial pieces of information: the size and sampling of the cutouts, i.e., how many pixels per cutouts and how many arcseconds per pixel. 
Answering these questions required to follow to rules: 1) one cannot modify the ground-based images before deconvolving them, doing so would lead to either a loss of information from the original image or a bias due to extrapolated information that was absent from the original image, and 2) for computational reasons, the deconvolution factor should be a power of 2. As explained in Section \ref{subsec:new_models}, the deconvolution factor is defined as the ratio between the sizes of the cutouts after and before deconvolution. Alternatively, it corresponds to the ratio of the pixel scale of the cutouts before and after deconvolution.

In our case, we had ground-based images coming from HAWK-I at 0.11"/pix, from FOURStar at 0.16"/pix and from WIRCam at 0.30"/pix. Since the NIRCam images had a resolution of 0.03"/pix, we had pixel scales ratios of 3.67, 5.33 and 10 for HAWK-I, FOURStar and WIRCam respectively. 
In order to respect the ``power of 2" rule, and since we did not want to over-sample the NIRCam images, we selected the power of 2 just below these ratios, hence a deconvolution factor of 2, 4 and 4 for HAWK-I, FOURStar and WIRCam respectively as mentioned in Table \ref{tab:models}.
The reason we chose a factor of 4 rather than 8 for the WIRCam model was that a factor of 8 would have required much longer computational time for a negligible improvement in image quality based on the relatively poor seeing of the original images. These factors meant that we had to resample the NIRCam cutouts to reach a pixel scale of $0.11^" / 2 = 0.055^"/px$, $0.16^" / 4 = 0.04^"/px$ and $0.30^" / 4 = 0.075^"/px$ for the HAWK-I, FOURStar and WIRCam model respectively. The left column of Figure \ref{appfig:mod_sim} shows the resample NIRCam cutout of a galaxy for the three models.

To select the size of the cutouts, we needed to ensure that 1) we had a single galaxy per cutout on average, 2) that the cutouts were large enough to fit the largest galaxies and some sky background, a reasonable size for the cutout should be $3-5$ arcseconds and, 3) for computational reasons, the size in pixel of the cutout should be a factor of 2. 
This led us to choose the NIRCam cutout sizes of 64x64, 128x128 and 64x64 for the HAWK-I, FOURStar and WIRCam model respectively leading to a 3.52", 5.12" and 4.80" field of view for the HAWK-I, FOURStar and WIRCam cutouts respectively. All of these values are summarized in Table \ref{tab:models} of Section \ref{subsec:new_models}.

Based on this, we were able to make the cutouts to prepare the training sample for our 3 near-IR models. As our goal was deconvolution, we added a layer of selection to remove compact galaxies. To do so, we measured the Gini coefficient ($G$) and the diameter containing 20\% of the Kron aperture flux of each source ($d_{20}$). 
To remove compact and/or point-like sources, we imposed $G < 0.56$ and $d_{20} > 3$ pixels.
Additionally, we removed sources with saturated (or close to saturation) pixels and when there were more than one galaxy in the cutouts, we ensured that the central source had a flux at least $5\times$ as great as the flux of the brightest neighboring galaxy. These additional cuts led our training samples to contain between $\sim 20,000$ to $\sim 25,000$ galaxies depending on the model, which was compatible with our target sample size from Section \ref{app:train_set}.

\subsection{Simulating ground-based images} \label{app:train_sim}

\begin{figure}[htb]
    \centering
    \includegraphics[width=0.86\linewidth]{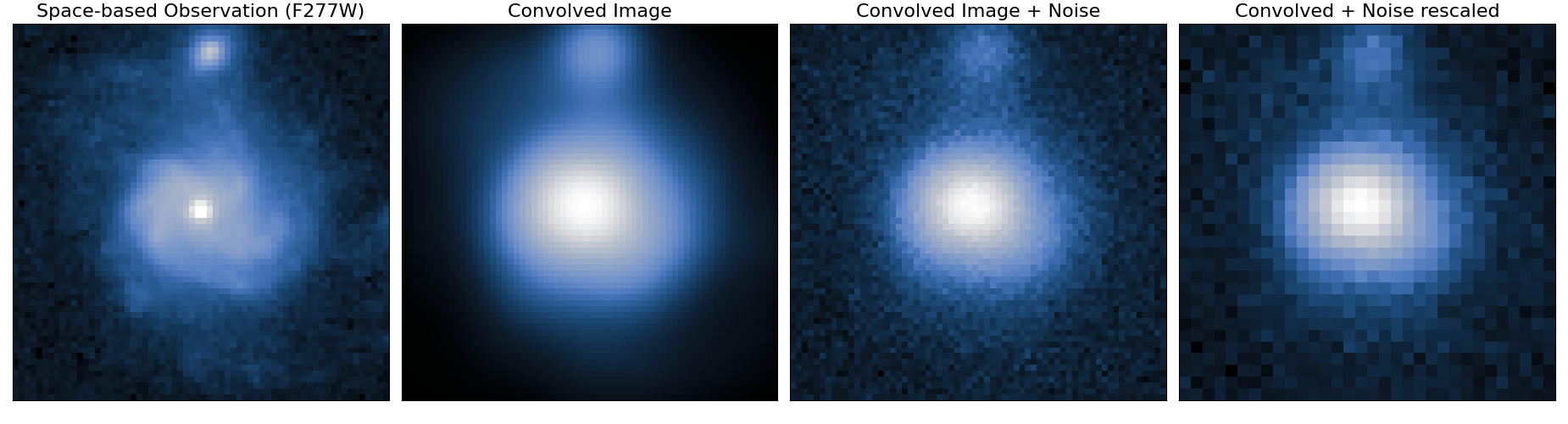}
    \includegraphics[width=0.86\linewidth]{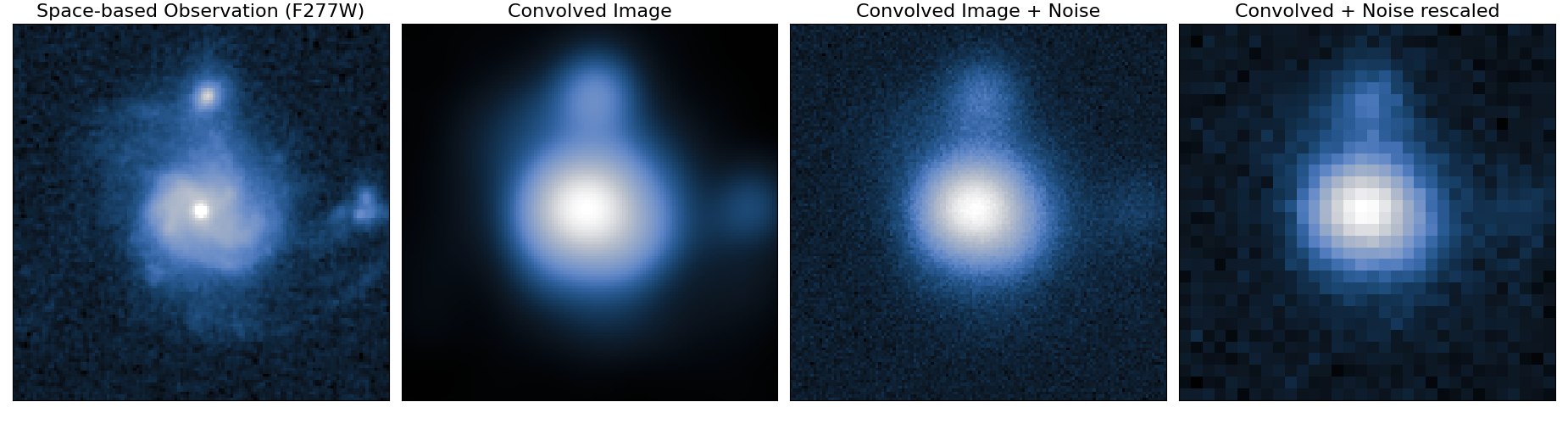}
    \includegraphics[width=0.86\linewidth]{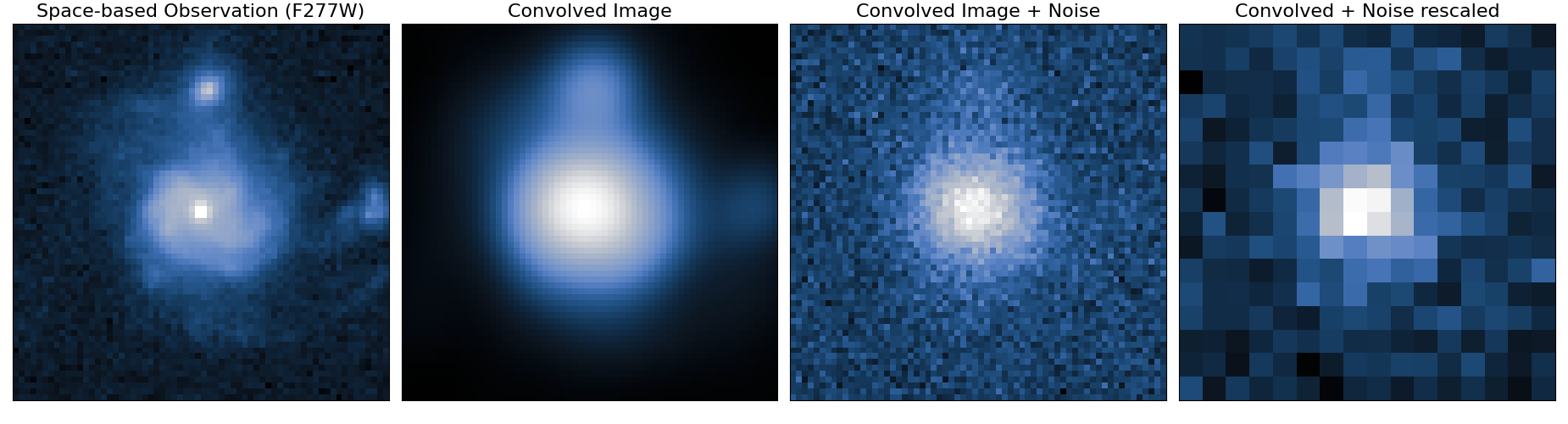}
    \caption{Illustration of our three step process to built simulated ground based images from the high-resolution space images for the HAWK-I model in the upper row, the FOURStar model in the middle row and the WIRCam model for the lower row. For each model, the space image (first column), was convolved with a customized PSF (second column), some Gaussian noise was added (third column) and they were resampled to get a simulated ground-based image (last column) close to the GOGREEN images. See Section \ref{app:train_sim} for additional details and Section \ref{app:train_prep} for the choice of the pixel scale and field-of-view of the NIRCam images.}
    \label{appfig:mod_sim}
\end{figure}

Once we had the cutouts of all the galaxies for our training sets, we simulated the corresponding ground-based images. 
These simulated images needed to be representative of the ground-based images we wanted to deconvolve. 
We built these simulated images in a three step process, first we convolved the high-resolution cutouts with a Gaussian PSF that had a full width at half maximum (FWHM) similar to the PSF of our ground-based images (see Table \ref{tab:instruments} for the FWHM of the GOGREEN PSFs). In our case, we chose a FWHM of 0.5", 0.7" and 0.75" for the HAWK-I, FOURStar and WIRCam model respectively as they are representative of the average PSF from each instrument.
After convolving the images, we added noise as our ground-based images are shallower than the space images. The noise images were built using the \texttt{datasets.make\_noise\_image} function from the \textsc{python} \texttt{photutils} package. We chose a Gaussian noise with a mean of 0. 
The standard deviation of the Gaussian noise was set so that the faintest selected galaxies had a S/N of 1. 
We added the same noise level to all the cutouts, independently of their intrinsic S/N. 
We chose a Gaussian noise as the noise measured in the ground-based images we wanted to deconvolve was Gaussian. However, this noise was also correlated as these real images were the result of a mosaic of several pointings. In Appendix \ref{app:test_grad}, we show that, even if the noise we added was not correlated, our deconvolution technique works extremely well with both correlated and uncorrelated noise.
The last step was to resample the images to the ground-based sampling. 
For the HAWK-I model, the cutouts were 64x64 and the deconvolution factor was 2, so the simulated images were resample to 32x32 pixels. 
For the FOURStar model, the cutouts were 128x128 and the deconvolution factor was 4, so the simulated images were resample to 32x32 pixels. 
For the WIRCam model, the cutouts were 64x64 and the deconvolution factor was 4, so the simulated images were resample to 16x16 pixels. 
Hence, all our simulated images had pixel scales corresponding to the ground-based images pixel scales showed in Table \ref{tab:models}.
In Figure \ref{appfig:mod_sim} we show the different steps leading to the simulated ground based images for the HAWK-I, FOURStar and WIRCam models respectively starting from the same galaxy. The left columns shows the original resampled NIRCam image (see Section \ref{app:train_prep} for why these cutouts are different between models), the second columns shows the NIRCam image after being convolved with the Gaussian PSF, the third columns shows the added Gaussian noise and the last column shows the simulated ground-based image after resampling. The models are then trained by comparing the images from the leftmost and rightmost columns.

\subsection{Running the training} \label{app:train_run}

With all the training sets ready, the last step was to run the training. To do so, we followed the instructions provided in the README file of \textsc{SUNet} on \textsc{GitHub} and adapted parameters like the image sizes when necessary. 
The training was a two step process, we first ran a code to perform a Tikhonov deconvolution on the simulated ground-based images that produced the images \textsc{SUNet} would actually be trained on. This first step could be performed locally as it did not require a lot of resources. However, the second step was the training itself and required GPUs to run properly. 
The training was decomposed into epochs that could be seen as small units of training. In total, one model required $\sim 500-600$ epochs to converge. We performed the training of the three models on two GPUs of the Pinnacles cluster at the Cyberinfrastructure and Research Technologies (CIRT) at the University of California, Merced. The two GPUs worked in parallel. Training one model took $\sim 2-3$ days. The training time increases exponentially with the deconvolution factor and the size of the cutouts.
Before each training, we randomly selected 100 galaxies and set them aside to be used to test the models. These tests allowed us to decide how many epochs to add or how to adapt our selection discussed in Section \ref{app:train_prep} when necessary. Once we were satisfied with the performance of the models, we used them on real GOGREEN data.

This technique can be used to train new customized models to deconvolve other datasets. For the deconvolution process to be optimized, the training set should be as close as possible to the data to deconvolve.

\section{Recovering color gradients from deconvolved images}  \label{app:test_grad}

While we know that, by design, the deconvolution process conserves the total flux of the galaxies, we do not know for sure that the color gradients are conserved. 
Indeed, this would imply that the three bands used to probe rest-frame U, V and J are deconvolved to a matching spatial resolution and more importantly in a way that conserves the flux distribution within the galaxies consistently between the bands.
Moreover, we use an optical model trained on F606W to deconvolve the I band and a near-IR model trained on F277W to deconvolve the J band, hence, assuming that the models are efficient at deconvolving different bands. While \cite{akhaury_ground-based_2024} showed that a model trained on F606W was extremely efficient at deconvolving other optical bands, including the I band by deconvolving ground-based images and measuring residuals, they did not measure gradients.

In order to test this, we decided to use high resolution images of galaxies and degrade them to simulate ground-based images, in a similar way we did to build the training set of our models (see Appendix \ref{app:train_sim}). 
Practically, we selected a random sample of 100 galaxies from our training set (in the COSMOS field) that we built to train our near-IR models. We chose galaxies that had a stellar mass and redshift consistent with our GOGREEN sample ($log(M_{*}/M_{\odot}) > 10.2$ and $1.0<z<1.4$). 
We used the same process as explained in Appendix \ref{app:train_sim} to build simulated ground-based images using the ACS\_WFC.F814W (\citealt{koekemoer_candels_2011}), NIRCam.F115W, and NIRCam.F277W (\citealt{franco_cosmos-web_2025}) filters: each cutout was convolved with a Gaussian PSF representative of the average PSF of our ground-based data in the I, J and K bands, we then added Gaussian noise and changed the pixel scales to match the requirement of the optical or near-IR deconvolution models.
We then deconvolved these images using the optical model and one of our near-IR models (we obtain similar results independently of the chosen near-IR model).

\begin{figure}[htb]
    \centering
    \includegraphics[width=0.47\linewidth]{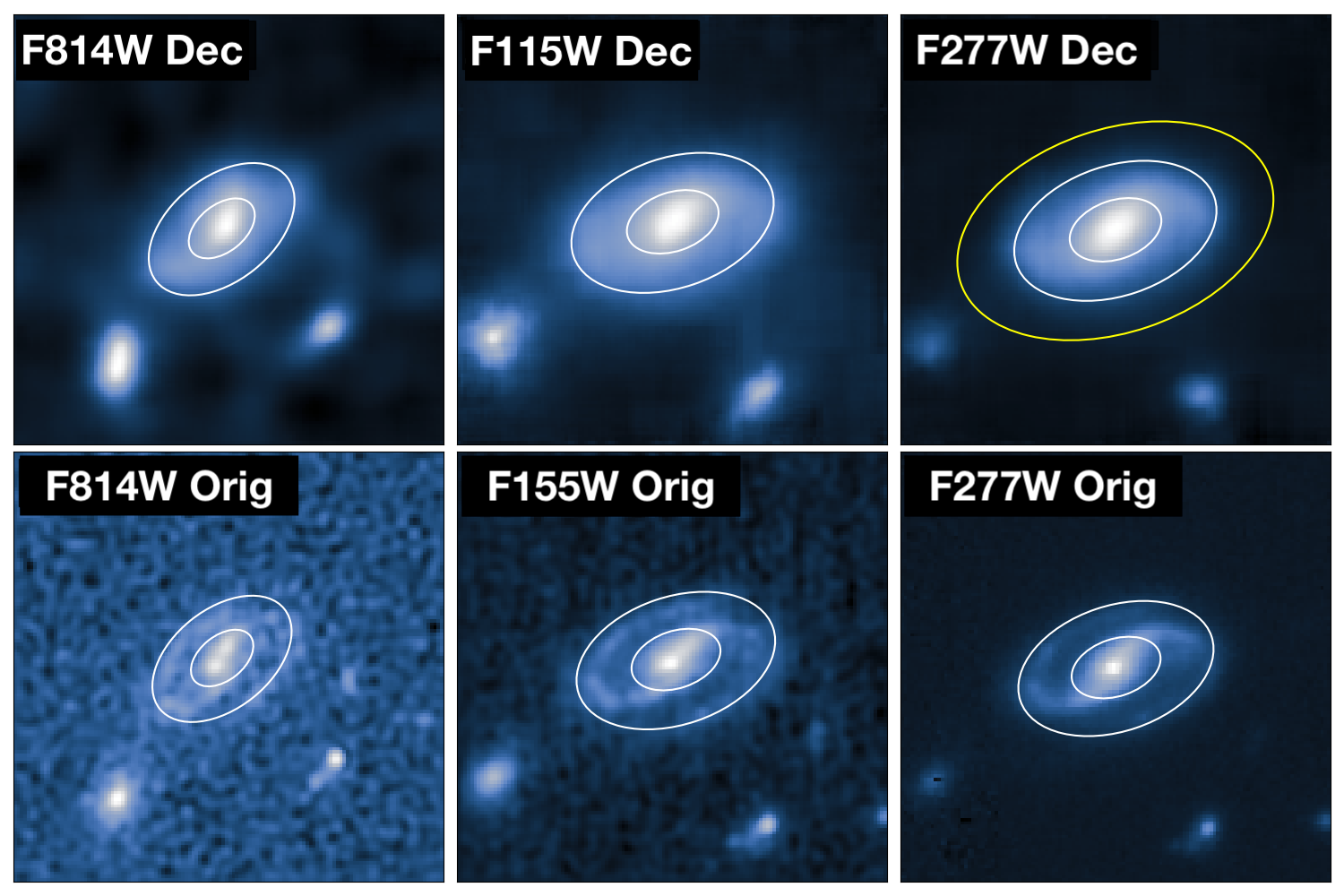}    
    \caption{Example of color measurements on a galaxy. Upper panels, from left to right: Cutouts of the galaxy after deconvolution in the F814W, F115W and F277W bands. Lower panels: Original cutouts after PSF-matching. The white ellipses correspond to the inner and outer apertures used to calculate the gradients. The yellow ellipse corresponds to the Kron aperture. The F814W cutout is angled (20$^{^\circ}$) compared to the two NIRCam cutouts.}
    \label{appfig:ex_test_eff}
\end{figure}

\begin{figure}[htb]
    \centering
    \includegraphics[width=0.99\linewidth]{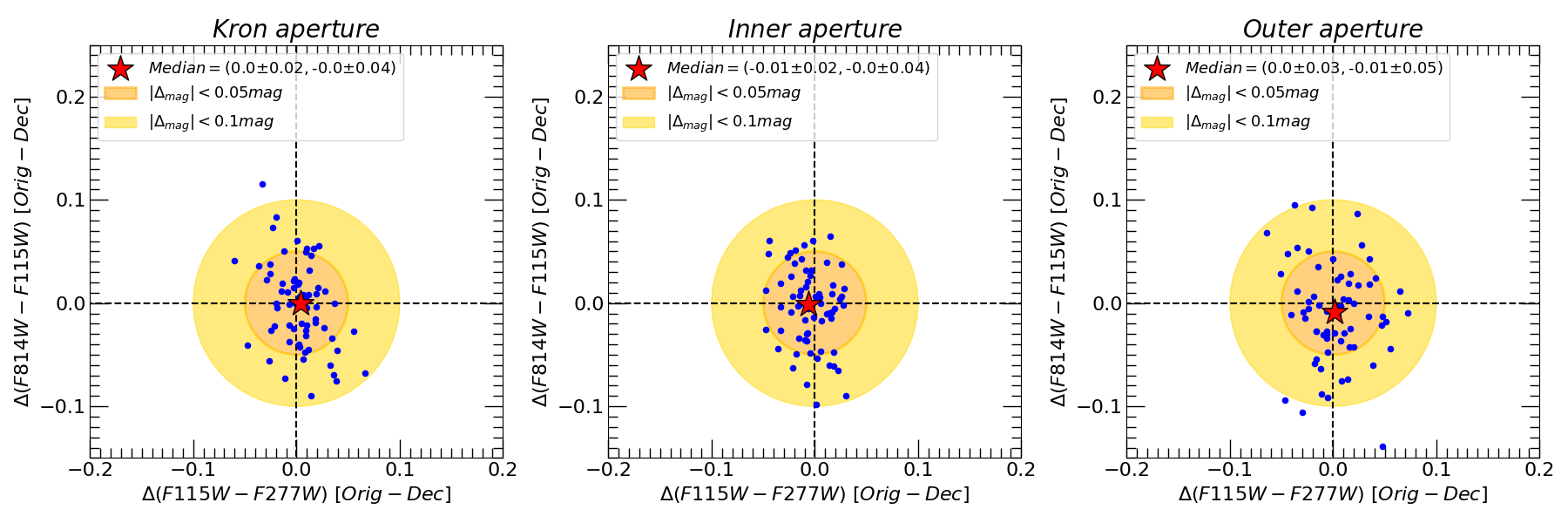}
    \caption{From left to right: Color recovery efficiency in the Kron, the inner and the outer apertures. Each panel shows how the (F814W $-$ F115W) and (F115W $-$ F277W) colors in the deconvolved images deviate from their "true" value measured in the original high-resolution images on the y and x axis respectively. We show the median value, the accompanying uncertainty is the standard deviation of the distribution along the corresponding axis.}
    \label{appfig:test_colors}
\end{figure}

\begin{figure}[htb]
    \centering
    \includegraphics[width=0.75\linewidth]{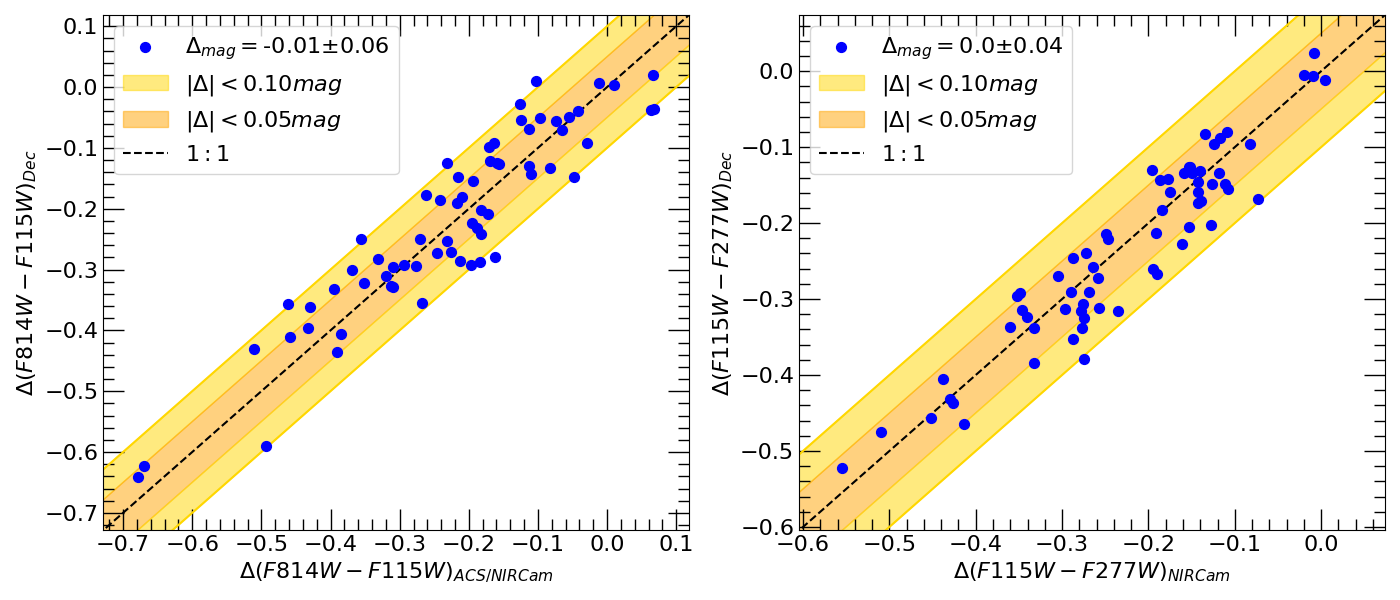}
    \caption{Gradient recovery efficiency along the (F814W $-$ F115W) axis (left panel) and the (F115W $-$ F277W) axis (right panel). Each panel shows the gradients measured in the original high resolution image on the x-axis and the gradients measured in the deconvolved images on the y-axis. The $\Delta_{mag}$ value corresponds to the median of the differences, the accompanying uncertainty is the standard deviation of the distribution of the differences.}
    \label{appfig:test_gradients}
\end{figure}

The last step was to perform the same color measurements as described in Section \ref{subsec:photo}: we measured the (F814W $-$ F115W) and (F115W $-$ F277W) colors in two elliptical apertures encompassing 50\% and 90\% of the Kron F277W flux (See Figure \ref{appfig:ex_test_eff} for an example). 
We used the same apertures when measuring the colors in the deconvolved images and in the original high-resolution images. We note that the original images were PSF-matched before performing the multi-band analysis.
When comparing the colors (see Figure \ref{appfig:test_colors}) and the gradients (see Figure \ref{appfig:test_gradients}) from the original images and from the simulated ground-based images, we found them to be in strong agreement, with the difference between the colors/gradients having a median close to zero and a standard deviation between 0.02 mag and 0.06 mag depending on the aperture and/or color/gradient.

Moreover, to take into account the fact that real ground-based images have a correlated noise, we added correlation in the noise when simulating our galaxies and the effects were negligible on the colors and gradients recovery efficiency.
Hence, with an average color recovery efficiency of 0.035 mag and an average gradient recovery efficiency of 0.05 mag, we can state that our deconvolution process is efficient at preserving and probing the color gradients of galaxies.

\end{document}